\documentclass[useAMS,usenatbib,minimal]{mn2e}

\usepackage{graphicx,amsmath,amssymb,natbib}
\usepackage{natbib}
\usepackage{enumerate}
\usepackage{url}
\newcommand {\reac}[6] {$\rm\,{}^{#2}\kern-0.8pt{#1}\,({#3}\,,{#4})
  \,{}^{#6}\kern-0.8pt{#5}\,$}
\title[The \texttt{COLIBRI} code for the TP-AGB evolution]
{Evolution of Thermally Pulsing Asymptotic Giant Branch Stars 
I. The \texttt{COLIBRI} Code}
\author[P. Marigo et al.]
{Paola Marigo$^{1}$\thanks{E-mail
paola.marigo@unipd.it}, Alessandro Bressan$^{2}$, Ambra Nanni$^{2}$, 
L\'eo Girardi$^{3}$, \newauthor
and Maria Letizia Pumo$^{1,3}$\\
$^{1}$Department of Physics and Astronomy G.\ Galilei, University of Padova,
        Vicolo dell'Osservatorio 3, I-35122 Padova, Italy\\
$^{2}$Astrophysics Sector, SISSA, Via Bonomea 265, I-34136 Trieste, Italy\\
$^{3}$Astronomical Observatory of Padova -- INAF,
            Vicolo dell'Osservatorio 5, I-35122 Padova, Italy}
\begin{document}

\date{Accepted 2013  xxx. Received 2013 January xxx; in original form 2013 February xxx}

\pagerange{\pageref{firstpage}--\pageref{lastpage}} \pubyear{2011}

\maketitle

\label{firstpage}

\begin{abstract}
We present the \texttt{COLIBRI} code for computing the 
evolution of stars along the TP-AGB phase.  Compared to purely synthetic TP-AGB
codes, \texttt{COLIBRI} relaxes a significant part of their analytic
formalism in favour of a detailed physics applied to a complete
envelope model, in which the stellar structure equations are integrated
from the atmosphere down to the bottom of the hydrogen-burning shell.
This allows to predict self-consistently: (i) the effective 
temperature, and more generally the {\em convective envelope and atmosphere
structures}, correctly coupled to the changes in the surface chemical
abundances and gas opacities; (ii) the conditions under which {\em sphericity effects} 
may significantly affect the atmospheres of giant stars; (iii) the {\em core mass-luminosity relation and
its possible break-down due to the occurrence of hot bottom burning}
(HBB) in the most massive AGB stars, by taking properly into account
the nuclear energy generation in the H-burning shell and in the
deepest layers of the convective envelope; (iv) the {\em HBB
nucleosynthesis} via the solution of a complete nuclear network
(including the pp chains, and the CNO, NeNa, MgAl cycles)
coupled to a diffusive description of mixing, suitable 
to follow also the synthesis of $^{7}$Li via 
the Cameron-Fowler beryllium transport mechanism; (v) the {\em intershell
abundances} left by each thermal pulse via the solution of a complete
nuclear network applied to a simple model of the pulse-driven
convective zone; (vi) the {\em onset and quenching of the third
dredge-up}, with a temperature criterion that is applied, at each
thermal pulse, to the result of envelope integrations at the stage of
the post-flash luminosity peak.

At the same time \texttt{COLIBRI} pioneers new techniques in the
treatment of the physics of stellar interiors, not yet adopted in full
TP-AGB models. It is the first evolutionary code ever to use accurate
{\em on-the-fly} computation of the {\em equation of
state} for roughly 800 atoms, ions, molecules, and of the Rosseland
mean {\em opacities} throughout the atmosphere and the deep
envelope. This ensures a complete consistency, step by step, of both
EoS and opacity with the evolution of the chemical abundances caused
by the third dredge-up and HBB. Another distinguishing aspect of 
\texttt{COLIBRI} is its high  computational speed, that allows
 to generate complete grids of TP-AGB
models in just a few hours. This feature  is absolutely
necessary for calibrating the many uncertain parameters and
processes that characterize the TP-AGB phase.

We illustrate the many unique features of \texttt{COLIBRI} by means of
detailed evolutionary tracks computed for several choices of model
parameters, including initial star masses, chemical abundances, nuclear
reaction rates, efficiency of the third dredge-up,
overshooting at the base of the pulse-driven convection zone, etc. 
Future papers in this series will deal with the calibration of all these
and other parameters  
using observational data of AGB stars in the Galaxy and 
in nearby systems, a step that is of paramount importance for producing reliable  
stellar population synthesis models of galaxies up to high redshift.
\end{abstract}


\section{Context and motivation}
The modelling of the Thermally Pulsing Asymptotic Giant Branch
(TP-AGB) stellar evolutionary phase plays a critical role in many
astrophysical issues, from the chemical composition of meteorites
belonging to the pre-solar nebula \citep[e.g.][]{Zinner_etal05}, 
up to the cosmological context of galaxy evolution in the high-redshift Universe 
\citep[e.g.][]{Maraston_etal06}. 
Indeed, luminous TP-AGB stars are potentially the dominant contribution to a galaxy's
flux, particularly at the red wavelengths and high redshifts that
are much of the focus of modern extragalactic astronomy. In spite of
its importance, the TP-AGB phase is still affected by large
uncertainties which uncomfortably propagate into the field of current
population synthesis models of galaxies that, for this reason, are strongly
debated \citep[e.g.][]{Conroy_etal09, Kriek_etal10, Zibetti_etal13}.

As a matter of fact, the evolution along TP-AGB phase 
is determined in a crucial way by processes which are challenging to 
model from first principles: turbulent convection, stellar
winds, and long-period variability. Also, these processes do not
take place in a steady and smooth way during the TP-AGB evolution,
but greatly vary in both character and efficiency over the single
thermal pulse cycles (TPC) -- the $10^2$ to $10^5$-yr long 
periods that go from
one He-shell flash, through quiescent H-shell burning, 
up to the next He-flash. 
Moreover, the rich nucleosynthesis in the
intershell convective region followed by recurrent dredge-up episodes, 
and the nuclear burning at the base of the convective envelope (hot-bottom burning, HBB) 
of the most massive TP-AGB stars ($M \ga 4\, M_{\odot}$), can
dramatically change the surface abundances, and hence the envelope
structure, over a timescale much  shorter than a single TPC.

The result is that the modelling of the TP-AGB phase is
quite difficult, time consuming, and affected 
by large uncertainties. Efforts to follow this phase
with ``full models'', which solve the time-dependent equations of
stellar structure with the aid of classical 1D stellar evolution
codes, are becoming increasingly successful thanks to the speeding-up of
modern processors, and to the particular care devoted to the 
nucleosynthesis \citep[e.g.][]{Ventura_etal02, Cristallo_etal09, Karakas_10}.
However, full TP-AGB models still meet three
fundamental difficulties. \\
(1) They are affected by quite subtle and nasty 
numerical uncertainties, that can greatly affect the predicted
efficiency of convective dredge-up episodes even within the same
set of models \citep{FrostLattanzio_96, Mowlavi_99a}.\\
(2) Full TP-AGB models need to resort to parametrized descriptions of crucial
processes (mass loss, convection, overshoot), with theoretical formulations
and ``efficiency parameters''  that may largely vary from study to study,
so that to date no universally accepted set of prescriptions  exists.
This intrigued situation is well exemplified by fact that, for instance, 
the so-called carbon-star mystery, pointed out by \citet{Iben_81} in the
far past, is now claimed to have been solved by full TP-AGB models
\citep{Stancliffe_etal05, WeissFerguson_09, Cristallo_etal11}.
However, it is somewhat disturbing to recognize that
the same observable, i.e. the carbon star luminosity 
function of carbon stars in the Large Magellanic Cloud, seems to be recovered
by different full TP-AGB models in which the third dredge-up takes
place with very different characteristics (in this respect, 
see Sect.~\ref{ssec_3dup} and Fig.~\ref{fig_3dup3z02}).\\
(3) The range of parameters to be covered, and
prescriptions to be tested, in order to obtain grids of TP-AGB models
that reproduce the wide variety of observational data for AGB stars in
resolved galaxies, is simply too large.

In this tricky context, a valuable contribution 
may be provided by the so-called
``synthetic models", in which the evolution from one thermal pulse 
to the next is described with analytical relations 
that synthesize the results of full models. 
Being very agile and 
hence suitable to explore wide ranges of parameters and prescriptions,
synthetic models can help to constrain the physical domain towards which full
models should converge in order to reproduce observations 
of TP-AGB stars (e.g. carbon star luminosity functions (CSLF), C/M ratios,
H-R diagrams, etc.). For instance, following  
the work of \citet{GroenewegendeJong_93}, 
based on synthetic models and focussed 
on the CSLF in the Large Magellanic Cloud, 
it became clear that the third dredge-up should not only be 
much more efficient, but also
start earlier, at fainter luminosities, than usually predicted by full 
TP-AGB models up to that time.

On the other hand,
synthetic models are
often criticised because they lack the accurate physics involved in
the evolution of these stars. Moreover, they are completely subordinate 
to the relations fitting the results of full AGB model calculations, 
which  severely limits their capability of exploring new evolutionary effects.
A notable example is the effective temperature, for which various formulas
have been proposed in the past in the usual form 
$T_{\rm eff}={\rm func}(L,M,Z)$, involving luminosity, stellar mass and 
metallicity.  Unfortunately, their validity is extremely narrow as they can 
apply only to oxygen-rich stars  (with surface C/O$<1$), 
hence being unable to account for the Hayashi limits of carbon stars. 
Moreover, these relations reflect the specific set of input physics adopted 
in the underlying full models, e.g. mixing-length parameter, gas opacities, 
equation of state, etc. 

If this criticism reasonably applies to the purely analytic TP-AGB models 
that rely on a mere compilation of fitting formulas 
\citep[e.g.][]{Hurley_etal00, Izzard_etal04, Izzard_etal06, 
Cordier_etal07},  it is not as well suited to the class of hybrid models 
\citep[e.g.][]{Marigo_etal96, Marigo_etal98, Marigo_etal99, Marigo_07, 
MarigoGirardi_07}, in which the analytic formalism 
is complemented with numerical integrations of the stellar
structure equations, carried out from the atmosphere down to the bottom
of the convective envelope.  In the latter case both the HBB nucleosynthesis 
and the basic changes in envelope structure -- including effective temperature and radius -- 
can be followed with the same richness of detail as in full models, but still
in a much quicker and more versatile way. 

It is not by accident that the crucial role of the surface C/O ratio and C-rich opacities in
determining the evolution of TP-AGB stars was established just with
the aid of these ``envelope-based models''  
\citep{Marigo_02, Marigo_07, Marigo_etal03, MarigoGirardi_07}. 
Although the same effect could have been assessed 
with the aid of full models,  the latter were fighting with so many numerical 
and physical difficulties related to the occurrence of the third dredge-up, 
that the key aspect of the C-rich opacities was  ignored, and likely forgotten, 
for long time in the field of AGB stellar evolution.
Since \citet{Marigo_02}, molecular opacities for C-rich mixtures
have been progressively adopted in full TP-AGB models 
\citep[e.g.][]{Kamath_etal12, VenturaMarigo_10, VenturaMarigo_09, 
WeissFerguson_09,  Cristallo_etal07}.

This example tells clearly that progresses in the description of the TP-AGB
phase do not rely only on full models, but they can come also from
other complementary approaches.

With this work  we go a few steps ahead in the development
of our ``envelope-based TP-AGB models''. 
We describe a code, called \texttt{COLIBRI}, that implements
a number of improvements which, effectively, make our models to perform much 
more like "almost-full" models than "improved synthetic" ones.
Among the most relevant points we mention: 
i) a spherically-symmetric deep envelope model extending from the atmosphere down 
to the bottom of the quiescent H-burning shell, so that the classical 
core-mass luminosity relation (CMLR)
is naturally predicted and not taken as an input prescription; ii) the first ever on-the-fly 
accurate calculation of molecular chemistry and  Rosseland mean opacities, fully 
consistent with the changing surface abundances, iii) a detailed HBB nucleosynthesis coupled 
with a diffusive description of convection, iv) a model for the pulse-driven convection 
zone to predict the chemical composition of the dredged-up material, and v) 
improved prescriptions to determine the onset and quenching of the third dredge-up.

Of course, in the development of the \texttt{COLIBRI} code 
full TP-AGB models still play a paramount role:
they are taken as a reference to check the accuracy of some basic predictions, 
and they are used to derive  quantitative information, via fitting relations, 
on those aspects  that the \texttt{COLIBRI} code cannot, by construction, 
address by itself like, for example, the evolution of the intershell 
convection zone during thermal pulses.  

In any case, all these aspects are treated fulfilling two extremely important conditions: 
a robust numerical stability which allows to follow the TP-AGB evolution until the 
complete ejection of the envelope, and a high computational speed which is 
kept comparable to the levels that made the success of the very first synthetic TP-AGB
models.  In this way the \texttt{COLIBRI} code is a tool perfectly suitable  
to perform a multi-parametric, but still accurate, 
calibration of the TP-AGB phase, our final goal. 

The plan of the paper is as follows. Section~\ref{sec_outline}
presents an outline of the \texttt{COLIBRI} code.
Section~\ref{sec_physmod} describes in detail all input physics and
the solution methods adopted to integrate the deep envelope model, and
to predict the nucleosynthesis in the pulse-driven convective zone and
during HBB.  Section \ref{sec_synthmod} summarises the analytic
ingredients of \texttt{COLIBRI}. Accuracy tests of \texttt{COLIBRI}
predictions against full stellar models are discussed in
Sect.~\ref{sec_tests}.  The present sets of TP-AGB evolutionary tracks
are introduced in Sect.~\ref{sec_tracks}, while the whole
Sect.~\ref{sec_results} is dedicated to illustrate several examples of
possible \texttt{COLIBRI} calculations.  Finally,
Sect.~\ref{sec_finalsum} closes the paper giving  
a r\'esum\'e of \texttt{COLIBRI}'s features, and briefly mentioning
current and planned applications.

\section{Overview of the COLIBRI code}
\label{sec_outline}

The \texttt{COLIBRI} code computes the TP-AGB evolution 
from the first thermal pulse up to the complete ejection of the stellar
mantle by stellar winds.
While maintaining a few basic features of our original TP-AGB model 
developed  and revised over the years 
\citep{Marigo_etal96, Marigo_etal98, Marigo_98, Marigo_etal99, 
MarigoGirardi_07}, we have introduced substantial improvements  that notably
enhance the predictive power of our TP-AGB calculations.
The main variables of the TP-AGB model, which are also frequently cited 
in the text, are operatively defined in Table~\ref{tab_mod}.

\texttt{COLIBRI} consists of three main components, that we conveniently
refer to as 1) the {\em physics module}, 2) the {\em synthetic module},
and 3) the {\em parameter box}.

The {\em physics module} involves all detailed input physics 
(equation of state,
opacities, nuclear reactions rates) and differential 
equations necessary to numerically integrate  
a stationary {\em deep envelope model}, 
extending from the atmosphere down to the 
bottom of the H-burning shell (see Sect.~\ref{sec_physmod}).
At each time step, 
the run of mass $M_r$, temperature $T_r$, pressure $P_r$, and 
luminosity $L_r$ is determined across the deep envelope during the quiescent interpulse
periods.
By adopting proper boundary conditions at the bottom of the convective envelope, we
obtain the effective temperature, and the luminosity provided by the hydrogen burning shell. 
In this way we are able to follow consistently the occurrence of HBB
in the most massive AGB stars, being responsible for the break-down of the 
CMLR (see Sect.~\ref{ssec_lum}), as well as a significant 
nucleosynthesis  (see Sect.~\ref{ssec_HBBnuc}).

The {\em synthetic module} contains the analytic formalism of the code, 
which includes
both fitting formulas that synthesize the results of full AGB models 
(e.g. the core mass-interpulse period relation, the core mass-intershell mass 
relation, the efficiency of the third dredge-up as a function of 
stellar mass and metallicity, etc.), 
and other auxiliary relations   (e.g. mass-loss prescription, 
period-mass-radius relations for variable AGB stars, etc.).
It is outlined in Sect.~\ref{sec_synthmod}.

The {\em parameters box} collects all free parameters that 
we think need to be calibrated (e.g. minimum base temperature for the
occurrence of the third dredge-up,
efficiency of mass loss, dependence on mass and metallicity, overshoot at the
base of the convective envelope) in order to reproduce basic observables.
Since a fine calibration of the  TP-AGB phase is not the primary  purpose of this paper,
the results presented here are obtained with a particular set of parameters,
as specified in Sect.~\ref{ssec_tpagbev}.
\begin{figure*} 
\begin{minipage}{0.49\textwidth}
\resizebox{\hsize}{!}{\includegraphics{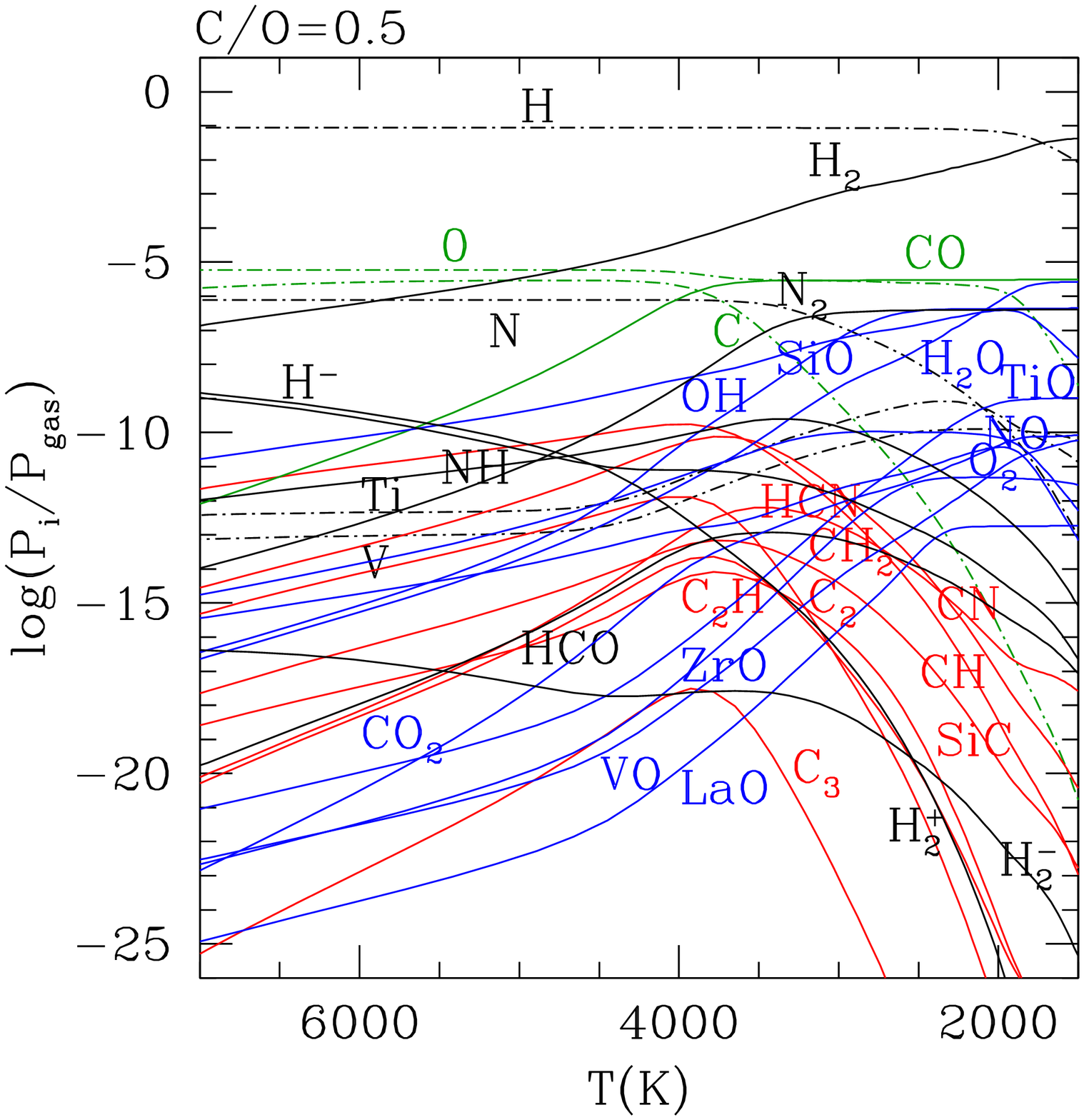}}
\end{minipage}
\begin{minipage}{0.49\textwidth}
\resizebox{\hsize}{!}{\includegraphics{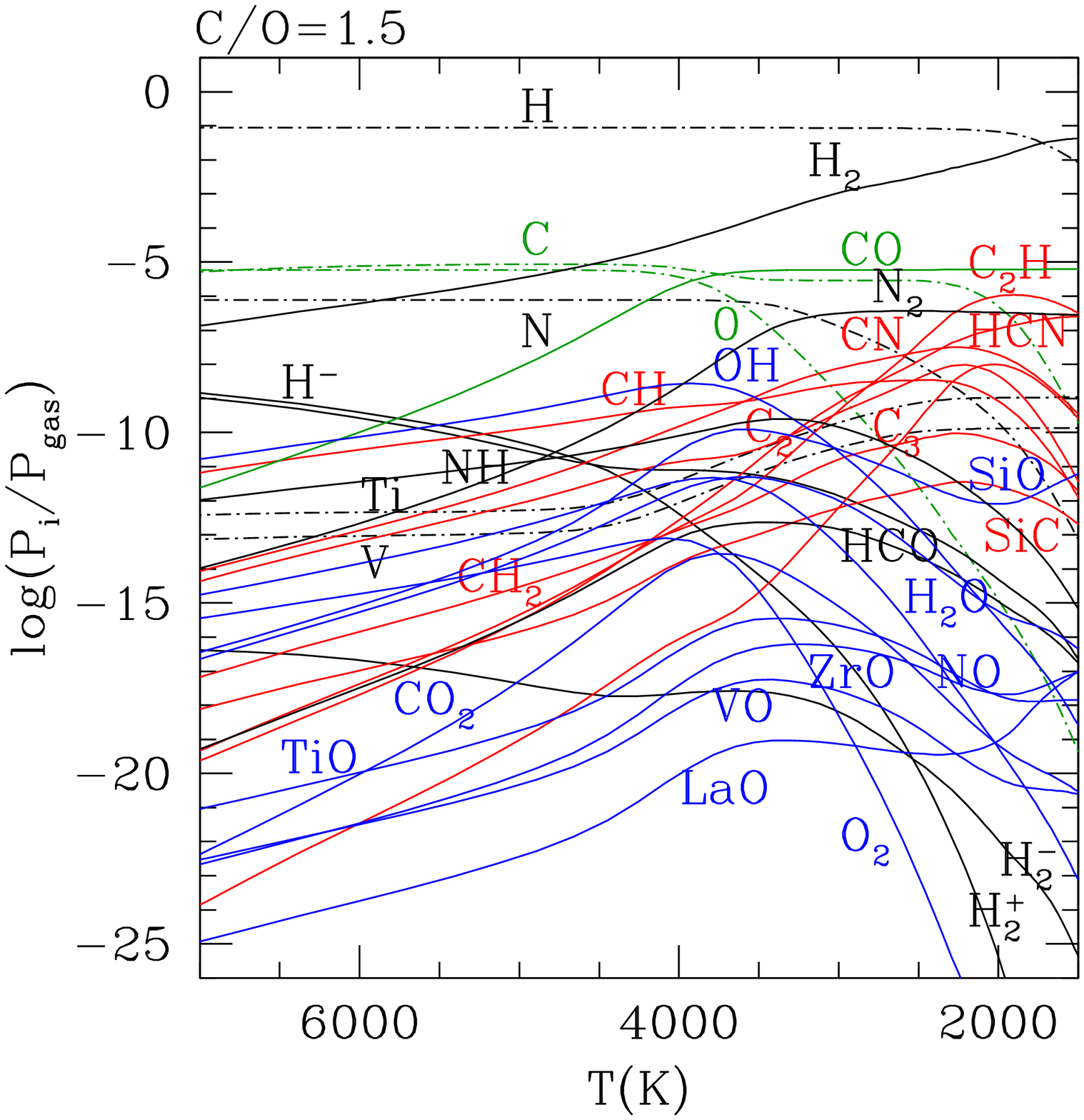}}
\end{minipage}
\caption{Partial pressures of a subset of atomic and molecular species computed 
with the \texttt{\AE SOPUS} code \citep{MarigoAringer_09} according to the temperature-pressure stratification of a complete envelope-atmosphere model with $\log (T_{\rm eff})=3.45$, 
$\log(L/L_{\odot})=3.7$, $M_{\rm i}=2M_{\odot}$, and solar metallicity  
$Z_{\rm i}=Z_{\odot}\!\simeq\!0.0152$ 
following the revision by \citet{Caffau_etal11}.
Two values of the C/O ratio have been considered, i.e. C/O$=0.5$ (left-hand side panel) and
C/O$=1.5$ (left-hand side panel). Note the abrupt change in the molecular equilibria of the O-bearing (blue) and C-bearing (red) molecules between the two cases, as well as the almost 
invariance of the abundance of the highly stable CO molecule.}
\label{fig_molec}
\end{figure*}

These three components clearly represent a sequence of decreasing
accuracy, and increasing uncertainty. While for most ingredients of
the physics module we rely on detailed and well-established
prescriptions, in the synthetic module we have to resort
to the results of various sets of full TP-AGB models in the literature 
that share a general agreement, but present also unavodaible 
differences due to specific model details.
The parameter box, instead, hides
a big deal of our ignorance about basic physical processes in AGB
stars. The coupling of these components, with very different degrees of
accuracy, is inescapable at this point. The situation resembles the
one that persists in practically all full stellar evolutionary codes
to date, in which rough descriptions for convective processes -- such
as the mixing length theory and overshooting -- are routinely adopted,
and anyhow being able to produce very useful results. 
Although we all know that ``fake physics'' is being used to some extent
in all these codes, it is also a matter of fact that, at some stages, 
these approximations have opened the way for advancing the theory 
of stellar evolution on other fronts.  
Our wish is that the same strategy can turn out to be useful
also for the TP-AGB phase.

\section{The physics module}
\label{sec_physmod}
\subsection{Equation of state}
\label{ssec_eos}
The equation of state (EoS) for temperatures  in the interval 
from $5\times 10^4$~K to $10^8$~K  is 
that of a fully-ionized gas, in the way 
described by \citet{Girardi_etal00}. 

For temperatures in the range from $5\times 10^4$~K to $10^3$~K 
all relevant thermodynamic quantities and their partial
derivatives (mass density, electron density,
mean molecular weight, entropy, specific heats, etc.)
are computed {\em on-the-fly}
with the \texttt{\AE SOPUS} code \citep{MarigoAringer_09}.
We briefly recall that \texttt{\AE SOPUS} solves the EoS for atoms and molecules in the gas
phase, under the assumption of an ideal gas in both 
thermodynamic equilibrium and instantaneous chemical equilibrium.
We consider the ionisation
stages from I to V  for all elements from C to
Ni (up to VI for O and Ne), and from I to III for heavier atoms
from  Cu to U. 
Saha equations for ionisation and dissociation 
are solved for $\approx 800$
species, including $\approx 300 $ atoms (neutral and ionised) 
from H to U,  and $\approx 500$ molecules.

An example of the EoS calculations across the outermost layers 
of a TP-AGB model is given in Fig.~\ref{fig_molec}, that also illustrates
the dramatic change in the equilibrium molecular chemistry 
as the surface C/O ratio passes from ${\rm C/O} < 1$, typical of M stars, 
to ${\rm C/O} > 1$, characteristic of C stars.

\subsection{Gas opacities}
\label{ssec_opac}
\begin{figure*} 
\begin{minipage}{0.49\textwidth}
\resizebox{\hsize}{!}{\includegraphics{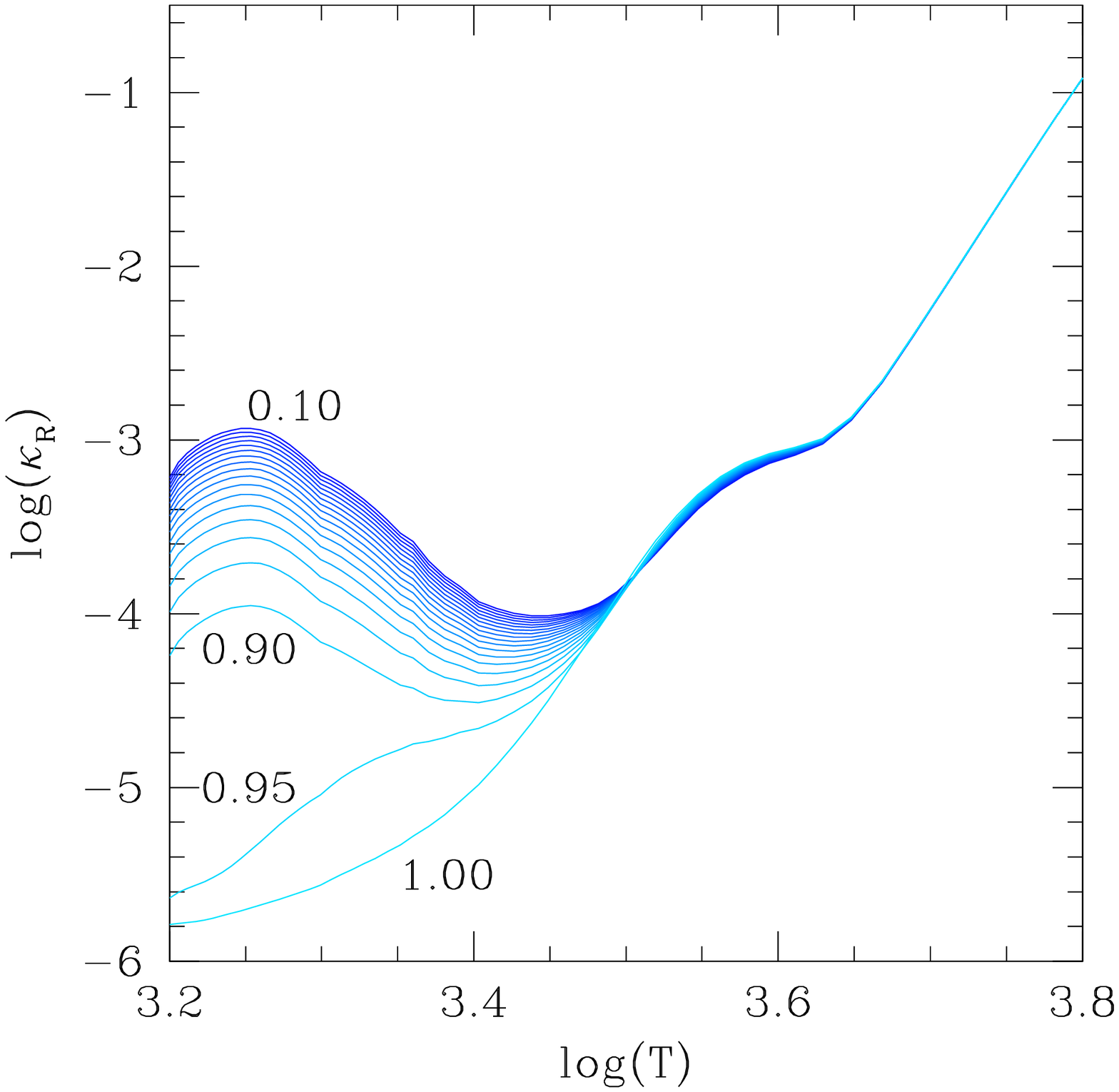}}
\end{minipage}
\begin{minipage}{0.49\textwidth}
\resizebox{\hsize}{!}{\includegraphics{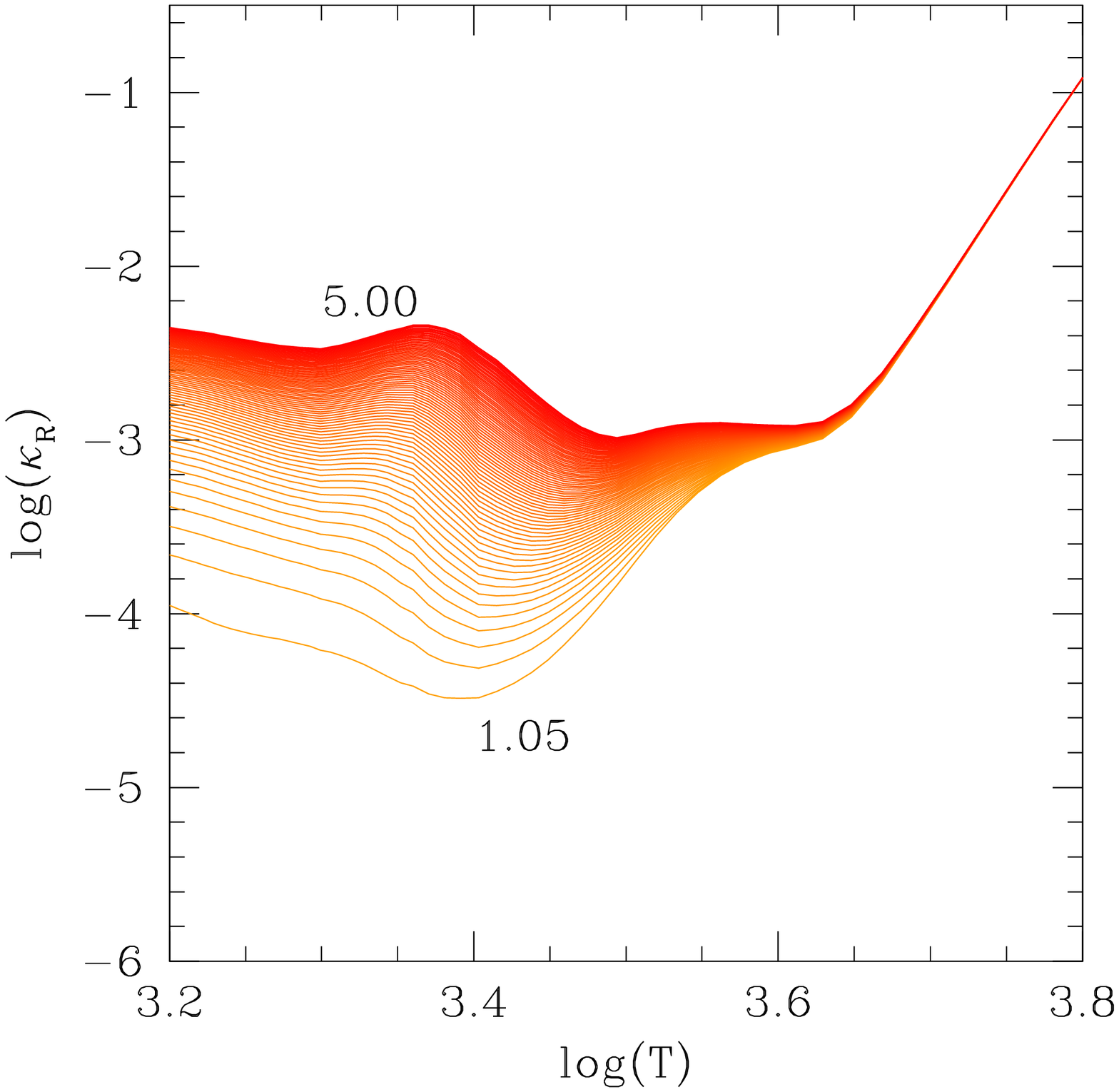}}
\end{minipage}
\caption{Rosseland mean opacities, computed  with the \texttt{\AE SOPUS} code 
\citep{MarigoAringer_09}, according to the temperature-pressure stratification 
of a complete envelope-atmosphere model with 
$\log (T_{\rm eff})=3.45$, $\log(L/L_{\odot})=3.7$, $M_{\rm i}=2M_{\odot}$, 
and solar metallicity  $Z_{\rm i}=Z_{\odot}\!\simeq\!0.0152$ according to \citet{Caffau_etal11}.
The C/O ratio is made to increase from 0.10 to 1.00 
(left-hand side panel), and from 1.05 to 5.00 (right-hand side panel) in steps of 0.05.  }
\label{fig_opac}
\end{figure*}
Rosseland mean gas opacities, in the whole temperature range 
$8.0 \le \log T \le 3.2$, are computed {\em on-the-fly}, i.e. contemporary
with the atmospheric and envelope integrations that constitute the 
kernel of our TP-AGB code.

We remark that this is the {\em first time ever} that  
accurate opacities are computed on-the-fly, just 
starting from the monochromatic absorption coefficients of 
the opacity sources, without interpolation in pre-exiting tables
of Rosseland mean opacities.

This choice is motivated by the demand of  
accurately describing the tight coupling 
of the opacity sources (mainly in the molecular
regime) with the frequent and significant changes in the envelope
chemical composition that characterise the TP-AGB phase.
In this way we avoid the loss in accuracy that one must otherwise pay when 
performing multi-dimensional interpolation.  

To this aim we have constructed a routine which, 
for any given set of chemical abundances 
of $92$ elements from H to U,  and a specified pair 
of state variables (e.g. gas pressure $P_{\rm g}$ and temperature $T$), 
makes direct calls to one of two opacity codes, depending on the temperature:
\begin{itemize}
\item The {\em Opacity Project}\footnote{We have used the OPCD$\_$3.3 open-source package available 
at the WEB page \url{http://cdsweb.u-strasbg.fr/topbase/op.html}}
(OP) \citep[OP;][]{Seaton05, Badnell_etal05}
for  $4.2 < \log T \le  8.0$;
\item The \texttt{\AE SOPUS}\footnote{The \texttt{\AE SOPUS} tool is accessible via the web interface 
at \url{http://stev.oapd.inaf.it/aesopus}} code \citep[][]{MarigoAringer_09} 
for  $3.2 \le \log T \le  4.2$.
\end{itemize}

The OP data provides the monochromatic opacities for several atoms
(H, He, C, N, O, Na, Mg, Al, Si, S, Ar, Ca, Cr, Mn, Fe, Ni) over a wide
range of values of temperature $T$ and electron density $N_{\rm e}$. 
We have employed the routines {\em mixv.f} and {\em opfit.f} 
to calculate the Rosseland
mean opacities on a pre-determined grid of OP$(T,\,N_{\rm e})$ meshes and 
then to interpolate to any specified values of $T$ and $\rho$.
Since the original OP version assumes a fixed mixture of elements
(i.e. scaled-solar chemical composition), 
we have suitably modified the OP routines
to compute the Rosseland mean for any chemical composition involving the 
$16$ species for which the OP monochromatic opacities are available.
This is an important improvement compared to the common practice in which 
the chemical parameters (besides the H or He abundances) 
are limited to few metal abundances.
For instance, the widely-used OPAL web tool \citep{RogersIglesias_96} allows 
the on-line computation and provides the interpolating
routines of Rosseland mean opacity tables with a fixed partition of metals, 
but for the abundances of two species (e.g. C and O), 
which are enhanced according to a specified grid of values. 
We notice that in this case, the possible depletion of a metal, 
due for instance to nuclear burning, cannot be considered.
At variance, the OP utility gives us an important flexibility in this
respect.

Suitably converted into
an internal routine of our \texttt{COLIBRI} code, for each pair of 
$P_{\rm g}$ and $T$,  \texttt{\AE SOPUS} calculates 
the monochromatic true absorption and scattering cross sections
due to a number of continuum and discrete processes, i.e.
bound-free absorption due to photoionisation, free-free absorption,
Rayleigh and Thomson scattering, collision-induced absorption, 
atomic bound-bound absorption and molecular absorption.
We note that the monochromatic cross sections for atoms 
(C, N, O, Na, Mg, Al, Si, S, Ar, Ca, Cr, Mn, Fe, Ni) are taken
from the OP database, thus assuring a complete consistency with the 
high-temperature opacities.
Then, after summing up all contributions,
the Rosseland mean (RM) opacity is computed.

The incorporation of \texttt{\AE SOPUS} in the \texttt{COLIBRI} code
allows us to follow accurately the changes in molecular opacities driven
by any variation of the  envelope composition, especially by the C/O ratio 
which plays the key role in determining the molecular chemistry 
\citep[see e.g.][]{MarigoAringer_09}.
The complex behaviour of the  RM opacities 
as a function of the C/O ratio is exemplified with the aid of
Fig.~\ref{fig_opac}. It turns out that  while 
the C/O ratio increases from $0.1$ to $0.9$ the opacity bump peaking 
at ($\log(T)\simeq 3.25-3.35$) -- mostly due to H$_2$O -- 
becomes more and more depressed because of the smaller 
availability of O atoms.
Then, passing from C/O~$=0.9$ up to  C/O~$=0.95$ the H$_2$O feature
actually disappears and $\kappa_{\rm R}$ drastically drops by more almost
 two orders of magnitude.
In fact, at this C/O value the chemistry enters in a transition region where 
most of both O and C atoms are trapped in the very stable CO
molecule at the expense of the other  molecular
species, belonging to both the O- and  C-bearing groups. 
 At C/O~$=1$ the RM opacity reaches its minimum throughout the
temperature range,
$3.2 \la \log(T) \la 3.4$, while a sudden upturn is
expected as soon as C/O slightly exceeds unity, as displayed by the
curve for C/O~$=1.05$ of Fig.~\ref{fig_opac} (right panel). This fact reflects
the drastic change in the molecular equilibria from the O- to the
C-dominated regime.
Then, at increasing C/O the opacity curves
move upward following a more gradual trend, which is related to the
strengthening of the C-bearing molecular absorption bands.

Note, however,  that the C-rich opacity does not rise 
linearly with C/O, 
but less and less steeply as the C/O ratio increases. 
This is mainly due to the underlying equilibrium chemistry of the most 
efficient absorbers, 
in particular of the CN and  HCN 
molecules, whose abundances are conditioned not only by 
the carbon excess (C-O), but also by the availability of the N atoms 
(having a fixed abundance in the case under consideration).
As we will see in Sect.~\ref{ssec_hayashi}, 
the non-linear dependence of the opacity on the C/O ratio 
impacts on the maximum extension  of the Hayashi lines 
for C stars towards lower effective temperatures.

\subsection{Nuclear reactions}
\label{ssec_nrat}

Our nuclear network consists of the p-p chains, the CNO tri-cycle, and the Ne-Na, Mg-Al chains, and
the most important $\alpha$-capture reactions, 
including explicitly $N_{\rm el} = 25$ chemical species:
$^1$H, $^2$H, $^3$He, $^4$He, $^7$Li, $^7$Be,
$^{12}$C, $^{13}$C, $^{14}$N, $^{15}$N, $^{16}$O, $^{17}$O, $^{18}$O,
$^{19}$F, $^{20}$Ne, $^{21}$Ne, $^{22}$Ne, $^{23}$Na, $^{24}$Mg, $^{25}$Mg, $^{26}$Mg,
$^{26}$Al$^m$,  $^{26}$Al$^g$, $^{27}$Al, $^{28}$Si. The latter nucleus acts as the ``exit element'', which
terminates the network.
In total we consider $42$ reaction rates, listed in Tab.~\ref{tab_rates}. 
For all of them we adopt analytic relations, with fitting coefficients taken from 
the JINA reaclib database \citep{Cyburt_etal10}.
 The alternative of using  detailed tables of reaction rates 
as a function of the temperature 
can be easily implemented in \texttt{COLIBRI}, and may
be done in future studies dedicated to nucleosynthesis calculations.

\begin{table}
 \centering
  \caption{Nuclear reaction rates adopted in this work.}
  \begin{tabular}{@{}ll@{}}
  \hline
\multicolumn{1}{c}{Reaction} & 
\multicolumn{1}{c}{Source} \\
\hline
\reac{p}{}{p}{\beta^+\,\nu}{D}{} & \citet{Cyburt_etal10} \\
\reac{p}{}{D}{\gamma}{He}{3} &  \citet{Descouvemont_etal04}\\
\reac{He}{3}{^{3}He}{\gamma}{2\,p + ^{4}\kern-0.8pt{He}}{} &  \citet{Angulo_99}\\
\reac{He}{4}{^{3}He}{\gamma}{Be}{7} &  \citet{Descouvemont_etal04}\\
\reac{Be}{7}{e^-}{\gamma}{Li}{7} &  \citet{CaughlanFowler_88}\\
\reac{Li}{7}{p}{\gamma}{^{4}\kern-2.0pt{He} + ^{4}\kern-2.0pt{He}}{} &  \citet{Descouvemont_etal04} \\
\reac{Be}{7}{p}{\gamma}{B}{8} & \citet{Angulo_99}  \\ 
\reac{C}{12}{p}{\gamma}{N}{13} & \citet{Angulo_99}  \\ 
\reac{C}{13}{p}{\gamma}{N}{14} & \citet{Angulo_99}  \\ 
\reac{N}{14}{p}{\gamma}{O}{15} & \citet{Imbriani_etal05}  \\ 
\reac{N}{15}{p}{\gamma}{^4He + ^{12}\kern-2.0pt{C}}{} & \citet{Angulo_99}  \\ 
\reac{N}{15}{p}{\gamma}{O}{16} & \citet{Angulo_99}  \\ 
\reac{O}{16}{p}{\gamma}{F}{17} & \citet{Angulo_99}  \\ 
\reac{O}{17}{p}{\gamma}{\,^4He + ^{14}\kern-2.0pt{N}}{} & \citet{Chafa_etal07}  \\ 
\reac{O}{17}{p}{\gamma}{F}{18} & \citet{Chafa_etal07}  \\ 
\reac{O}{18}{p}{\gamma}{\,^4He + ^{15}\kern-2.0pt{N}}{} & \citet{Angulo_99}  \\ 
\reac{O}{18}{p}{\gamma}{F}{19} & \citet{Angulo_99}  \\ 
\reac{F}{19}{p}{\gamma}{\,^4He + ^{16}\kern-2.0pt{O}}{} & \citet{Angulo_99}  \\ 
\reac{F}{19}{p}{\gamma}{Ne}{20} & \citet{Angulo_99}  \\ 
\reac{Ne}{20}{p}{\gamma}{Na}{21} & \citet{Angulo_99}  \\ 
\reac{Ne}{21}{p}{\gamma}{Na}{22} & \citet{Iliadis_etal01}  \\ 
\reac{Ne}{22}{p}{\gamma}{Na}{23} & \citet{Hale_etal02}  \\ 
\reac{Na}{23}{p}{\gamma}{\,^4He + ^{20}\kern-2.0pt{Ne}}{} & \citet{Hale_etal04}  \\ 
\reac{Na}{23}{p}{\gamma}{Mg}{24} & \citet{Hale_etal04}  \\ 
\reac{Mg}{24}{p}{\gamma}{Al}{25} & \citet{Iliadis_etal01}  \\ 
\reac{Mg}{25}{p}{\gamma}{Al^g}{26} & \citet{Iliadis_etal01}  \\ 
\reac{Mg}{25}{p}{\gamma}{Al^m}{26} & \citet{Iliadis_etal01}  \\ 
\reac{Mg}{26}{p}{\gamma}{Al}{27} & \citet{Iliadis_etal01}  \\ 
\reac{Al^g}{26}{p}{\gamma}{Si}{27} & \citet{Iliadis_etal01}  \\ 
\reac{Al}{27}{p}{\gamma}{\,^4He + ^{24}\kern-2.0pt{Mg}}{} & \citet{Iliadis_etal01}  \\ 
\reac{Al}{27}{p}{\gamma}{Si}{28} & \citet{Iliadis_etal01}  \\ 
\reac{He}{4}{2\,^{4}He}{\gamma}{C}{12} & \citet{Fynbo_etal05}  \\ 
\reac{C}{12}{^{4}He}{\gamma}{O}{16} & \citet{Buchmann_96}  \\ 
\reac{N}{14}{^{4}He}{\gamma}{F}{18} & \citet{Gorres_etal00}  \\ 
\reac{N}{15}{^{4}He}{\gamma}{F}{19} & \citet{Wilmes_etal02}  \\ 
\reac{O}{16}{^{4}He}{\gamma}{Ne}{20} & \citet{Angulo_99}  \\ 
\reac{O}{18}{^{4}He}{\gamma}{Ne}{22} & \citet{Dababneh_etal03}  \\ 
\reac{Ne}{20}{^{4}He}{\gamma}{Mg}{24} & \citet{Angulo_99}  \\ 
\reac{Ne}{22}{^{4}He}{\gamma}{Mg}{26} & \citet{Angulo_99}  \\ 
\reac{Mg}{24}{^{4}He}{\gamma}{Si}{28} & \citet{CaughlanFowler_88}  \\ 
\reac{C}{13}{^{4}He}{n}{O}{16} & \citet{Angulo_99}  \\ 
\reac{O}{17}{^{4}He}{n}{Ne}{20} & \citet{Angulo_99}  \\ 
\reac{O}{18}{^{4}He}{n}{Ne}{21} & \citet{Angulo_99}  \\ 
\reac{Ne}{21}{^{4}He}{n}{Mg}{24} & \citet{Angulo_99}  \\ 
\reac{Ne}{22}{^{4}He}{n}{Mg}{25} & \citet{Angulo_99}  \\ 
\reac{Mg}{25}{^{4}He}{n}{Si}{28} & \citet{Angulo_99}  \\ 
\hline
\end{tabular}
\label{tab_rates}
\end{table}

\subsection{The atmosphere model}
\label{ssec_atmo}
For given chemical composition of the gas, an atmosphere model  is generally 
specified by three stellar parameters, e.g:
total mass $M$, luminosity $L$, and radius $R$.
The effective temperature derives from the Stefan-Boltzmann law
$L=4\pi R^2 \sigma T_{\rm eff}^4$. 
In our TP-AGB code the atmospheric structure can be 
obtained  by choosing among two different options, namely:
i) static plane-parallel atmosphere, and ii) static spherically symmetric
atmosphere.
\subsubsection{Plane-parallel atmospheres}
\label{sssec_ppatmo}
The plane-parallel grey atmosphere model is described by a 
temperature stratification given
by a modified Eddington approximation for radiative transport:
\begin{equation}
T^{4} = \frac{3}{4} T_{\rm eff}^{4} \left [  \tau +q\left ( \tau  \right ) \right ]
\label{eq_ttau}
\end{equation}
where $\tau(r)$ is the optical depth defined by the differential equation
\begin{equation}
d\tau = - \kappa \rho dr
\label{eq_tau}
\end{equation}
with the boundary condition $\tau(+\infty)=0$. 
Here $\kappa$ is the opacity which is usually described by 
the Rosseland mean, and $\rho$ is the mass density.
The quantity $q(\tau)$ in the right-hand side of Eq.~(\ref{eq_ttau}) is the Hopf function. 

Under the plane-parallel assumption the variations across the atmospheres 
of mass, radius, and luminosity can be neglected so that we have
\begin{equation}
\nonumber
M_r \approx M,\; \; \; \; \; \; r \approx R,\; \; \; \; \; \; L_r \approx L.
\end{equation}

Let us denote with $\tilde{\tau}$ the optical depth of the photosphere 
(approximately $2/3$), and $r_{\tilde{\tau}}$  its
radial coordinate.
In the plane-parallel approximation, it defines the radius
of the star, i.e. $R=r_{\tilde{\tau}}$, and the corresponding temperature 
$T_{\tilde{\tau}}$ coincides with the effective temperature $T_{\rm eff}$, 
defined by the Stefan-Boltzmann law 
$T_{\textup{eff}} =  \left (L/4\pi \sigma R^2 \right )^{1/4}$.

Combining the equations of mass continuity, hydrostatic equilibrium
and Eq.~(\ref{eq_tau}), we obtain the atmospheric equation for the total pressure 
\begin{equation}
\frac{d\tau }{d P} = \frac{\kappa R^2}{G M}
\label{eq_dtaudp}
\end{equation}
where $P= P_{\rm gas} + P_{\rm rad}$ includes the contributions from gas 
and radiation and obeys the boundary condition that $P_{\rm gas}=0$ for 
$\tau=0$. The integration of Eq.~(\ref{eq_dtaudp}) is accomplished by a
standard extrapolation-interpolation procedure, from $\tau=0$ to $\tau=\tilde{\tau}$. 
The solution is obtained through iteration on the total pressure 
$P$. Starting from the top of the atmosphere, with $P=P_{\rm rad}$ and
$\tau=0$, we integrate Eq.~(\ref{eq_dtaudp}) inward with a sequence 
of extrapolation-interpolation steps. The adopted scheme is a combination of a
third-order  Adams-Bashforth predictor followed by a fourth-order  
Adams-Moulton corrector \citep[chapter XVI of ``Numerical Recipes'';][]{Press_etal88}.
In brief, for a given increment $\Delta P$,  to proceed from the mesh-point 
$j$ to mesh-point $j+1$, we first extrapolate the optical depth 
$\tau_{j+1}^{\rm extr}$ with the predictor part,
using the known value $\tau_j$. Then, we use the corrector
to interpolate the derivative at $j+1$, and hence to obtain 
the value $\tau_{j+1}^{\rm int}$.
The integration step is considered successful if the   
extrapolated $\tau_{j+1}^{\rm extr}$ and interpolated $\tau_{j+1}^{\rm int}$ 
values agree to within a given tolerance, normally set to $10^{-4}$ for the logarithmic optical
depth. Otherwise, the integration
step is repeated halving the pressure step-width $\Delta P$.

\subsubsection{Spherically-symmetric atmospheres}
We have implemented the spherical-symmetry geometry following the
formalism described in \citet{Lucy76}, but with the addition that
the mass above the atmosphere is not neglected compared to that of 
the entire star.
Introducing the variable $z=r/R$, 
the temperature stratification accounts for the geometrical dilution 
of the radiation field and is given by:
\begin{equation}
T^{4} = \frac{3}{4} T_{\rm eff}^{4} 
\left [\tilde\tau + \frac{4}{3} W \right]\,,
\label{eq_spheatm}
\end{equation}
where 
\begin{equation}
W = \frac{1}{2} \left ( 1 - \frac{\sqrt{z^2 - 1}}{z} \right )
\end{equation}
is the dilution factor; $\tilde\tau$ is the optical depth defined by the
differential equation:
\begin{equation}
\frac{d\tilde\tau}{dz} = -\frac{\kappa \rho R}{z^2}\,.
\end{equation}

In this case, the radial extension of the atmosphere is not neglected, and
$r=R_0$ refers to the maximum outer radius of the atmosphere, where
by definition $\tau(R_0)=0$ and $P_{\rm gas}(R_0)=0$. 
Since in principle these two boundary conditions 
are met for $r \rightarrow + \infty$, we define the 
outer boundary $R_0$ of the atmosphere the radial coordinate of the 
point at which  $P_{\rm gas} = 10^{-4}$ dyne cm$^{-2}$.
\noindent
The parameter
\begin{equation}
\delta R = \frac{R_0-R}{R} 
\label{eq_dr}
\end{equation}
quantifies the geometrical extension of the atmosphere.

In an extended atmosphere an effective temperature cannot be 
uniquely defined; therefore we refer to it as the photospheric 
temperature obeying the relation
\begin{equation}
T_{\textup{eff}} = T(\tilde{\tau}) = \left ( \frac{L}{4\pi \sigma R^2} \right )^{1/4} \,\,\,\,\,\,\,{\rm and} \,\,\,\,\,\,\,\tilde{\tau}=2/3
\label{eq_tef}
\end{equation}
which is formally analogous to that of a compact atmosphere star.

In summary, together with the auxiliary relation Eq.~(\ref{eq_spheatm}),
our extended atmosphere model requires the integration
of three differential equations for the unknowns
optical depth $\tau$,  non-dimensional radial coordinate $z=r/R$, 
and mass coordinate $m$, which are conveniently expressed in the form $d \tau /d\log P $,  $d z /d\log P $, and
$d \log m /d\log P $, where the total pressure $P$ is  
the independent variable.

For any  given atmosphere model specified by a choice of 
$L$, $M$, $T_{\rm eff}$ (hence with $R$ known from Eq.~\ref{eq_tef}), 
and chemical composition, we proceed as follows.
We make an initial guess of the ratio $R_0/R$.
Then the differential equations, 
reduced to a finite-difference form, 
are solved starting from the  provisional outermost point at 
$r=R_0$, with the boundary conditions 
\begin{equation}
\tau(R_0)=0,\;\;\;\; m(R_0)=M,\;\;\;\; P(R_0)=P_{\rm rad}, 
\end{equation}
and proceeding inward  by using the same  extrapolation-interpolation 
method already described in Sect.~\ref{sssec_ppatmo}, but this time extended
to the three differential equations in the unknowns $\tau$,  $r$, and $m$.
Integration is stopped when the photosphere at $\tau = \tilde{\tau}$ 
is reached.
In general the temperature at the photospheric layer, $T_{\tilde{\tau}}$,
 will differ from $T_{\rm eff}$ given by
Eq.~(\ref{eq_tef}), so we adopt
a new value for  $R_0/R$ 
and integrate another atmospheric structure.
The procedure is repeated until the 
$\left | \log(T_{\tilde{\tau}})  - \log(T_{\textup{eff}})\right| < \varepsilon$,
where the tolerance $ \varepsilon$ is normally set to $10^{-4}$. 

\subsection{The quiescent interpulse phases}
\subsubsection{The deep envelope model}
\label{ssec_envmod}
 
In synthetic AGB models $L$, $T_{\rm eff}$, and 
the temperature at the base of the convective envelope, $T_{\rm bce}$,
are usually obtained with the aid of 
formulas that fit the results of full models calculations
\citep[e.g.][]{Hurley_etal00,Izzard_etal04, Izzard_etal06, Cordier_etal07}.
In \texttt{COLIBRI} the approach is completely different:
during the quiescent interpulse periods the four stellar structure equations 
(i.e. mass continuity, hydrostatic equilibrium, energy transport, and
energy balance)  are integrated from the photosphere down to the bottom 
of the quiescent H-burning shell, a region
which we globally refer to as {\em deep envelope}.

The energy balance equation reads 
\begin{equation}
\frac{\partial l}{\partial m } = \varepsilon _{\rm nuc} + \varepsilon _{\rm g}
- \varepsilon _{\rm \nu}\,,
\label{eq_energ}
\end{equation}
where the right-hand side member 
accounts for the energy contributions/losses from    
nuclear, gravitational, and neutrino sources, with rates 
(per unit time and unit mass) 
$\varepsilon _{\rm nuc}$, 
$\varepsilon _{\rm g}$, and $\varepsilon _{\rm \nu}$, 
respectively.

The efficiency of nuclear energy generation is computed as 
$\varepsilon _{\rm nuc} = \varepsilon _{\rm pp} + \varepsilon _{\rm CNO}$, 
that is including the contributions of the p-p chains and CNO cycles.
The corresponding nuclear reaction rates are listed 
in Table~\ref{tab_rates}.
 
In our deep envelope model we assume $\varepsilon _{\rm \nu}=0$, which is a
safe approximation since thermal neutrinos mainly 
come from the degenerate core.

The gravitational energy generation, given by
\begin{equation}
\varepsilon _{\textup{g}} = - T \, \frac{\partial S}{\partial t}\,,
\end{equation}
where $S$ is the gas entropy and $t$ denotes the time variable,
is computed in the {\em stationary} wave approximation 
\citep{Weigert_66, Iben_77}:
\begin{equation}
\frac{\partial S}{\partial t} = \frac{\mathrm{d} M_{\textup{c}}}{\mathrm{d} t} \: \frac{\partial S}{\partial m}
\end{equation}
where $T$ is the local temperature, $\partial S/\partial m$ is the local
derivative of entropy with respect to mass, and  
$\mathrm{d} M_{\textup{c}}/{\mathrm{d} t}$ denotes the rate at which the mass 
coordinate of the centre of the hydrogen-burning shell advances outward.

The rate of displacement of the H-burning shell actually  measures the 
growth rate of the core mass and it is computed with
\begin{equation}
\frac{\mathrm{d} M_{\textup{c}}}{\mathrm{d} t} = \frac{q}{X_{\textup{env}}} L_{\textup{H}}
\end{equation}
where  $L_{\textup{H}}$ is the total luminosity  produced by the  
{\em radiative} portion of the hydrogen burning shell, $X_{\textup{env}}$ 
corresponds to the hydrogen abundance
(in mass fraction) in the convective envelope, and 
$q = 1.05\times 10^{-11} + 0.017\times 10^{-11}\,\log(Z)\;
[M_{\odot}\,L_{\odot}^{-1}\,{\rm yr}^{-1}]$  \citep{Wagenhuber_96}.

\paragraph*{Method of solution.}
\label{sssec_envsol}
Since  we deal with a set of four stellar structure equations, 
we need to set up four  boundary conditions to close the system.

The first pair of boundary conditions applies to the surface, and corresponds to
the photospheric values of radius and temperature, 
$r(\tilde{\tau})$, and $T(\tilde{\tau})$,
provided by the atmosphere model  (either in the plane-parallel
or spherically-symmetric assumption as described in Sect.~\ref{ssec_atmo}):
\begin{equation}
\label{c1}
T(\tilde{\tau}) =  T_{\rm eff}\,,
\end{equation}
\begin{equation}
\label{c2}
r(\tilde{\tau})  =    R\,.
\end{equation}

The second pair of boundary conditions applies to the interior.
Moving inward across the {\em deep envelope},
the bottom of the H-burning shell corresponds to the radiative 
layer where the hydrogen abundance first goes to zero ($X=0$).
We choose the mass coordinate of the corresponding mesh, $m(X=0)$, 
to identify a key parameter of the AGB evolution, the core mass $M_{\rm c}$.

The third boundary condition is therefore:
\begin{equation}
\label{c3}
m({X=0}) = M_{\rm c}\,. 
\end{equation} 
The fourth inner boundary condition is given by the temperature $T_{\rm c}$ at 
the bottom  of the H-burning shell:
\begin{equation}
\label{c4}
T({X=0}) = T_{\rm c}\, .
\end{equation} 
Full stellar AGB models calculated with \texttt{PARSEC} 
show that $T_{\rm c}=T(M_ {\rm c}, Z_{\rm i})$ is a well-behaving, increasing function of the core mass, 
with some moderate dependence on metallicity.  After the first sub-luminous thermal pulses, 
in the full-amplitude regime $T_{\rm c}$ is found to vary within  a narrow range 
(i.e. $\log(T_{\rm c})\approx 7.9-8.0$),  reflecting  the thermostatic property of the shell-hydrogen burning  
(mainly via  the CNO cycle), occurring at a well-defined temperature.
This fact makes the boundary condition Eq.~(\ref{c4}) a robust choice, only little dependent 
on technical and model details.

In summary, 
Eqs.~(\ref{c1}), (\ref{c2}), (\ref{c3}), and (\ref{c4}) 
provide the four boundary constraints
necessary to determine the entire structure of the {\em deep envelope}.
The total pressure $P$ is chosen as the independent variable, and
 the four differential equations of the stellar structure 
are suitably expressed in the form $d\log m /d\log P $,  $d\log r /d\log P $,
$d\log l /d\log P $, and  $d\log T /d\log P$. 
Inward numerical integrations 
are carried out using an Adams-Bashforth-Moulton  
extrapolation-interpolation scheme, that combines
a third-order predictor with a fourth-order corrector.
The procedure is formally the same as that described 
in Sect.~\ref{sssec_ppatmo}, but applied to the four equations
in the unknowns $m$, $r$, $l$, $T$. 
The integration accuracy is usually set to $10^{-4}$ 
for all logarithmic variables. 

We adopt a very fine mass resolution,
the width of the innermost shells (where the structural
gradients become extremely steep)  typically amounting to 
$10^{-7} - 10^{-8} M_{\odot}$.
The chemical composition is assumed homogeneous throughout
the convective envelope (possible deviations for specific elements are discussed 
in Sect.~\ref{ssec_HBBnuc}).
Once in the deep interior the radiative temperature gradient 
falls below the adiabatic one and the energy transport becomes radiative,
a chemical profile is built with abundances  that change 
with mass in direct proportion to the rate of energy generation by
the hydrogen-burning reactions, until hydrogen vanishes 
The procedure is the same as that described by \citet{Iben_77}.

The integration method just illustrated is adopted to obtain 
the atmosphere-envelope structure at the quiescent stage just preceding each
thermal pulse.  In particular, this yields the quiescent pre-flash luminosity
maximum, $L_{\rm Q}$.
To follow the subsequent structural variations, driven by the occurrence 
of thermal pulses, we proceed as follows. Let us denote with 
\begin{equation}
\phi\equiv t/\tau_{\rm ip}
\label{eq_phi}
\end{equation}
 the  pulse-cycle phase, where $\tau_{\rm ip}$ is the interpulse period and 
$t$ is the current time, counted from the stage of 
quiescent pre-flash luminosity maximum, such that $\phi =0$ at $t=0$, 
 and $\phi =1$ at $t=\tau_{\rm ip}$ (and $L=L_{\rm Q}$).
According to \citet{WoodZarro_81} and 
\citet{WagenGroen_98} 
the star luminosity as a function of the pulse-cycle phase, $L(\phi)$, 
when normalized to $L_{\rm Q}$, 
has a very well-known and almost universal form 
(${\rm f}(\phi) = L(\phi)/L_{\rm Q}$), 
independent of $Z_{\rm i}$ 
 \citep[][see their equation 15]{WagenGroen_98}.
Therefore, 
once we determine $L_{\rm Q}$ at $\phi =1$ by solving 
the complete set of stellar equations, then
the structure of the envelope over the next thermal TPC 
(for each value of the phase $0\le \phi< 1$)  
is obtained iteratively in a similar fashion, but this time
adopting $L=L(\phi)={\rm f}(\phi)\:L_{\rm Q}$, and fulfilling three out 
of four boundary conditions. 
While the first pair, Eqs.~(\ref{c1}) and (\ref{c2}), 
is the same for any value of $\phi$, the third boundary
condition depends on phase of the pulse cycle.

Following the thorough analysis by \citet{WagenGroen_98},
in the initial phases of a TPC, for
$0\le \phi \la 0.1$, (that include the so-called ```rapid dip'', 
``rapid peak'' and part of the ``slow dip'',  i.e. from A to D in their figure 1), 
the H-shell is extinguished, while the He-shell is on.
During these very short-lived stages, immediately
after the onset of a TP, we adopt $M(R_{\rm c}) = M_{\rm c}$ 
(Eq.~\ref{c5}) as
the third boundary condition for the envelope integrations.
More details can be found in Sect.~\ref{ssec_TP}.
At later stages, for $0.1 < \phi \le 1$ (i.e. from D to A'), 
when the helium burning drops and the quiescent H-shell recovers
 becoming the dominant  energy source, the third boundary condition 
is again given by $m({X=0}) = M_{\rm c}$ (Eq.~\ref{c3}).

It is worth remarking that the integration 
of the {\em deep envelope} allows us to predict the integrated luminosity provided  
by the quiescent H-burning shell, both in the relatively simple case of low-mass TP-AGB stars 
(in which the H-burning shell is completely radiative and thermally decoupled from the
convective envelope),  and in the more complex case of  intermediate-mass TP-AGB  stars 
experiencing HBB (in which the bottom of
the convective envelope lies inside the H-burning shell, providing an extra-luminosity 
$\Delta L_{\rm HBB}$ contribution  above the classical CMLR).
Section~\ref{sec_tests} is devoted to compare and test  our results against  those from 
various sets  of full AGB models in the literature.

Another important implication is that our method assures a correct 
treatment of HBB, i.e. a full consistency  between energetics 
and associated nucleosynthesis.
In other words, the rates of variation of the surface chemical abundances 
caused by HBB (i.e via the CNO, NeNa, and MgAl cycles) 
are precisely those that correspond to the luminosity contribution
$\Delta L_{\rm HBB}$.
Despite being a basic requirement \citep{MarigoGirardi_01}, 
the strict coupling between the 
consumption of the nuclear fuel and the chemical composition changes, 
are in general not fulfilled by analytical approximations of HBB, 
often adopted in synthetic TP-AGB models.
\subsubsection{Nucleosynthesis in convective envelope layers}
\label{ssec_HBBnuc}

Besides being an important energy source for AGB stars 
with $M_{\rm i} > 3-4\, M_{\odot}$, HBB significantly alters the 
chemical composition of their envelopes through the nuclear reactions
(pp chains, and CNO, NeNa, MgAl cycles) taking place in the innermost
convective layers \citep[e.g.][]{Boothroyd_etal95, ForestiniCharbonnel_97,
Marigo_01, Karakas_10, Ventura_etal11}.

In \texttt{COLIBRI} the HBB nucleosynthesis is treated in detail.
Once the structure of the convective envelope is determined, as explained 
in Sect.~\ref{ssec_envmod},
nucleosynthesis occurring in the convective envelope is treated in detail, 
by coupling 
nuclear burning to a diffusive description of convection.
In a one-dimensional, spherically-symmetric system the conservation equation 
for an arbitrary chemical species $i$, locally defined at the 
Lagrangian coordinate $m_r$, reads
\begin{eqnarray}
\label{eq_difmix}
\frac{\partial Y_i}{ \partial t}\Bigl\lvert_{m_r} & = & \frac{1}{\rho r^2} \frac{\partial}{\partial r} 
\left( r^2 \rho D \frac{\partial Y_i }{\partial r} \right)\\
\nonumber
& & \pm \sum_{j} Y_j \lambda_k(j) \pm \sum_{j\ge k} Y_j Y_k r_{j k}\,,
\end{eqnarray}
where $Y_i = X_i/A_i$ (in units of mole/mass) is the ratio between 
the abundance (in mass fraction) of the nucleus $i$ and its atomic weight $A_i$.
The term on the left-hand side gives the local rate of change of abundance of
element $i$ at the coordinate $m_r$, which is due to two different 
processes, namely: mixing and nucleosynthesis.

On the right-hand side of Eq.~(\ref{eq_difmix}) the first term
is  the mixing contribution, that is the local abundance variation 
produced by the convective motions in the gas. In our approach
convection is treated as a diffusion process, with the 
diffusive coefficient approximated as
\begin{equation}
\label{eq_difco} 
D = \frac{1}{3} v_{\rm conv} l_{\rm conv}\,,
\end{equation}
where $v_{\rm conv}$ and $l_{\rm conv}$ denote the velocity and the mean-free path 
of the convective eddies, respectively.
Both quantities are computed in the framework of the  standard mixing length theory \citep{mlt_58}.
The mixing length $l_{\rm conv}$ is assumed linearly proportional to the pressure scale height, $H_{\rm p}$,
with the proportionality coefficient $\alpha_{\rm MLT}=1.74$, as derived from a recent calibration of the
solar model \citep{Bressan_etal12}.
The convective velocity is obtained from the only real root of the ``cubic equation'' 
\citep[equation $14.82$, Vol.~I of ``Principles of Stellar Structure'';][]{CoxGiuli_68}, under the condition that the total energy flux is specified. 

The second and third terms on the right-hand side of Eq.~(\ref{eq_difmix}) describe the 
abundance change due to nuclear reactions involving the species $i$, being related to 
single-body decays (with rates $\lambda$) and two-body reactions (with rate $r$), respectively. 
As usual, the negative (positive) sign is used to denote destruction (production) of the species $i$.

\paragraph*{Method of solution.}
The convective envelope is divided into a number $N_{\rm mesh}$  of concentric shells, 
so as to ensure smooth enough variations of the physical variables 
(radius, temperature, density, etc.) between consecutive mesh points. 
 For instance, in the deepest zones, where nuclear burning takes place the temperature 
difference of consecutive shells is chosen $\delta\log(T) = 0.01-0.02$ dex. 

We deal with a system of coupled, non-linear, partial differential
equations, given by Eq.~(\ref{eq_difmix}), for each chemical species at all mesh points.   
The equations are first converted to finite central-difference equations and the quadratic terms,
$Y_j Y_k$, are linearized according to \citet{ArnettTruran_69}. To estimate the diffusion coefficient
between two shells, $D_{k \pm 1/2}$, we adopt the prescription proposed by 
\citet[][]{Meynet_etal04}:
\begin{equation}
D_{k \pm 1/2} =\frac{D_{k \pm 1} D_k}{f D_{k \pm 1} + (1-f) D_k}
\end{equation}
with $f=0.5$, which appears to be  more physically sound than adopting a simple arithmetic 
mean.

Following the scheme proposed by \citet{Sackmann_etal74}, we set up a matrix equation $A=Y\,b$ in the unknown abundances $Y_{i,k}^{n+1}$
at the time $n+1$, where $i=1,\dots,N_{\rm el}$ denotes the element,
 and $k=1,\dots, N_{\rm mesh}$ refers to the mesh-point.
$A$ is the $(N,N)$ matrix of the coefficients with
$N=N_{\rm el}\times N_{\rm mesh}$. Since we assume that each species is coupled to all others 
at the same mesh point and to its own abundance at adjacent mesh-points, the matrix $A$ has 
a band-diagonal structure with $k_l=N_{\rm el}$ sub-diagonals and  $k_u=N_{\rm el}$ super-diagonals
(hence the band width is $k_l+k_u+1$). This property is taken into consideration to reduce 
the computing-time requirement of the adopted numerical algorithm.
The $(N,1)$ matrix $b$ contains the known terms, which  depend on the chemical abundances  across the 
envelope, $Y_{i,k}^{n}$, at the previous time $n$.  

Finally, the system is solved by means of a fully implicit method that, when applied to
diffusion problems, proves to yield robust results in terms of numerical stability 
and accuracy \citep[see the thorough analysis in][]{Meynet_etal04}.
Compared to explicit and ``Crank-Nicholson'' methods the great advantages of the implicit technique are that  
i) we are not forced to stick to the ``Courant condition'', 
that imposes short integration time steps to assure stability,
ii) in most cases it does not yield unphysical solution (e.g. negative abundances), and iii)  
the conservation of the mass, i.e. the normalization condition of the abundances, 
at each mesh-point is reasonably fulfilled, typically not exceeding 
$\simeq 10^{-5}$.

Fortran routines taken from the LAPACK\footnote{LAPACK is a freely-available copyrighted library of Fortran 90 with subroutines for solving
the most commonly occurring problems in numerical linear algebra. 
It can be obtained via \url{http://www.netlib.org/lapack/}}
software package are employed to 
get the numerical solution of the matrix equation, which is accomplished through three main steps, namely: 
1) LU decomposition\footnote{In linear algebra LU decomposition 
factorizes a matrix as the product of a lower (L) triangular matrix and an 
upper (U) triangular matrix.} 
of the matrix $A$, which is conveniently stored in a compact form so as to
get rid of most of the useless null terms outside the main diagonal band; 
2) solution of the system of linear equations by partial pivoting, and  3) iterative improvement of the solution.
The latter step attempts to refine the solution by reducing the backward errors (mainly due to
round-off and truncation errors) as much as possible. 

\subsubsection{Time integration}
To follow the time evolution along the TP-AGB phase 
we proceed as follows.
Each interpulse period is divided into a suitable number, $N_{\phi}$,
of phase intervals, $\Delta\phi_j=(t_{\rm j+1}-t_j))/\tau_{\rm ip}= 
\Delta t_j/\tau_{\rm ip}$,
so as to assure a good sampling of the complex luminosity variations
driven by the pulse (see Eq.~(\ref{eq_phi}) and Sect.~\ref{ssec_envmod}).
This defines a first guess of the time step. 
A subsequent adjustment 
may be done by imposing the condition that the time step
does not exceed a given limit, i.e. $\zeta (M-M_{\rm c})/\dot M$, 
where $(M-M_{\rm c})/\dot M$ is a measure of the time-scale 
required to expel the envelope at the current mass loss rate $\dot M$.  
The coefficient $\zeta$ is normally set to $10^{-3}$. This condition
determines a sizable reduction of the time step in the last evolutionary
stages, when the super-wind regime of mass loss is attained.

Once $\Delta t_j$ is fixed, 
the increment of  the core mass  and the decrease of the total mass are 
predicted with the explicit Eulerian method:
\begin{eqnarray*}
M_{\rm c, j+1} & = &  M_{\rm c, j} 
+ (q  L_{\textup{H}}/ X_{\textup{env}})_j \, \Delta t_j \\
M_{j+1} & = & M_{j} -  \dot M_j \, \Delta t_j 
\end{eqnarray*}
At this point all other variables (e.g. $T_{\rm eff}$, $L$, 
$T_{\rm bce}$, and  
chemical abundances in case of HBB, etc.) at the time $t_{\rm j+1}$ 
are obtained from envelope integrations with the new values 
$M_{\rm c,j+1}$ and $M_{j+1}$.  

With the current set of prescriptions, 
typical values of $N_{\phi}$ over one TPC range from few 
to several hundreds, depending on stellar parameters 
and evolutionary status.

\subsection{The thermal-pulse phases}
\label{ssec_TP}
In addition to the quiescent interpulse phases (see Sect.~\ref{ssec_envmod}), 
we carry out envelope integrations to test whether 
appropriate thermodynamic conditions  exist for the  occurrence of the third dredge-up.
This approach replaces the use of the parameter  $M_{\rm c}^{\rm min}$, i.e.
the minimum core mass for the third dredge-up 
(see Sect.~\ref{ssec_tbdred}), used in previous models \citep{Marigo_etal96, 
Marigo_etal98, MarigoGirardi_07}.
Also, we set up a nuclear network to follow the synthesis 
of C, O, Ne, Na, and Mg in the flash-driven convective zone, which determines 
the chemical composition of the dredged-up material.
All details are given in Sect.~\ref{ssec_pdcz}.  
\subsubsection{Onset and quenching of the third dredge-up}
\label{ssec_tbdred}
We follow the method first proposed by \citet{Wood_81} and later adopted 
by  \citet{Marigo_etal99} to predict {\em if} and {\em when} the third dredge-up
may take place  during the TP-AGB evolution of a star of given current mass and 
chemical composition. We refer to the quoted papers for all details, and
recall here the basic scheme.

The technique makes use of  suitable envelope integrations at the stage 
of post-flash luminosity maximum, $L_{\rm P}$,  when the envelope is close to hydrostatic and 
thermal equilibrium \citep{Wood_81}.
TP-AGB models show that $L_{\rm P}$ is essentially controlled by the
core mass of the star, in analogy with the existence of the CMLR 
relation during the quiescent interpulse periods for low-mass AGB stars.
Following \citet[][]{Wood_81} and \citet{BoothroydSackmann_88b}, 
at the post-flash luminosity peak  the
nuclearly processed material involved in the He-shell flash is pushed
out and cooled down to its minimum temperature over the flash-cycle,
$T_{\rm N}^{\rm min}$, approaching a limiting characteristic value,
as the thermal pulses reach the full-amplitude regime.
This latter typically lies in the range 
$\log(T_{\rm N}^{\rm min})\approx 6.5-6.7$ \citep{BoothroydSackmann_88b, 
Karakas_etal02}, being little dependent on chemical composition 
and core mass.  
At the same time the envelope
convection reaches its maximum inward penetration (in mass fraction)
and the maximum base temperature, $T_{\rm bce}^{\rm max}$.

Hence it is reasonable to assume that the third dredge-up takes place if,  
at the stage of post-flash luminosity maximum, the condition 
$T_{\rm bce}^{\rm max} \ge T_{\rm N}^{\rm min}$
is satisfied.

\begin{figure*} 
\resizebox{0.8\hsize}{!}{\includegraphics{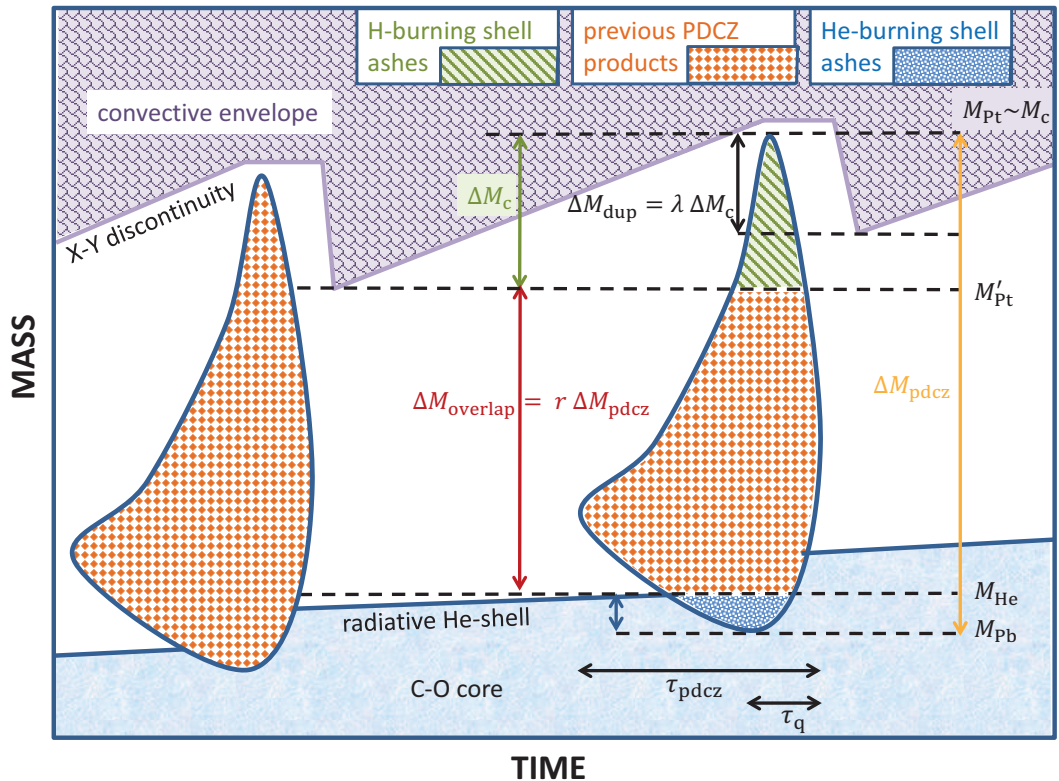}}
\caption{Sketch of the Kippenhahn diagram showing the evolution of 
the inner layers of a TP-AGB star during and between 
two consecutive thermal pulses. Mass boundaries and relevant quantities
(e.g. the degree of overlap $r$ and the efficiency of the third dredge-up $\lambda$)
are indicated. We refer to Table~\ref{tab_mod} for operative definitions.
Note that mass and time coordinates are not on real scales,  for graphical clarity.
The hatched areas over the later PDCZ 
correspond to the three-zone stratification of the material just before the
development of the convective pulse, containing from top to bottom:
the ashes left by the H-burning shell, the products of the previous 
PDCZ, and the hashes left by the He-burning shell. 
On the abscissa we show the lifetime 
of the convective pulse $\tau_{\rm PDCZ}$, and the  quenching time 
$\tau_{\rm q}$ counted from the maximum extension of the PDCZ. 
}
\label{fig_pdcz}
\end{figure*}

Operatively, let us denote with $T_{\rm dup}$ the parameter 
representing the minimum temperature that the envelope 
base must exceed to activate the third dredge-up, that is:
\begin{equation}
\label{eq_tbdred}
T_{\rm bce}^{\rm max} \ge T_{\rm dup}\,.
\end{equation}
In order to check it, at each thermal pulse, 
we integrate our envelope model described in Sect.~\ref{ssec_envmod}.
These numerical integrations are 
computed under particular conditions\footnote{The absence of nuclear energy sources in the envelope 
implies that the system of the stellar structure can be reduced 
from four to three equations (following \citet{Wood_81} the local luminosity 
is reasonably constant across the envelope, $l = L$), so that 
we need to specify three boundary conditions,
i.e. two at the photosphere Eqs.~(\ref{c1})-(\ref{c2}), and one at the core border 
Eq.~(\ref{c5}).}, namely: i) we set 
$\varepsilon _{\rm nuc} = \varepsilon _{\rm pp} + \varepsilon _{\rm CNO} = 0$,
since at this stage the H-burning shell is extinguished;
ii) the two inner boundary conditions Eqs.~(\ref{c3}) and (\ref{c4}) are replaced 
with 
\begin{equation}
\label{c5}
M(R_{\rm c}) = M_{\rm c}\,.
\end{equation}
This condition means that the mass of the degenerate core is equal to the 
mass contained inside the radius of a {\em warm} white dwarf,
$R_{\rm c}= \delta \times R_{\rm WD}$.
In the latter expression $R_{\rm WD}$ is the radius of a zero-temperature 
white dwarf (WD) with mass $M=M_{\rm c}$, while the coefficient 
$\delta> 1$ accounts for the fact that the nearly isothermal degenerate core is warm, i.e. 
it has  a non-zero temperature.  To compute $R_{\rm c}$ 
we follow the same prescriptions as in \citet{Marigo_etal99},
and adopt the  $M_{\rm c} - L_{\rm P}$ relation of \citet{WagenGroen_98}.

Then, for given stellar mass, core mass, surface chemical composition,
and peak-luminosity $L_{\rm P}$, envelope integrations are performed
iterating on the effective temperature, $T_{\rm eff}$, until when
$M(R_{\rm c})=M_{\rm c}$.  At this point, the structure of
the envelope is entirely and uniquely determined.

Since the typical values of $T_{\rm N}^{\rm min}$ may vary 
between  different sets of models (reflecting  its dependence 
on the adopted input physics and on the description of convection),  
we take $T_{\rm dup}$ as a free parameter.
An advantage is that with the condition given by
Eq.~(\ref{eq_tbdred})  we can also 
test the eventual quenching of the third dredge-up due, for instance, 
to a drastic reduction of the envelope mass, without the need for 
another external assumption (see Sect.~\ref{ssec_3dup}).
For the present set of TP-AGB models we have adopted 
the temperature parameter $\log(T_{\rm dup})=6.40$.

\subsubsection{Pulse-driven nucleosynthesis}
\label{ssec_pdcz}
We have developed a simplified model to predict the intershell 
chemical composition produced by the flash-driven nucleosynthesis, 
using an approach similar in some aspects 
to those proposed by \citet{IbenTruran_78}, \citet{Mowlavi_99a, Mowlavi_99b},
and \citet{DenissenkovHerwig_03}.

The assumed scheme for the pulse-driven convection zone (PDCZ) is
sketched with the aid of a Kippenhahn diagram in Fig.~\ref{fig_pdcz},
showing the time evolution of the PDCZ borders from its appearance to its
final quenching.
Several relevant variables are defined in Table~\ref{tab_mod}.

\begin{table*}
 \centering
  \caption{Characteristic quantities of the TP-AGB model}
  \begin{tabular}{ll}
\hline 
$Z_{\rm i}$ & initial (zero-age-main-sequence) metallicity (mass fraction) \\
$Y_{\rm i}$ & initial (zero-age-main-sequence) helium abundance (mass fraction)\\
$X_{\rm i}$ & initial (zero-age-main-sequence) hydrogen abundance (mass fraction)\\
$Z$              & current metallicity (mass fraction)\\
$M_{\rm c}$ & current core mass $\equiv$ mass of the H-exhausted core \\
$M_{\rm c, 1}$ & core mass at the first thermal pulse \\
$M_{\rm c, nodup}= M_{\rm c, 1}+\displaystyle{\int_{0}^{t}\frac{d M_{\rm c}}{dt'} dt'}$ & 
core mass in absence of the third dredge-up, where $t=0$ is the time of the first TP. \\
$M_{\rm i}$ & initial stellar mass at the zero-age main sequence \\
$M_{1}$ & stellar mass at the first thermal pulse \\
$M$       & current stellar mass  \\
$T_{\rm bce}$ & temperature at the base of the convective envelope\\
$\tau_{\rm ip}$ & interpulse period\\
$\phi\equiv t/\tau_{\rm ip}$ $(0 \le \phi \le 1)$ & pulse-cycle phase; 
the time $t=0$ refers to the quiescent pre-flash luminosity maximum. \\
\hline 
\multicolumn{2}{c}{Quiescent interpulse evolution} \\
  \hline
$\Delta M_{\rm c, tpc}$ & core mass growth over one  interpulse period\\
$\Delta M_{\rm c} = M_{\rm c} - M_{\rm c, 1}$ & cumulative core mass growth since the $1^{\rm st}$ TP\\
$\Delta M_{\rm c, nodup} = M_{\rm c,nodup} - M_{\rm c, 1}$ & 
cumulative core mass growth in absence of the third dredge-up \\
   \hline
 \multicolumn{2}{c}{Pulse-driven convective zone} \\
\hline
$M_{\rm Pt}$ & mass coordinate of the top of the current PDCZ 
at its maximum extension \\
$M_{\rm Pt}^{\prime}$  &mass coordinate of the top of the previous PDCZ at its maximum extension \\
$M_{\rm He}$ & mass coordinate of the He-exhausted core \\
$M_{\rm Pb}$  &mass coordinate of the bottom of the current PDCZ  at its maximum extension \\
$f_{\rm ov}$ & parameter to mimic overshoot applied to the bottom of the PDCZ \\
$\Delta M_{\rm pdcz}$  &PDCZ mass at its maximum extension \\
$\tau_{\rm pdcz}$ & total duration of the PDCZ \\
$\tau_{\rm q}$ & quenching time since maximum extension \\
$T_{\rm pdcz}^{\rm max}$ & maximum temperature reached in a TP at the inner border of the PDCZ \\
$\rho_{\rm pdcz}^{\rm max}$ & maximum density reached in a TP at the inner border of the PDCZ \\
\hline
 \multicolumn{2}{c}{The third dredge-up} \\
\hline
$M_{\rm c}^{\rm min}$  & minimum core mass for the occurrence of the third dredge-up \\
$M_{\rm c}^{\rm 3dup}$ & actual core mass at the first episode of  the third dredge-up \\
$T_{\rm N}^{\rm min}$ & minimum temperature reached by the pulse at the stage
of post-flash luminosity maximum \\
$T_{\rm dup}$  & minimum temperature at the base of the convective envelope for the occurrence of the third dredge-up \\
$\Delta M_{\rm dup}$ & dredged-up mass at a given thermal pulse \\
$\Delta M_{\rm overlap}=M_{\rm Pt}^{\prime}-M_{\rm He}$ & overlap mass between two consecutive PDCZs \\
$\lambda = \displaystyle{\frac{\Delta M_{\rm dup}}{\Delta M_{\rm c, tpc}}}$ & efficiency of the third dredge-up\\
$r=\displaystyle{\frac{\Delta M_{\rm overlap}}{\Delta M_{\rm pdcz}}}
$ & degree of overlap between two consecutive PDCZ \\
\hline
\end{tabular}
\label{tab_mod}
\end{table*}

At the onset of each TP
the quantities $\Delta M_{\rm pdcz}$, 
$\tau_{\rm pdcz}$, $\tau_{\rm q}$, $T_{\rm pdcz}^{\rm max}$,
$\rho_{\rm pdcz}^{\rm max}$ 
are preliminarily computed with the aid of analytic relations as a function 
of the core mass and metallicity, that  can be obtained as fits to full AGB models 
(see Sect.~\ref{sec_synthmod} for more details). For the present work we use mainly the results
by \citet{IbenTruran_78}, \citet{Wagenhuber_96}, 
\citet{Karakas_etal02}, \citet{Straniero_etal03}. 

A nuclear network is set up which includes 
the triple-$\alpha$ reaction and the most important $\alpha$-captures 
listed in Table~\ref{tab_rates}.
Among them we consider the main reactions which may
be important as neutron sources: 
\reac{C}{13}{^{4}He}{n}{O}{16}, 
\reac{O}{17}{^{4}He}{n}{Ne}{20}, 
\reac{O}{18}{^{4}He}{n}{Ne}{21},  
\reac{Ne}{21}{^{4}He}{n}{Mg}{24}, 
\reac{Ne}{22}{^{4}He}{n}{Mg}{25}, and 
\reac{Mg}{25}{^{4}He}{n}{Si}{28}.

At time $t=0$, just before the development of a TP, 
the chemical composition of the region over which  
the flash-driven convection will extend,
is assumed to be stratified over three zones:
\begin{enumerate}[a)]
\item  $M_{\rm Pt}- M_{\rm Pt}^{\prime}$ containing the ashes, 
with abundances $\{X_{\rm Hb}\}$, left by the quiescent radiative 
H-shell over the previous interpulse period;
\item $M_{\rm Pt}^{\prime} - M_{\rm He}$  containing the nuclear products
of the PDCZ developed during the {\em previous} TP;
\item $M_{\rm He}- M_{\rm Pb}$  containing the products of 
radiative He burning.
\end{enumerate}
For simplicity each of the three zones is assigned an average chemical
composition, though a chemical profile exists in the a) and c) regions where 
nuclear burning has occurred in radiative conditions.

Denoting with $X^{\rm s}$ the homogeneous surface abundances,  
the composition of the hydrogen free layer left 
by the H-burning shell
is estimated following the 
indications by \citet{Mowlavi_99a, Mowlavi_99b}, which can be 
summarised as follows:
\begin{itemize}
\item all hydrogen is burnt into helium: 
$X_{\rm Hb}({\rm H})=0$;
\item all available CNO isotopes are converted into $^{14}$N: 
$X_{\rm Hb}(^{14}{\rm N})=14 \times \sum_{i=12}^{i=18}
X_{i}^{\rm s}/A_i$ (where $A_i$ is the mass number);
\item all  $^{22}$Ne is burnt into $^{23}$Na by the NeNa chain:
$X_{\rm Hb}(^{22}{\rm Ne})=0$;
\item the abundance of $^{23}{\rm Na}$  is computed with:\\
$X_{\rm Hb}(^{23}{\rm Na})=
f_{\rm Na}[23/22\times X^{\rm s}(^{22}{\rm Ne})+  X^{\rm s}(^{23}{\rm Na})]$.
\end{itemize}
The factor $f_{\rm Na}$ accounts for the possible destruction of
$^{23}$Na by proton captures at $T > 6\times10^7$ K.
Its value typically ranges from  $f_{\rm Na}=1$ (no destruction) down to
$f_{\rm Na}=0.2$ \citep[see figure A.3 in][]{Mowlavi_99a}.
For the present set of calculations we have adopted $f_{\rm Na}=1$.
The effects of the Mg-Al chain on the resulting $X_{\rm Hb}$ abundances 
is not considered in this work, and it will be implemented in a future
study. 

During each TP we follow 
the progressive development of pulse convection  
and related nucleosynthesis, over the  duration $\tau_{\rm pdcz}$. 
The process is divided into two consecutive phases:

\begin{enumerate}[I.]
\item from the onset of the PDCZ at time $t=0$ up to maximum extension 
 at time $t=\tau_{\rm pdcz}-\tau_{\rm q}$;
\item from maximum PDCZ extension to final pulse quenching at
time  $t=\tau_{\rm pdcz}$, with duration $\tau_{\rm q}$.
\end{enumerate}
The PDCZ is resolved both in time and in space. The 
 entire duration $\tau_{\rm pdcz}$ is subdivided 
in typically $\simeq 100$ time steps, while at each time a suitable grid 
of mass meshes is set up across the current PDCZ,
with a maximum mass resolution of $\simeq 10^{-4}\,M_{\odot}$. 
The evolution of  $T_{\rm pdcz}^{\rm max}$ and 
$T_{\rm rho}^{\rm max}$ over $\tau_{\rm pdcz}$, and the
temperature and density stratifications across the PDCZ mass 
are described on the basis of detailed calculations of thermal
pulses \citep[][and private communications]{Wagenhuber_96, WagenGroen_98}.
Illustrative examples are discussed later, in Sect.~\ref{ssec_xpdcz}.

\begin{figure*}
\begin{minipage}{0.48\textwidth}
\resizebox{0.85\hsize}{!}{\includegraphics{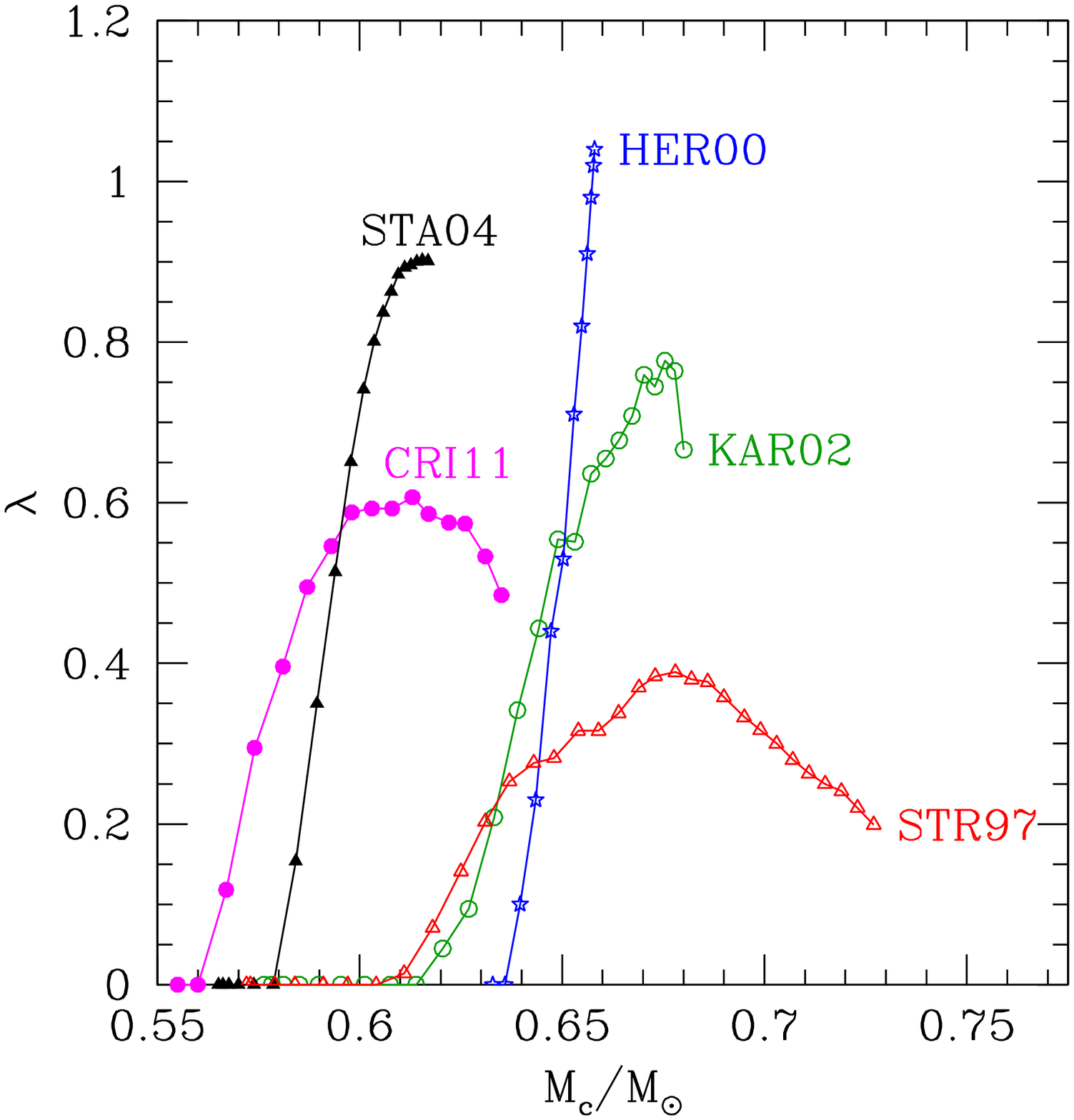}}
\end{minipage}
\hfil
\begin{minipage}{0.48\textwidth}
\resizebox{0.85\hsize}{!}{\includegraphics{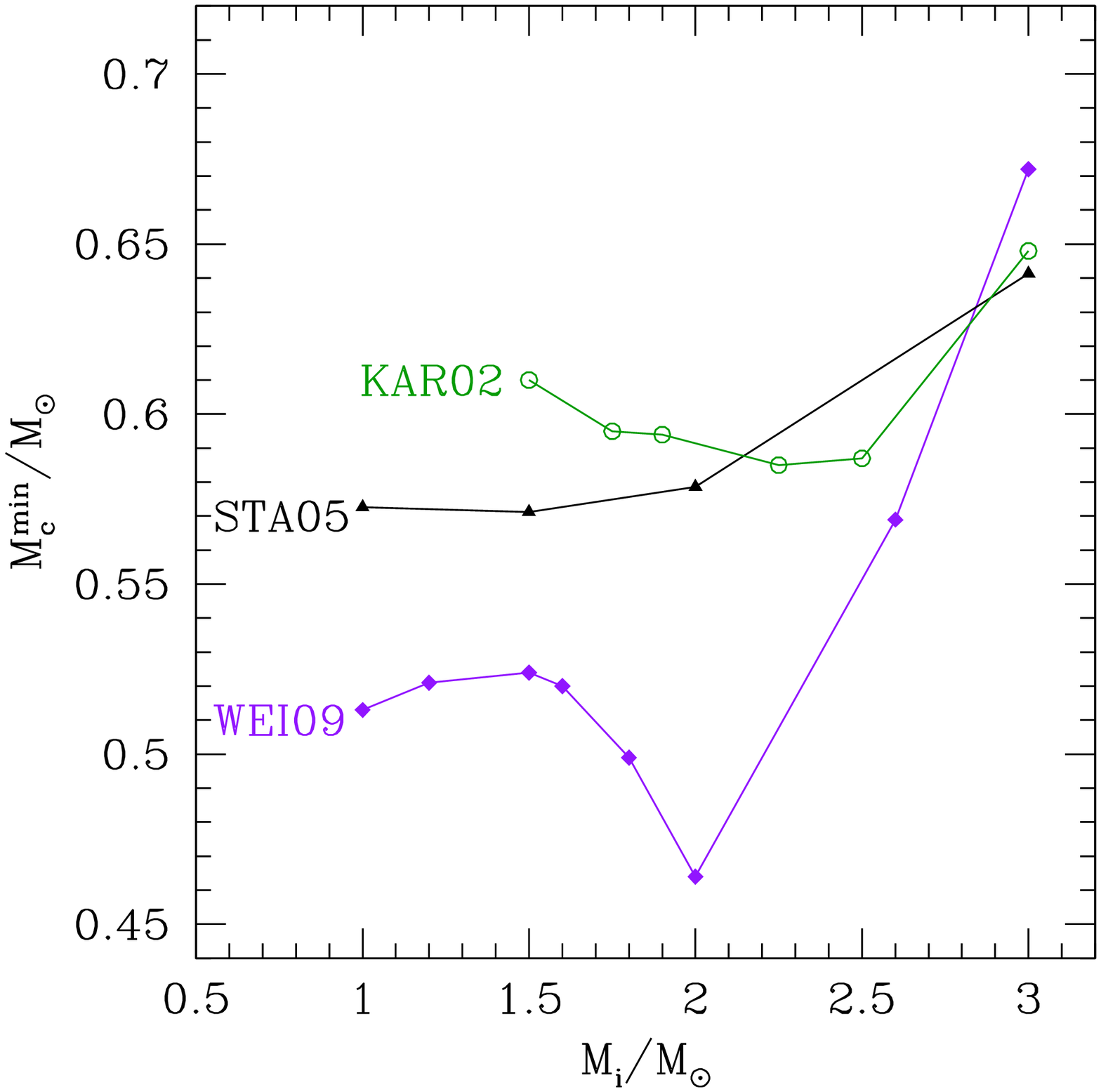}}
\end{minipage}
\caption{{\em Left panel}: Efficiency $\lambda$ of the third dredge-up 
as a function of the current core mass during the TP-AGB evolution 
of a stellar model with initial mass $M_{\rm i}=3.0 \, M_{\odot}$ and 
metallicity $Z_{\rm i}=0.02$. {\em Right panel}:
Minimum core mass $M_{\rm c}^{min}$ for the third dredge-up 
as a function of the stellar mass for TP-AGB models with initial 
metallicity $Z_{\rm i}=0.008$.
Predictions from full AGB calculations of various authors are compared, 
namely: CRI11 \citep{Cristallo_etal11}; WEI09 \citep{WeissFerguson_09}; 
STA05 \citep{Stancliffe_etal05}; STA04 \citep{Stancliffe_etal04}; 
KAR02 \citep{Karakas_etal02}; 
HER00 \citep{Herwig_00}; STR97 \citep{Straniero_etal97}. 
Note the large differences from author to author
both in $\lambda$ and in $M_{\rm c}^{\rm min}$.
 }
\label{fig_3dup3z02}
\end{figure*}

During the phase I the evolution of the PDCZ is followed by cycling 
over the sequence of steps: 
nucleosynthesis $\rightarrow$ homogenization $\rightarrow$ expansion/recession$\rightarrow$ homogenization.
At each time step, starting from the current PDCZ bottom 
(with mass coordinate $m_{\rm Pb}$) up 
to the current PDCZ top border (with mass coordinate $m_{\rm Pt}$)
the nuclear network is solved locally in each mesh point.
 
A homogeneous chemical composition is assigned to the PDCZ 
by mass-averaging the mesh abundances. 
Then, the PDCZ is made expand
i.e. inner/upper borders of the PDCZ are shifted 
inward/outward, and elements of new material, stratified according to
the initial composition, are engulfed. 
Eventually, a new PDCZ composition is obtained by averaging the abundances 
with weights proportional to the masses of the corresponding meshes.

The entire process, i.e.
convective burning followed by expansion and homogenization, 
is iterated until the maximum extension is reached, i.e. 
$m_ {\rm Pb}= M_{\rm Pb}$ and $m_{\rm Pt} = M_{\rm Pt}$, and the mass contained in the PDCZ
is equal to $\Delta M_{\rm pdcz}$. At this point $t=\tau_{\rm pdcz}-\tau_{\rm q}$.

The quenching phase II is described by a similar scheme, except that
now the PDCZ convection retreats and the inner/upper borders are shifted 
outward/inward until $t=\tau_{\rm pdcz}$.
The nuclear network is integrated over the pulse quenching
phase and a final homogeneous chemical composition is obtained.
This sets the chemical mixture of the material that may be
brought up to the surface by the subsequent third dredge-up phase.  

Despite its simplicity the PDCZ model yields results that nicely agree
with those of full TP-AGB computations.  
A detailed discussion of the predictions and their main dependencies 
is given in Sect.~\ref{ssec_xpdcz}.
\section{The synthetic module}
\label{sec_synthmod}
Most analytical ingredients of the \texttt{COLIBRI} code are formulas
accurately fitting the results of full AGB models covering  
wide ranges of initial stellar mass and metallicity.
The formulas are taken either from the extensive compilations by 
\citet{Wagenhuber_96, WagenGroen_98, Karakas_etal02, 
Izzard_etal04, Izzard_etal06}, and other sources 
\citep{Straniero_etal03}, or they
are directly derived from AGB model data sets 
by using standard $\chi^2$-minimization techniques.
New fits can be found in Appendix~\ref{app_fit}.
 
Importantly, all these analytic relations include 
a metallicity dependence, and take into account 
the peculiar behaviour of the first sub-luminous pulses while 
approaching the full-amplitude regime.

Among the most important prescriptions we mention 
the flash-driven luminosity variations as a function of the pulse-cycle phase 
\citep{WagenGroen_98},  
the core mass-interpulse period relation \citep{WagenGroen_98}, 
the maximum mass of the
PDCZ and its duration, the maximum temperature attained at the bottom 
of the PDCZ during a TP \citep[][]{KarakasLattanzio_07}\footnote{AGB models by \citet{KarakasLattanzio_07} are 
available for download at \url{http://www.mso.anu.edu.au/~akarakas/model_data/}},
the efficiency $\lambda$  of the third dredge-up \citep{Karakas_etal02}.

Due to their particular relevance, below we will discuss in more detail 
a few analytic relations adopted in the present version of \texttt{COLIBRI}.    
\subsection{The third dredge-up: the need for a parametric description}
\label{ssec_3dup}
It is common practice describing the third dredge-up by means of two characteristic quantities, namely:
\begin{itemize}
\item $M_{\rm c}^{\rm min}$: the minimum core mass for the onset of the third dredge-up;
\item $\lambda= \displaystyle\frac{\Delta M_{\rm dup}}{\Delta M_{{\rm c, tpc}}}$: the efficiency of the third dredge-up, defined as the fraction of the core-mass growth over 
the interpulse period that is dredged-up to the surface at the next TP.
\end{itemize}

Compared to earlier computations, recent full TP-AGB evolutionary models 
have allowed a wide exploration of the third dredge-up 
characteristics as a function of stellar mass 
and metallicity \citep[e.g.][]{Karakas_etal02, 
Herwig_00, Herwig_04a, Herwig_04b, WeissFerguson_09, Cristallo_etal11}.
A few general trends can be extracted from these calculations. 

The efficiency $\lambda$ is expected to increase with stellar mass $M$, 
such that TP-AGB stars  with initial masses  $M > 3\,M_{\odot}$ are predicted to reach 
$\lambda \simeq 1$, which implies no, or very little, core mass growth. 
Lower metallicities favour an earlier onset of 
the third dredge-up and a larger efficiency,  
resulting in an easier formation of low-mass carbon stars.
Full TP-AGB models exist which are found to reproduce, or at least to be reasonably 
consistent with, basic observables, such as the luminosity functions of carbon stars in the 
Magellanic Clouds \citep[e.g.][]{Stancliffe_etal05, WeissFerguson_09, Cristallo_etal11}.   

Together with these improvements, present TP-AGB models also document 
that the third dredge-up is plagued by 
severe theoretical uncertainties. They are due mainly to our 
still deficient knowledge of convection and mixing, as well to a nasty 
sensitivity of the depth of the third dredge-up  to technical and 
numerical details (see \citealt[][]{FrostLattanzio_96}, and \citealt[][]{Mowlavi_99b}
for thorough analyses).

As a consequence we still lack a robust assessment 
for  $M_{\rm c}^{\rm min}$ and $\lambda$, and these parameters
are found to vary considerably from author to author even for the same 
combination ($M_{\rm i}$,\,$Z_{\rm i}$) of initial stellar mass and metallicity. 
The theoretical dispersion is exemplified 
in Fig.~\ref{fig_3dup3z02}. The dynamical 
ranges of the parameters covered by the various sets of computations  
are large, amounting to almost a factor of 3 for the maximum $\lambda$
attained in a ($M_{\rm i}=3.0 \, M_{\odot}$, $Z_{\rm i}=0.02$) model, and more than
$\simeq 0.1\, M_{\odot}$ for $M_{\rm c}^{\rm min}$  for the 
($M_{\rm i}=2.0 \, M_{\odot}$, $Z_{\rm i}=0.008$) case.
It is clear that these variations propagate dramatically in terms of the
predicted stellar properties: significant differences are expected in
the luminosities spanned during the C star phase,
the final masses, the chemical yields, etc.
The situation appears even more unclear considering, for instance, that  
two independent sets of calculations, i.e. \citet{Stancliffe_etal05} 
and \citet{WeissFerguson_09}, with largely different  
predictions for $M_{\rm c}^{\rm min}$ 
(see the right-hand side panel of Fig.~\ref{fig_3dup3z02}) are
found by the authors to recover the same observable, i.e. the carbon star
luminosity function in the LMC. This uncomfortable convergence of the results 
is likely due to the combination of other critical parameters 
(e.g. efficiency $\lambda$, and  mass loss). 
In fact, it is differences in details of the chosen input physics, 
such as the treatment of convective boundaries and the inclusion 
or not of overshoot, that produces most of the variations seen in full models, 
such as those shown in Fig.~\ref{fig_3dup3z02}.

All these reasons amply justify the approach of taking $\lambda$ and 
$M_{\rm c}^{\rm min}$ (or, in alternative,  $\lambda$ and  the temperature parameter
$T_{\rm dup}$; see Sect.~\ref{ssec_tbdred}), as 
free parameters, and to calibrate them with the largest possible set of observations to reduce
the likely degeneracy between different factors.

\subsection{Properties of the pulse-driven convection zone}
\begin{figure} 
\resizebox{\hsize}{!}{\includegraphics{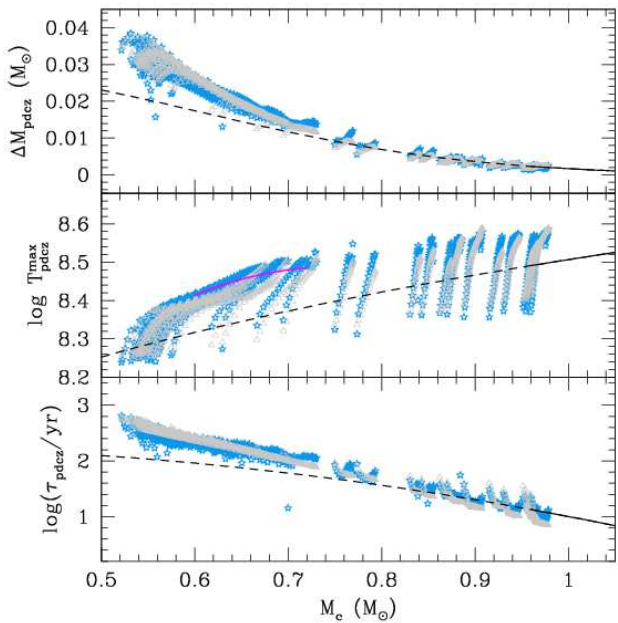}}
\caption{Characteristic quantities of the pulse-driven convection
zone, i.e. maximum mass, maximum bottom temperature, and 
duration, as a function of the core mass. 
A large set of $1658$  models from \citet{KarakasLattanzio_07},
 corresponding to various choices of stellar mass and metallicity,
is plotted (blue stars), together with our synthetic predictions 
(grey triangles) for the same stellar parameters. 
Other fitting relations are shown for comparison.
The \citet{Straniero_etal03} relation (magenta solid line) for 
$T_{\rm pdcz}^{\rm max}$ corresponds to a model with 
$M_{\rm i}=2.0\, M_{\odot},\, Z_{\rm i}=0.02$ and the core mass increasing from 
$0.60\, M_{\odot}$ to $0.72\, M_{\odot}$. The 
\citet{IbenTruran_78} relations (black line) are strictly valid for
high core masses  $(0.96 \la M_{\rm c}/M_{\odot} \la 1.33)$ 
(solid line portion), but they have been extrapolated to lower 
$M_{\rm c}$ (dashed line portion) for illustrative purpose only.}
\label{fig_pdczfit}
\end{figure}

In Fig.~\ref{fig_pdczfit} we show three key quantities of the 
PDCZ as a function of the core mass (starred symbols), 
as predicted by \citet{Karakas_etal02, KarakasLattanzio_07} for five values of the
initial metallicity 
$(Z_{\rm i}=0.0001,\,Z_{\rm i}=0.004,\,Z_{\rm i}=0.008,\,Z_{\rm i}=0.012,\,{\rm and}\,Z_{\rm i}=0.02)$.
Superimposed we plot the results obtained with the analytic relations (grey triangles) 
for the same stellar parameters ($M_{\rm i}$, $M_{\rm c}$, and $Z_{\rm i}$) 
as in the original full computations,
The fitting relations behave well all over the core-mass range covered by the full models.
The formulas and their coefficients are given in Appendix~\ref{app_fit}.

For comparison we draw two more relations taken from  
literature, namely \citet[][black line]{IbenTruran_78} 
and \citet[][magenta solid line]{Straniero_etal03}.
We have extrapolated the \citet{IbenTruran_78} relations over the whole
$M_{\rm c}$ range, but one should consider that they were originally 
derived from the high core mass $(0.96 \la M_{\rm c}/M_{\odot} \la 1.33)$ 
AGB models of \citet{Iben_77}.  We see that
for $M_{\rm c} \ga 0.85\, M_{\odot}$  the 
\citet{IbenTruran_78} 
relations for $\Delta M_{\rm pdcz}$ and 
$\tau_{\rm pdcz}$ are in general agreement
with the average trend predicted by the recent AGB computations of
\citet{KarakasLattanzio_07}. The earlier results of \citet{Iben_77} for $T_{\rm pdcz}^{\rm max}$
are systematically lower by up to $0.6-0.8$ dex.

The other relation proposed by \citet{Straniero_etal03}, on the basis
of their full AGB calculations, appears to be consistent with the 
\citet{KarakasLattanzio_07} results inside its validity range,
(i.e. $0.6 \la M_{\rm c}/M_{\odot} \la 0.7$). However, we notice that 
it does not allow to describe
the initial rise of the temperature typical of the first pulses.

\section{Tests: COLIBRI vs full stellar models}
\label{sec_tests}
\subsection{Effective temperature and convective-base temperature}
\label{ssec_teftest}
As a first test we compare the effective temperatures 
obtained with \texttt{COLIBRI} from envelope integrations 
(the method is outlined in Sect.~\ref{ssec_envmod}),  
against the predictions of full stellar models computed
with \texttt{PARSEC} \citep{Bressan_etal12}.
A detailed discussion is given in Appendix~\ref{sec_dtests}.

Figure~\ref{fig_dtefzvar} quantifies  the comparison in relation to 
the quiescent stage just preceding the occurrence of the $1^{\rm st}$ thermal
pulse for several values of stellar masses and metallicities.  
We see that the differences are in most cases  quite low,  
amounting to few tens of degrees, well below the typical observational errors 
for  $T_{\rm eff}$ of AGB stars, about equal to $\pm (100-200)$ K.

The results shown in the two panels of Fig.~\ref{fig_dtefzvar} 
differ in the chemical distributions of metals  assumed in \texttt{COLIBRI}.
They are usually expressed  in terms of the ratios $X_i/Z$, 
where $X_i$ denotes the fractional mass  of a given metal $i$. 
While in one case (top panel) both  EoS and opacities are computed with the 
\texttt{\AE SOPUS} and {\em Opacity Project} codes adopting, for each model,
the actual set of surface abundances predicted by \texttt{PARSEC} at the $1^{\rm st}$ TP, 
in the other case (bottom panel) the mixtures are assumed to be all scaled-solar
for any metallicity, i.e.  $X_i/Z=X_{i,\odot}/Z_{\odot}$ for each metal $i$.

\begin{figure} 
\begin{minipage}{\textwidth}
\resizebox{0.45\hsize}{!}{\includegraphics{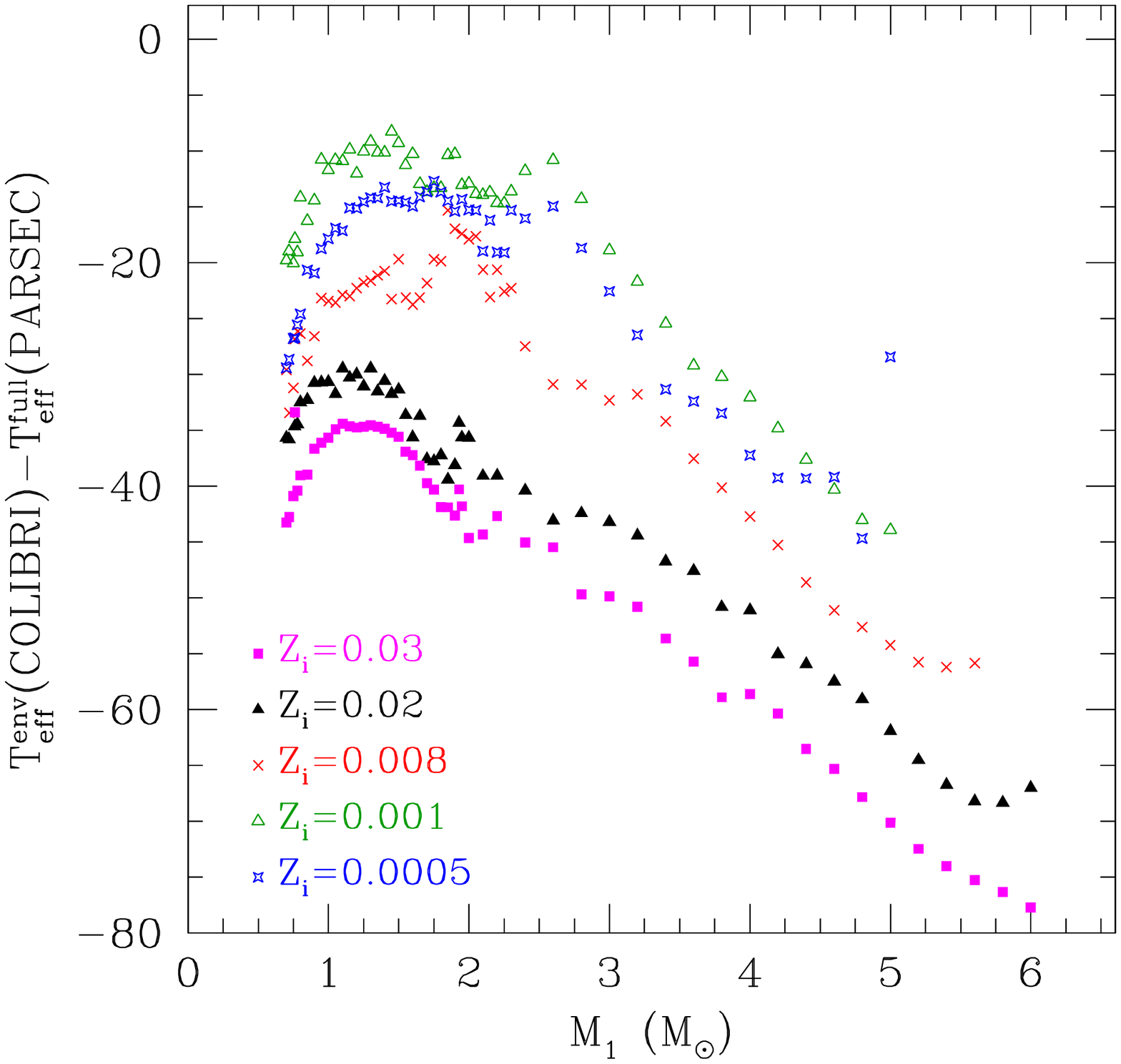}}
\end{minipage}
\vfill
\begin{minipage}{\textwidth}
\resizebox{0.45\hsize}{!}{\includegraphics{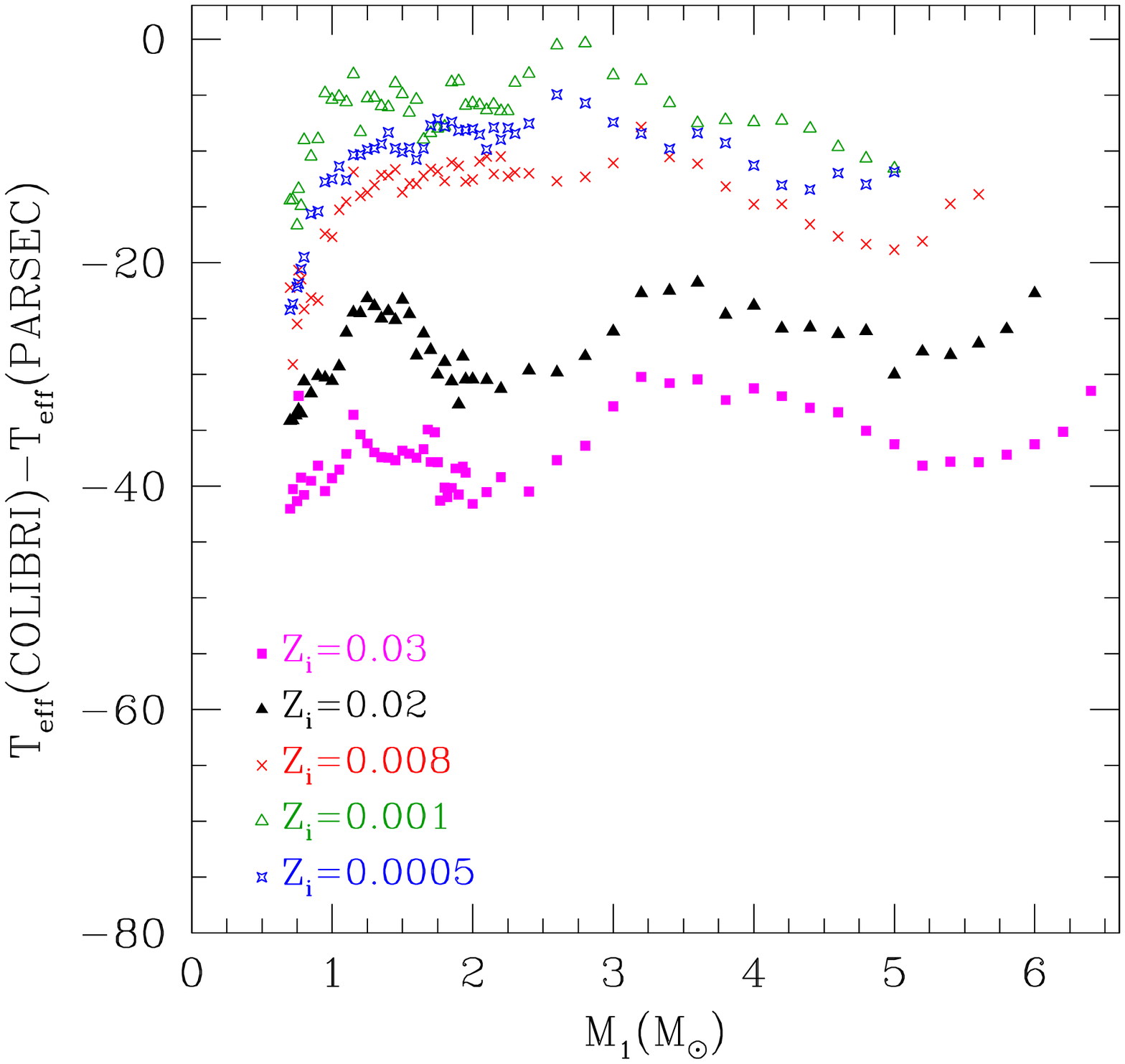}}
\end{minipage}
\caption{Differences in the predicted effective temperature between the
envelope-integration method adopted in the \texttt{COLIBRI} code 
and the \texttt{PARSEC} full evolutionary calculations. All models refer to the
pre-flash luminosity maximum just before  the $1^{\rm st}$ TP for 
several choices of the initial stellar mass and metallicity.
{\em Top panel}: The \texttt{COLIBRI} results are obtained 
with on-the-fly \texttt{\AE SOPUS} and {\em Opacity Project} 
computations for the EoS and opacities, 
consistently coupled to the actual chemical abundances across the ``deep'' 
envelope. {\em Bottom panel}: The \texttt{COLIBRI} predictions
are derived adopting a distributions of metals
frozen to the scaled-solar ratios for all metallicities, i.e. 
$X_i/Z=X_{i,\odot}/Z_{\odot}$, 
as assumed in \texttt{PARSEC}.}
\label{fig_dtefzvar}
\end{figure}
\begin{figure}
\begin{minipage}{\textwidth}
\resizebox{0.45\hsize}{!}{\includegraphics{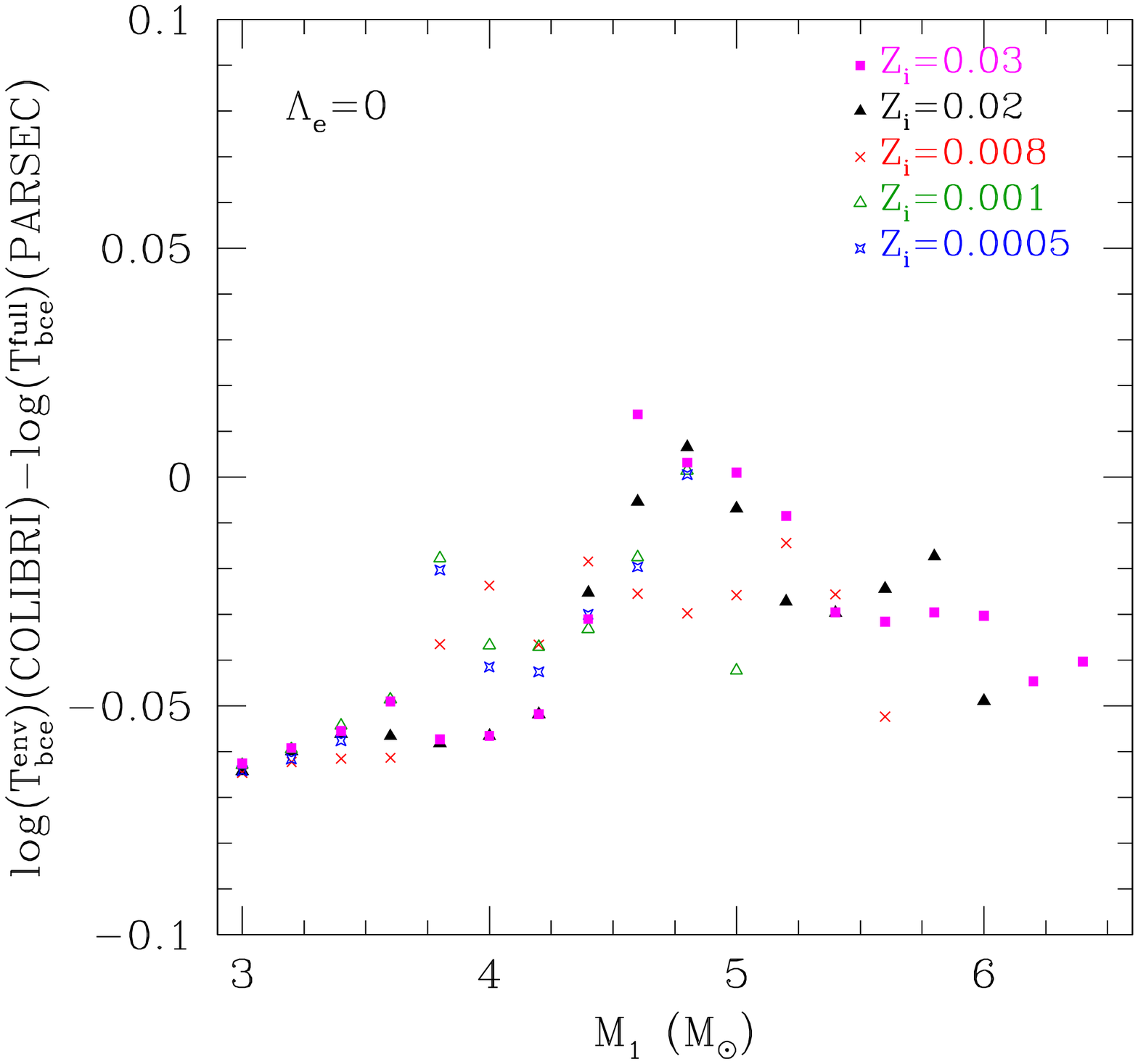}}
\end{minipage}
\vfill
\begin{minipage}{\textwidth}
\resizebox{0.45\hsize}{!}{\includegraphics{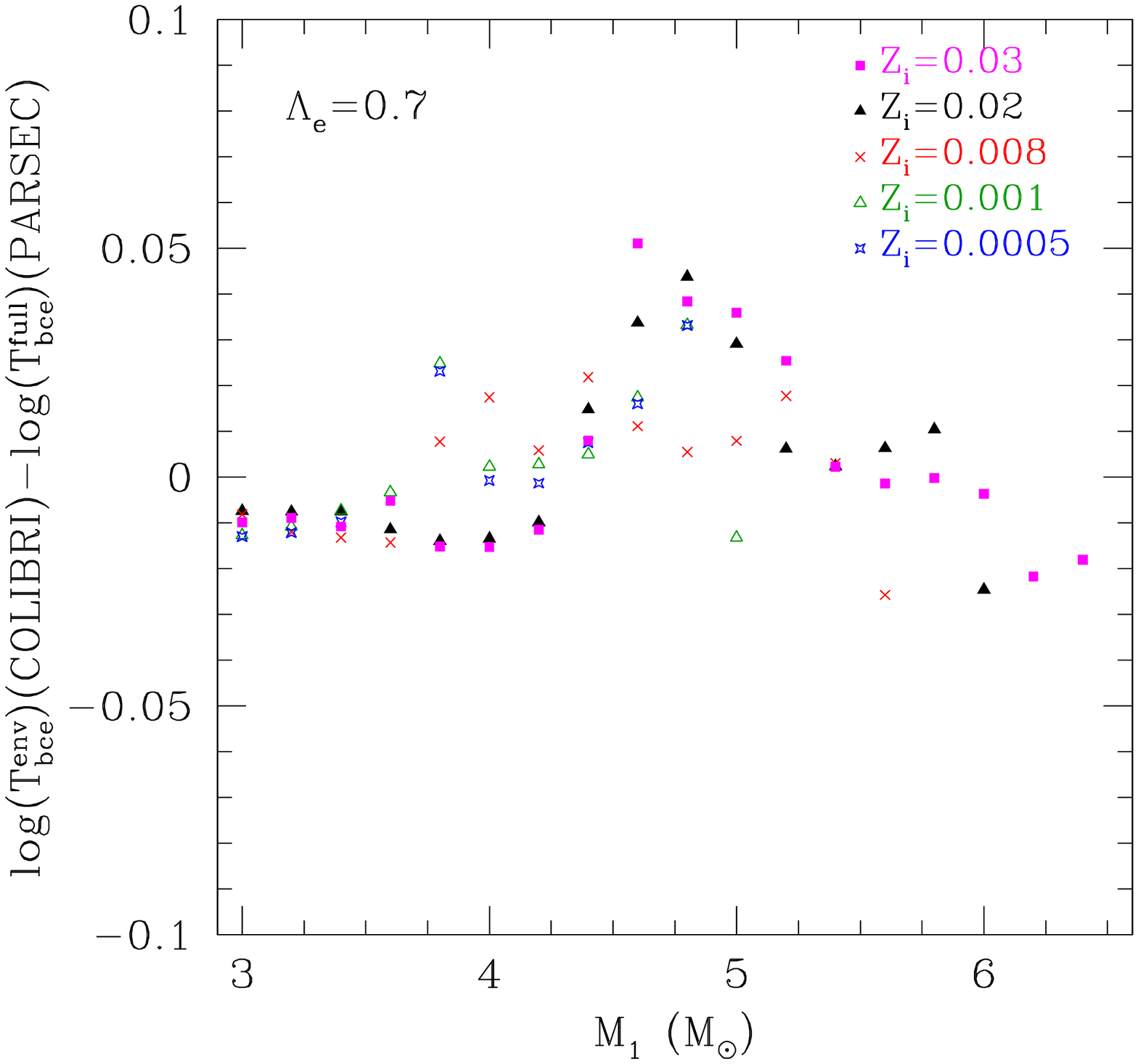}}
\end{minipage}
\caption{Differences in the predicted temperature at the base of the convective 
envelope  between the envelope-integration method adopted in the \texttt{COLIBRI} code 
and the \texttt{PARSEC} full evolutionary calculations. All models refer to the
pre-flash luminosity maximum just before  the $1^{\rm st}$ TP for several choices of the 
initial stellar mass and metallicity.
{\em Top panel}: 
\texttt{COLIBRI} models are computed adopting the classical
Schwarzschild criterion\protect\footnotemark ~to define the inner border of the convective envelope.
{\em Bottom panel}: The \texttt{COLIBRI} predictions
are derived with convective overshoot at the base of the 
envelope, extending over a distance $l_{\rm ov} = \Lambda_{\rm e}\times H_P$
(where $H_P$ is the pressure-scale height), with $\Lambda_{\rm e}=0.7$
as assumed in \texttt{PARSEC}.}
\label{fig_dtbot}
\end{figure}

In principle,  the former case is the correct one as it couples 
consistently EoS and opacities with the current metal abundances, that may
have varied with respect to the values at the zero-age main sequence, following
the $1^{\rm st}$ and second dredge-up processes.
On the other hand, the latter case, which is also adopted 
in the \texttt{PARSEC} models and, more generally,  by most full stellar codes,
neglects the variation of the elemental ratios,  e.g. the lowering of the C/O, due 
to mixing episodes prior to the TP-AGB phase.

It follows the accuracy degree of \texttt{COLIBRI} against \texttt{PARSEC } 
is best represented by the temperature differences in the bottom 
panel of Fig.~\ref{fig_dtefzvar}, since the same metal  ratios, 
$X_i/Z=X_{i,\odot}/Z_{\odot}$, are assumed in both sets of computations.
In fact, passing from the top to the bottom panel of Fig.~\ref{fig_dtefzvar} 
it is evident that the agreement between the \texttt{COLIBRI} and \texttt{PARSEC} 
predictions improves, particularly for models of larger
masses which are most affected by the second dredge-up. 
A more detailed discussion of this aspect and other related effects 
can be found in Appendix~\ref{ssec_teff}. 

The temperature at the base of the convective envelope, $T_{\rm bce}$,
provides an additional  test for our envelope-integration method, and
it is particularly relevant for massive AGB models ($M > 4\,M_{\odot}$)
as it measures the efficiency of HBB.
As analysed in Appendix~\ref{ssec_tbot}, the results are affected by several technical
details not dealing with the envelope integration method,  such as differences in the 
operative definition of the convective border,  inclusion or not
of convective overshooting, assumed metal partitions,
adopted equation of state, high-temperature opacities, etc.
All these aspects, together with the fact that the base of the convective
envelope may fall inside a region characterized by an extremely steep 
temperature gradient, concur to somewhat amplify the differences in $T_{\rm bce}$.

Figure~\ref{fig_dtbot} shows the temperature differences between \texttt{COLIBRI} and
\texttt{PARSEC } predictions for initial masses $M_{\rm i} \ge 2.6\, M_{\odot}$ and various 
metallicities. Two cases are considered in the \texttt{COLIBRI} 
definition of the innermost stable mesh-point of the convective envelope, namely:  the strict application of the
Schwarzschild criterion (top panel), and the inclusion 
of convective overshoot by the same amount as adopted in  \texttt{PARSEC} (bottom panel).
In both cases the differences remain fairly small, i.e.
$|\log(T_{\rm bce}^{\rm full}) - \log(T_{\rm eff}^{\rm env})| < 0.05$ dex.
\footnotetext{According to the Schwarzschild criterion the
border of a convective region is the layer at which the equality 
$\nabla_{\rm rad}=\nabla_{\rm ad}$ holds, where the $\nabla_{\rm rad}$, 
$\nabla_{\rm ad}$ denote the radiative temperature gradient 
and the adiabatic temperature gradient, respectively.}

\begin{figure*} 
\begin{minipage}{0.48\textwidth}
\resizebox{0.9\hsize}{!}{\includegraphics{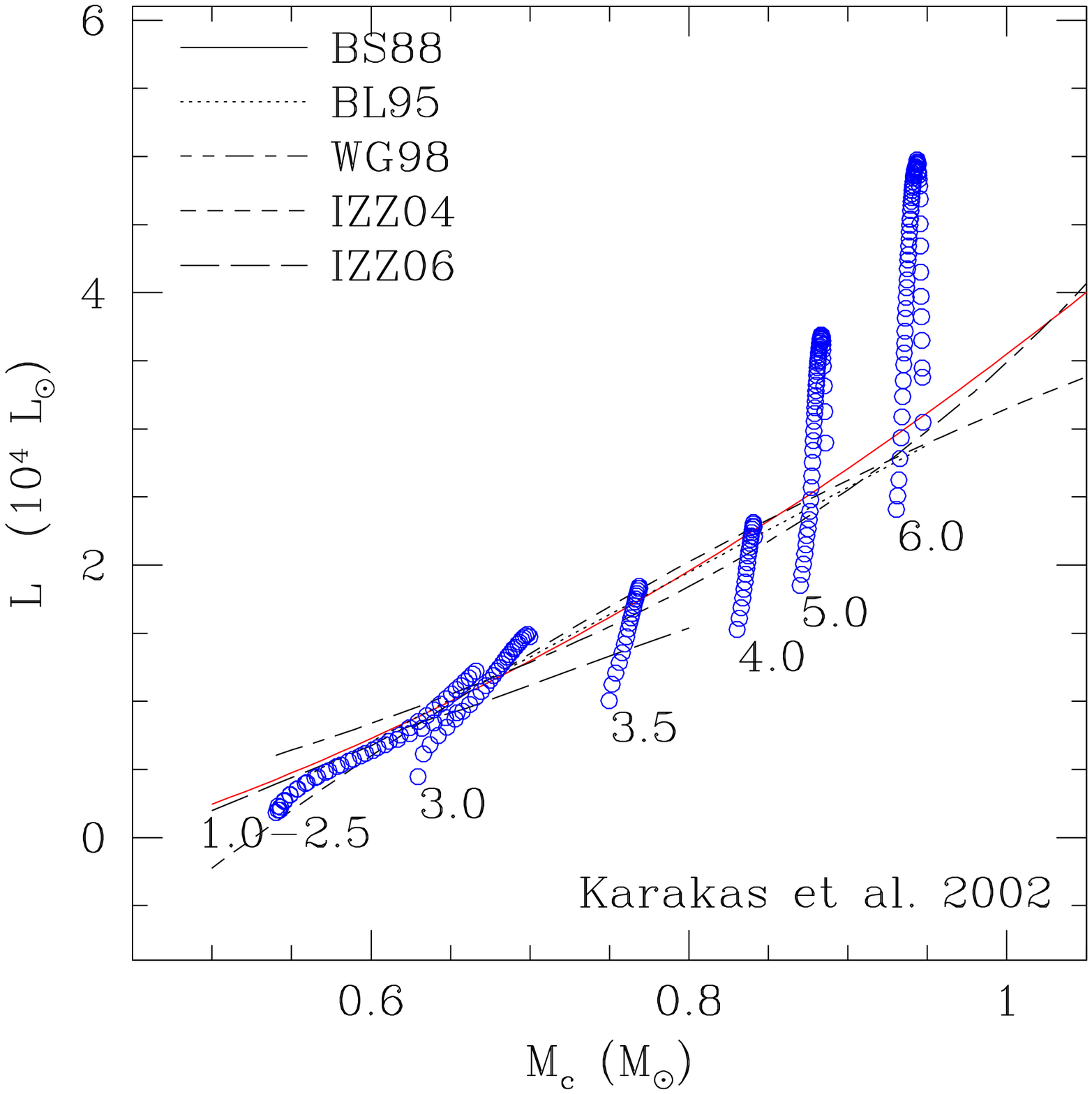}}
\end{minipage}
\hfill
\begin{minipage}{0.48\textwidth}
\resizebox{0.9\hsize}{!}{\includegraphics{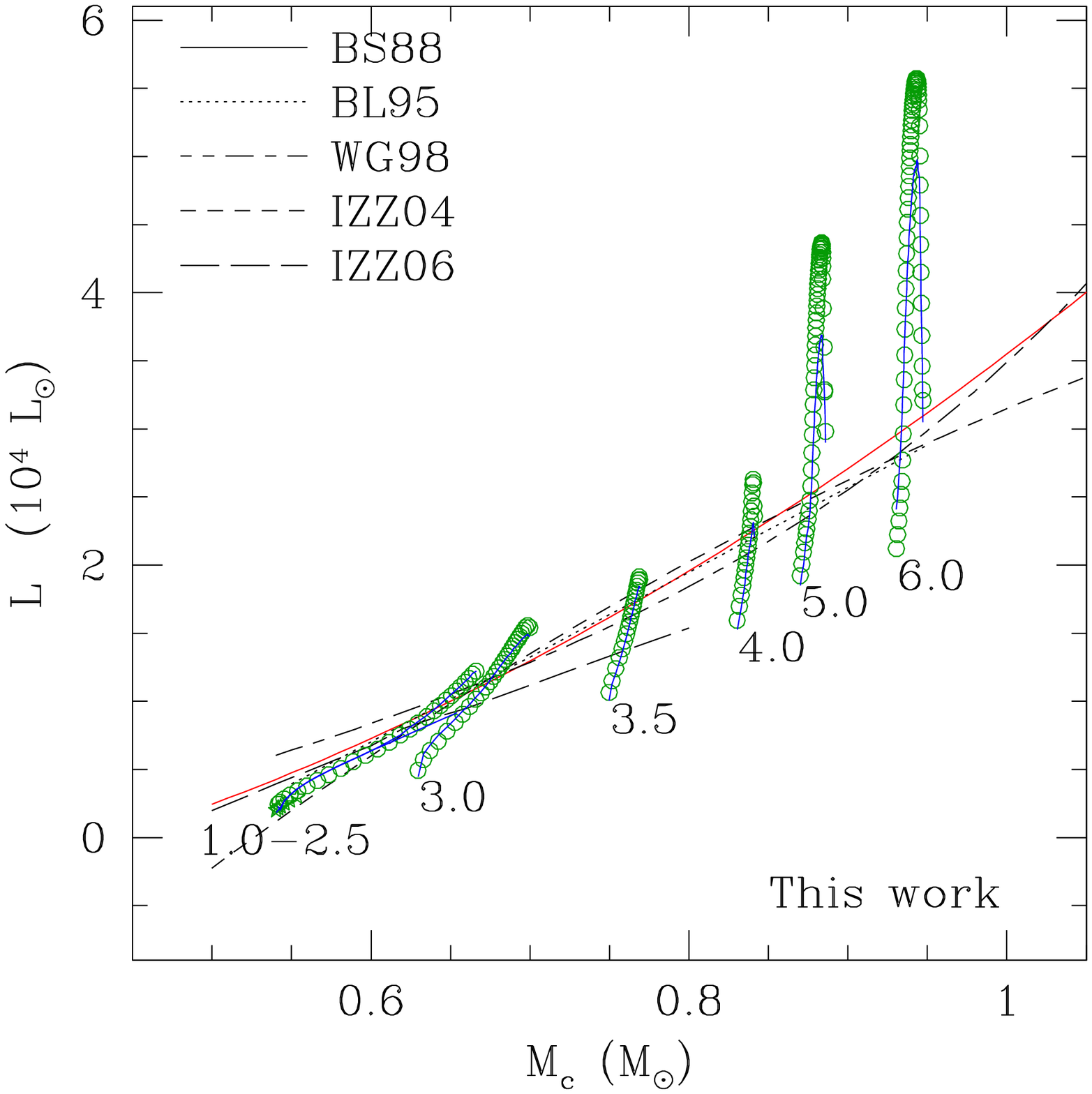}}
\end{minipage}
\caption{Stellar luminosity as a function of the core mass 
at the quiescent pre-flash stage preceding each thermal pulse,
for two sets of TP-AGB models with initial metallicity $Z_{\rm i}=0.008$.
{\em Left-hand side panel}: full TP-AGB models
by \citet{Karakas_etal02} (empty blue circles).
{\em Right-hand side panel}: predictions of our {\em deep envelope} integrations 
that include the H-burning shell (empty green circles). 
To facilitate comparison we overplot the \citet{Karakas_etal02} 
predictions (solid blue lines).
The initial stellar masses (in $M_{\odot}$) are quoted nearby the 
corresponding sequences.
A few CMLRs from various authors are also plotted, namely:
\citet[][BS88]{BoothroydSackmann_88a}, \citet[][BL95]{Bloecker_95},
\citet[][WG98]{WagenGroen_98}, \citet[][IZZ04]{Izzard_etal04}, 
\citet[][IZZ06]{Izzard_etal06}.
Note the effect of HBB which makes more massive TP-AGB models
to deviate significantly from the CMLRs towards higher luminosities.
See text for more details.}
\label{fig_cmlr}
\end{figure*}

In conclusion our tests indicate that:
\begin{itemize}
\item the agreement in effective temperatures between our envelope
integrations and full stellar modelling is extremely good,  with differences
$|T_{\rm eff}^{\rm env} - T_{\rm eff}^{\rm full} |< 40$ K and in many cases
practically negligible; 
\item  the differences $T_{\rm eff}^{\rm env} - T_{\rm eff}^{\rm full}$ are always negative
and tend to systematically decrease at lower metallicity, suggesting that
they are likely related to the elemental abundances and the way they are treated 
in the EoS and opacity computations. Indeed cooler $T_{\rm eff}^{\rm env}$ 
compared to  $T_{\rm eff}^{\rm full}$
are partly explained by the differences in  the assumed $X_i/Z$ used in the 
EoS and opacities, i.e. actual chemical abundances in \texttt{COLIBRI} against  
frozen scaled-solar ratios adopted by \texttt{PARSEC}.
\item A very good agreement is found also for $T_{\rm bce}$ (within $0.05$ dex), which strongly supports the ability
of our envelope-integration method to account correctly for the occurrence of HBB 
in more massive AGB models.
\end{itemize}

\subsection{Quiescent luminosity on the TP-AGB}
\label{ssec_lum}
Thanks to the extension of the {\em deep envelope} model to include the H-burning shell,
 we can predict the luminosity during the quiescent stages without 
adopting any auxiliary CMLR, as usually done in 
synthetic  TP-AGB  models \citep[e.g.][]
{Hurley_etal00, Izzard_etal04, Izzard_etal06, Cordier_etal07, MarigoGirardi_07}.

Figure ~\ref{fig_cmlr} shows  the pre-flash luminosity as a function of the core mass
 for two sets of TP-AGB models with initial metallicity $Z_{\rm i}=0.008$ and a few values
of the initial stellar mass, computed by \citet{Karakas_etal02}, 
and with the \texttt{COLIBRI} envelope-integration technique adopting the same stellar parameters (e.g. total mass, core mass, dredged-up mass, 
mixing-length parameter, and initial metallicity).
Considering that the two sets of calculations differ both in technical details 
(e.g. solution method of the stellar structure equations, zone-meshing, etc.) 
and in the input physics (e.g. EoS, opacities, nuclear reaction rates, etc.)
the overall agreement is quite striking. We derive two main implications: 
i) in absence of HBB,  i.e. for TP-AGB models with smaller cores  
($M_{\rm c} \la 0.75\,M_{\odot}$)  and less massive envelopes 
 ($M_{\rm env} \la 2.5\,M_{\odot}$), the CMLR is a robust prediction 
of the theory (essentially reflecting the thermostatic character of the H-burning shell), 
ii) in our {\em deep envelope} integrations the treatment 
of the H-burning energetics is reliable. 

In fact, in the range $0.5\, M_{\odot} \la M_{\rm c} \la 0.7\, M_{\odot}$ 
our predictions for the pre-flash luminosity maximum 
recover the \citet{Karakas_etal02} results remarkably well,  and more generally
the classical CMLRs \citep[e.g.][red line]{BoothroydSackmann_88a}.  
The brightening of the tracks beyond the CMLR, as shown by \citet{Karakas_etal02} models 
with  $M_{\rm c} \la 0.75\, M_{\odot}$ and $M \la 3.5\, M_{\odot}$, is driven by the occurrence of a deep 
third dredge-up. This effect is discussed in Sect.~\ref{sssec_lum3dup}.

At  larger core masses, $M_{\rm c} \ga 0.75\,M_{\odot}$ (see the models with initial
masses $M_{\rm i} = 4,\,5,\,6\, M_{\odot}$ in Fig.~\ref{fig_cmlr}), HBB is expected
to produce the break-down of the CMLR:  similarly to the tracks by \citet{Karakas_etal02},
the \texttt{COLIBRI} sequences  with $M \ge 4\,M_{\odot}$ exhibit a steep luminosity increase at almost constant core mass ($\lambda \simeq 1$ in these models).
After reaching a maximum, the luminosity starts to decline quickly from pulse to
pulse until the CMLR is recovered again. The luminosity peak and the subsequent 
decrease are controlled by the onset of the super-wind phase, which determines a rapid
reduction of the envelope mass, hence the weakening and eventual extinction of HBB.
We note that the \texttt{COLIBRI} tracks with HBB reach higher luminosity maxima 
than the \citet{Karakas_etal02} models with the same initial masses, a circumstance
that confirms the sensitivity of the HBB process on the adopted input physics and 
details of the convection treatment \citep{Ventura_etal05}.
 
\subsubsection{The effect of deep third dredge-up}
\label{sssec_lum3dup}
Full AGB calculation indicate that the occurrence of deep dredge-up events make
the models brighter than expected by the CMLR \citep{Herwig_etal98, Mowlavi_99b,
Karakas_etal02}, due to the intervening non-linear relation between  
the core mass and the core radius.

\begin{figure} 
\resizebox{1.0\hsize}{!}{\includegraphics{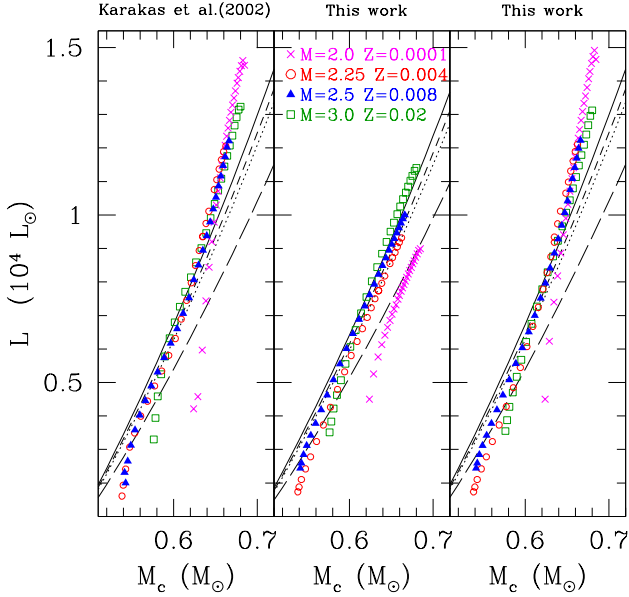}}
\caption{
Quiescent pre-flash luminosity maximum 
as a function of the core mass for a few combinations of 
initial masses (in $M_{\odot}$) and metallicities $Z_{\rm i}$, as indicated
in the middle panel.
{\em Left panel}:
Results from full TP-AGB models taken from \citet{Karakas_etal02, KarakasLattanzio_07}
Note the steep increase of the luminosities beyond the
CMLRs, which is interpreted as an effect driven by the third dredge-up. 
{\em Middle panel}: Results from deep envelope integrations described
in Sect.~\ref{ssec_envmod}. The boundary condition Eq.~(\ref{c4}) 
is used with the temperature corresponding to the true core mass, 
$M_{\rm c}$. {\em Right panel}:
Results from deep envelope integrations described
in Sect.~\ref{ssec_envmod}, but with the boundary condition Eq.~(\ref{c4}) 
evaluated using a fictitious core mass given by Eq.~(\ref{eq_mcfic}).
For comparison, the analytic CMLR of \citet{BoothroydSackmann_88a} 
is plotted for metallicities $Z_{\rm i}=0.02$ (solid line), $Z_{\rm i}=0.008$ (short-dashed line),
$Z_{\rm i}=0.004$ (dotted line), and $Z_{\rm i}=0.0005$ (long-dashed line).}
\label{fig_cmlr3dup}
\end{figure}

To account for this effect we have analysed a large number of full TP-AGB models
from \citet{Karakas_etal02}. 
These models are characterised by a large range of dredge-up efficiencies, from
$\lambda \approx 0$ to $\lambda \approx 1$, depending on stellar mass and metallicity.

We find that, in presence of dredge-up, 
the quiescent pre-flash luminosity $L_{\rm Q}$ of a TP-AGB model with a
core mass $M_{\rm c}$  is well recovered with our envelope-integration 
method by applying the boundary condition 
for the core temperature (Eq.~\ref{c4}) in the form
$T_{\rm c}=T(M_ {\rm c}^{\rm fict})$,
where we introduce a fictitious core mass
\begin{equation}
M_ {\rm c}^{\rm fict} = M_{\rm c} + \xi\,(M_{\rm c,nodup} - M_{\rm c, 1})\,,
\label{eq_mcfic}
\end{equation}
with the multiplicative factor $\xi \simeq 0.3-0.4$.

The variable $M_{\rm c, nodup}$ has been already used in past
  synthetic TP-AGB  models 
\citep[e.g.][]{Hurley_etal00, Izzard_etal04, Izzard_etal06}.
It was introduced to account for effects due to an increase 
in core degeneracy during the quiescent interpulse growth, so that
stars with the same core mass, 
but different dredge-up histories, may have different quiescent luminosities.
Since in \texttt{COLIBRI} the integrations of stellar structure 
are performed down to the bottom of the H-burning shell, 
for the electron-degenerate core beneath it
we need to resort to a parametrized description.   
The variable $M_{\rm c, nodup}$ is a suitable choice  
for the case under consideration.

The results are illustrated in Fig.~\ref{fig_cmlr3dup},  where   
the \texttt{COLIBRI} tracks computed with Eq.~(\ref{eq_mcfic}) setting $\xi =  0.3$
(right-hand side panel)  are compared to the original sequences 
\citet{Karakas_etal02} (left-hand side panel). 
Despite the simple formulation of the corrective term in 
Eq.~(\ref{eq_mcfic}), 
the agreement is quite satisfactory.

It is also instructive to look at the middle panel of Fig.~\ref{fig_cmlr3dup} 
showing the \texttt{COLIBRI} predictions for $\xi=0$, i.e. without the effect 
of the third dredge-up.
In this case all the tracks comply with the classical CMLR by
\citet{BoothroydSackmann_88a}, and reproduce quite well  the dimming of the quiescent 
luminosity at decreasing metallicity. As a matter of fact, the TP-AGB models from which
\citet{BoothroydSackmann_88a} derived their analytic  CMLR
were characterised by rather shallow, in most cases absent,  
convective dredge-up events, and  were mostly limited to the first 
few thermal pulses. This fact explains why the over-luminosity effect 
due to the third dredge-up does not show up in the  
\citet{BoothroydSackmann_88a} models.

It follows that the very nice accordance between the CMLR of
\citet{BoothroydSackmann_88a} and the \texttt{COLIBRI} predictions for $\xi=0$
adds a further confirmation on the validity of our envelope-integration method 
in terms of the H-burning energetics  (see also Sect.~\ref{ssec_lum}).

\subsection{Computational agility}
\label{ssec_cpu}
A key feature of the \texttt{COLIBRI} code is the computational agility,
that is kept to competitive levels despite the several numerical operations
performed at each time step,
i.e. iterative solution of the atmosphere and envelope structures, 
integration of nuclear networks,  {\em on-the-fly} computation of the EoS and Rosseland mean 
opacities across all meshes.

\begin{figure} 
\resizebox{\hsize}{!}{\includegraphics{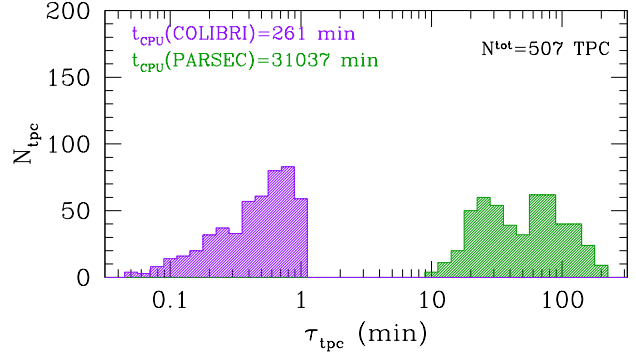}}
\caption{Distributions of CPU times relative to one thermal pulse cycle,
either with  the \texttt{COLIBRI} code  (purple histogram), or with  
the \texttt{PARSEC} stellar evolution code (green histogram).
In both cases the  sample consists of $N=507$ complete pulse cycles.}
\label{fig_cpu}
\end{figure}

Figure~\ref{fig_cpu} compares the performance of  
the \texttt{COLIBRI} and the \texttt{PARSEC} codes, in terms of the typical 
CPU time required to compute one thermal pulse cycle, i.e. 
the time interval between two consecutive pre-flash luminosity maxima.
The two histograms correspond to the distributions of 
$N^{\rm tot}_{\rm tpc}= 507$ thermal pulse cycles followed over a 
wide range of initial stellar masses $(0.6 M_{\odot}\, \la M_{\rm i} \la 6 M_{\odot})$,
and metallicities  $(0.0005 \le Z_{\rm i} \le 0.07)$.

The difference in CPU time\footnote{In our discussion we
refer to the CPU time taken by a typical $2.2$-GHz processor.} 
requirements is noticeable.
The \texttt{COLIBRI} distribution shows a broad peak at 
$ \tau_{\rm tpc}\sim 30-40$ s, and a low tail extending down to 
$3-4$ s. The median of the distribution is $\tilde{\tau}_{\rm tpc}\simeq 14$ s.
Bins at longer $\tau_{\rm tpc}$ are populated by TP cycles referring 
to i) the last TP-AGB stages in which the high mass-loss rates impose 
the reduction of the evolutionary time steps, and ii) more massive 
AGB stars experiencing both the third dredge-up and HBB, with consequent
intensive computing of EoS and opacities to follow the continuous changes 
in the envelope chemical abundances.

The \texttt{PARSEC} distribution is located over much
longer time scales, with  $\tau_{\rm tpc}$  ranging from  $\approx 10$ min to
$\approx 200$ min. The median of the \texttt{PARSEC} distribution 
is $\tilde{\tau}_{\rm tpc} \simeq 29$ min.

\begin{figure*}  
\begin{minipage}{0.48\textwidth}
\resizebox{0.8\hsize}{!}{\includegraphics{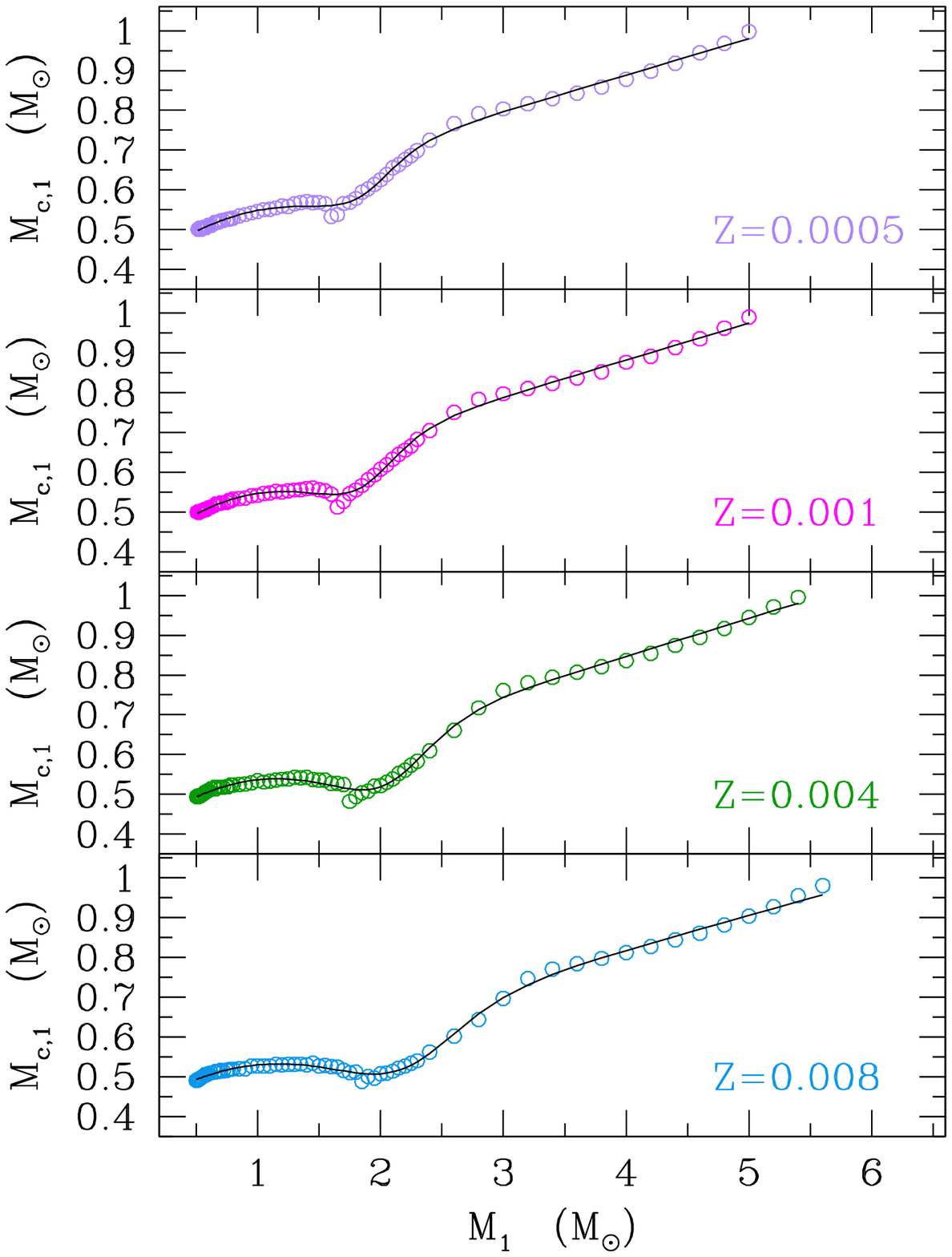}}
\end{minipage}
\hfill
\begin{minipage}{0.48\textwidth}
\resizebox{0.8\hsize}{!}{\includegraphics{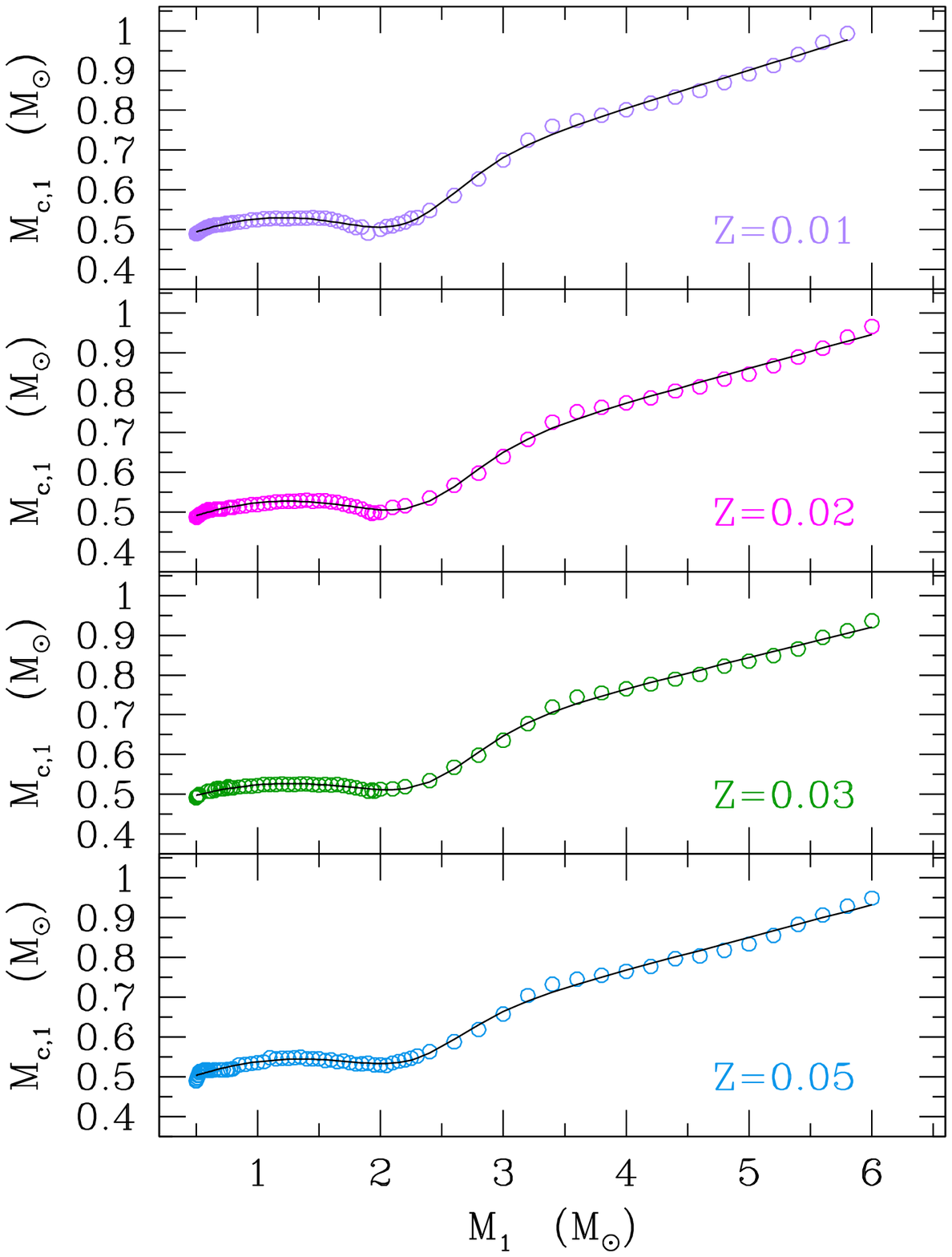}}
\end{minipage}
\caption
{Core mass as a function of the stellar mass at the stage of the pre-flash luminosity maximum,
just preceding the occurrence of the $1^{\rm st}$ thermal pulse. 
 The data, extracted from the  \texttt{PARSEC} database of stellar models 
\citep{Bressan_etal12}, are shown for eight choices of the
initial metallicities, as indicated. In each panel the solid line is the fit obtained with Eq.~(\ref{eq_mc1}),
and the coefficients given in Table~\ref{tab_mc1}.}
\label{fig_mc1}
\end{figure*}

In any case, the gain in terms of CPU time with \texttt{COLIBRI} is sizable:
the integrated CPU time to compute  $N^{\rm tot}_{\rm tpc}= 507$ thermal
pulse cycles is roughly $4$ hours for \texttt{COLIBRI} and $\simeq 21$
days for \texttt{PARSEC}.

While we acknowledge that the continuing  increase 
in computing speed of modern computers enables present-day 
full evolution codes to compute extended grids of TP-AGB tracks,
we should also realize that performing a
multi-parametric fine calibration of the 
uncertain processes/assumptions is extremely more  demanding 
in terms of computational agility and numerical stability, characteristics that
do not ordinarily apply to the full approach.

Processes and assumptions that are known to dramatically affect the 
TP-AGB evolutionary phase are, for instance, mass loss, 
third dredge-up, nucleosynthesis, convection efficiency, overshooting, 
initial chemical abundances, etc.
For each of them, we could single out more than one characteristic 
parameter,  depending on the theoretical picture 
one aims investigating at.
A dozen parameters may represent a reasonable estimate of the number
of factors one should take into consideration for an extensive 
analysis.

To get an order of magnitude of the time requirements, let us consider 
our specific working case.
At present we are dealing with $14$ metallicity sets (limited to the scaled-solar
compositions, other sets are planned), 
from very low to super-solar $Z$. From the \texttt{PARSEC} database of models, 
we extract the initial conditions at the first TP for $65-70$ values 
of the initial stellar mass (on average), 
from $\simeq 0.5 \, M_{\odot}$ to $\simeq 5-6 \, M_{\odot}$.
The fine grid in mass is important to allow for the construction of
accurate and detailed stellar isochrones.

The total number of TP-AGB tracks to be calculated
is $951$. With the set of parameters adopted in this exploratory
work, all the TP-AGB tracks followed by \texttt{COLIBRI} cover $14293$ 
thermal pulse cycles, for a true CPU time of 
$7854\,{\rm s} \simeq 5.2$ days.

With the conservative  assumption that the \texttt{PARSEC} code 
takes a computing time $\approx 100$ longer (probably more), 
the whole TP-AGB tracks would be ready after $\simeq 520$ days, that 
is $\approx 1.5$ yr.  These are likely  optimistic estimates,
considering that the current \texttt{PARSEC} distribution of CPU times   
is biased towards shorter values  since, in general, each evolutionary
track includes the first few
TPs, that usually involve a lighter computational effort compared 
to the later, well-developed TPs. Moreover, the PARSEC tracks are
calculated at constant mass, while the inclusion of a mass-loss prescription
would certainly impose a further reduction of the time steps, hence an increase
of the CPU time. 

It is also worth noting that the computing time request 
is expected to increase with the stellar mass, given that the pace at which 
TPs  take place correlates with the core mass, while HBB gets  stronger. 
In a recent study \citet{Siess_10}  reported that 
$\approx 6$ months of CPU time were required 
by his full evolution code to follow the whole Super-AGB phase 
of just one model with strong HBB. Of course, this may not be the same 
for other full codes,  but a trend of increasing computational cost with
the stellar mass is of general validity.

In any case, we emphasize here that 
what makes the computational effort particularly challenging for full models 
is the calibration process.
In fact, promptly producing extended and dense (in mass and metallicity) 
sets  of TP-AGB tracks is a necessary requisite to build accurate 
stellar isochrones spanning the whole relevant ranges of ages and 
metallicities. In turn, the stellar isochrones are the building blocks of
population synthesis simulations of galaxies including AGB stars,  
which can be readily put in direct comparison with observations. 
Possible discrepancies between predictions and observed data will 
bring the work-flow back to the theoretical side, and new sets of TP-AGB tracks
with a different set of input assumptions should be put in execution.  
This calibration cycle may be repeated several times before 
a satisfactory match between models and observations is attained.

Even before starting the calibration loop, in this preliminary and
exploratory phase, we have already computed
ten complete grids, for a total of $9510$ TP-AGB tracks, each time
changing a technical/physical parameter (e.g. an efficiency mass-loss
parameter, the mass meshing, the time-step regulation,  
or a subset of nuclear reaction rates).
It seems realistic that many more iterations, maybe hundreds, 
are necessary for an adequate global calibration.
As a consequence, numerical stability and computational agility are 
essential conditions, both fully met by our \texttt{COLIBRI} code.  

\section{Evolutionary tracks}
\label{sec_tracks}

We consider $14$ sets of stellar tracks covering a wide range of the initial metallicity, namely 
for $Z_{\rm i}=$0.0001, 0.0005, 0.001, 0.004, 0.006, 0.008, 0.01, 0.014, 0.017, 
0.02, 0.03, 0.04, 0.05, and 0.06 with initial scaled-solar abundances of 
metals. The reference solar mixture is that recently revised  by \citet{Caffau_etal11},  
corresponding to a Sun's metallicity $Z\!\simeq\!0.0152$.
\subsection{Up to the onset of the TP-AGB}
The evolution prior to the TP-AGB phase, from the pre-main sequence to the 
occurrence of the first TPs, is computed at constant mass 
with the \texttt{PARSEC} code, 
as described in the paper by \citet{Bressan_etal12} to which we refer for all details. 
We recall here only a few relevant points.
For each value of $Z_{\rm i}$, the initial helium abundance is determined
by the $Y_{\rm i}=0.2845+1.78\,Z_{\rm i}$ enrichment law.
The energy transport in the convective regions is described according
to the mixing-length theory of \citet{mlt_58}. The mixing length parameter 
$\alpha_{\rm MLT}$ is fixed by means of the solar model calibration, 
and turns out to be $\alpha_{\rm MLT}=1.74$. 
The PARSEC tracks include overshoot 
applied to the borders of both convective cores
and envelopes, with overshooting scales that vary
with the stellar mass as described  in \citet{Bressan_etal12}.   
Envelope overshoot is discussed also in Sects.~\ref{ssec_teftest} and \ref{ssec_tbot}, 
in relation to the accuracy checks performed on \texttt{COLIBRI} results. 

\begin{figure}  
\begin{minipage}{\textwidth}
\resizebox{0.48\hsize}{!}{\includegraphics{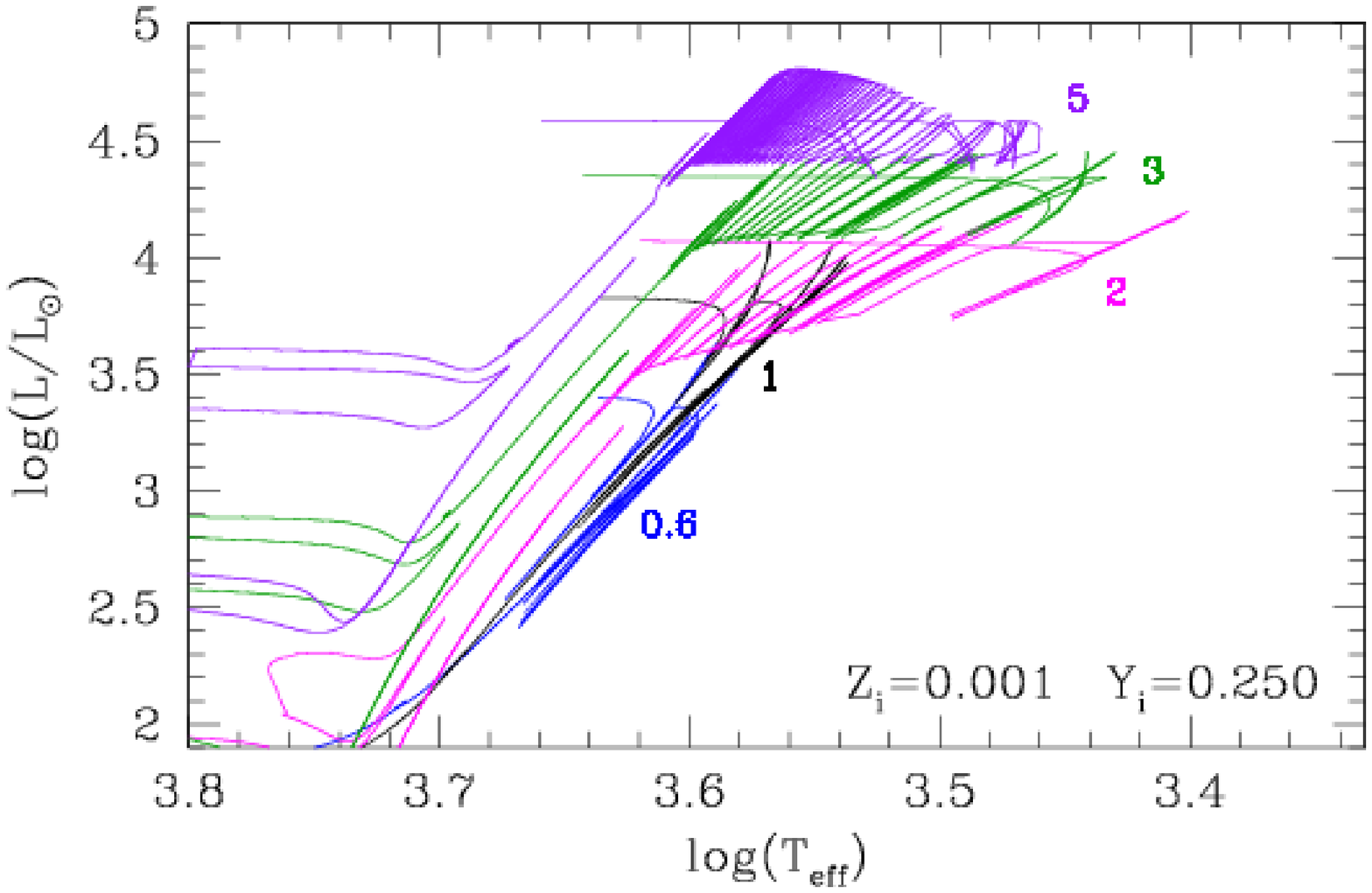}}
\end{minipage}
\hfill
\begin{minipage}{\textwidth}
\resizebox{0.48\hsize}{!}{\includegraphics{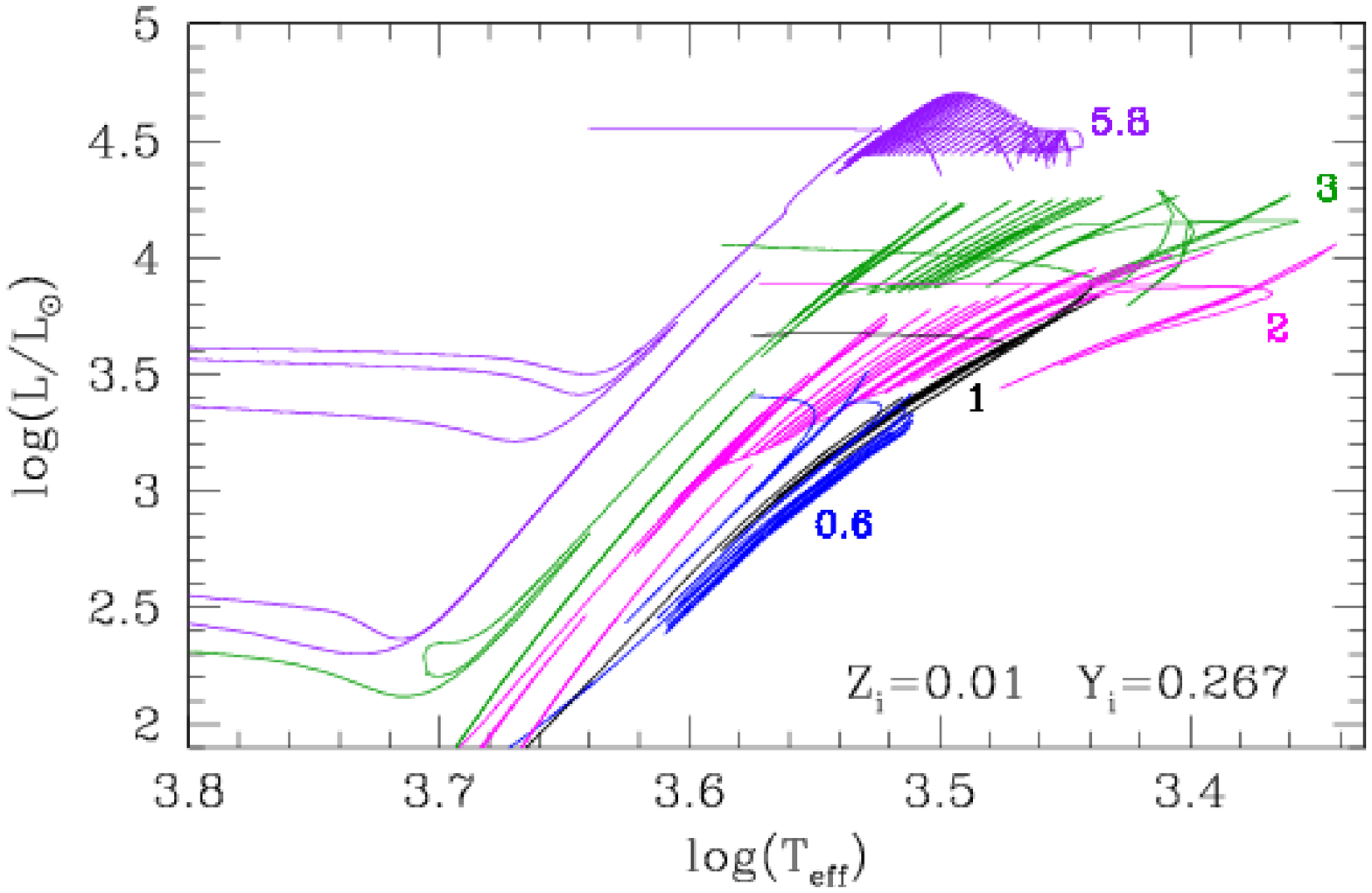}}
\end{minipage}
\caption
{Selected evolutionary tracks of low- and intermediate-mass stars, 
zooming in their coolest and brighter parts in the 
H-R diagram, for different values of the stellar mass 
at the onset of the TP-AGB phase 
(indicated in $M_{\odot}$ nearby the corresponding track), and for two
choices of the initial chemical composition.
The plots include the entire TP-AGB tracks calculated with \texttt{COLIBRI}, 
and a portion of the previous
evolution computed with \texttt{PARSEC}. Note the smooth transition
from \texttt{PARSEC} to \texttt{COLIBRI}.}
\label{fig_hr}
\end{figure}

\begin{figure*}  
\begin{minipage}{0.49\textwidth}
\resizebox{\hsize}{!}{\includegraphics{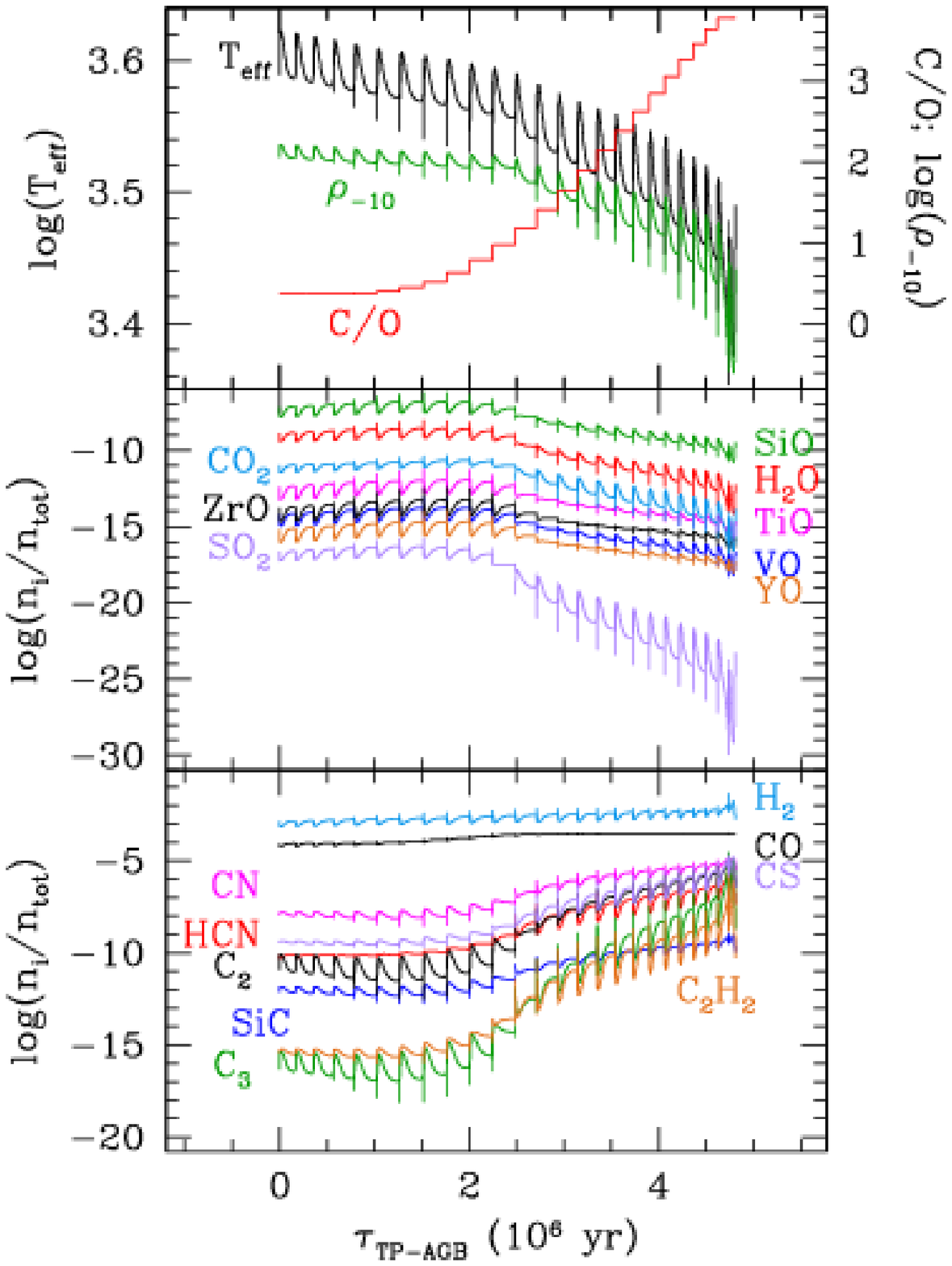}}
\end{minipage}
\hfill
\begin{minipage}{0.49\textwidth}
\resizebox{\hsize}{!}{\includegraphics{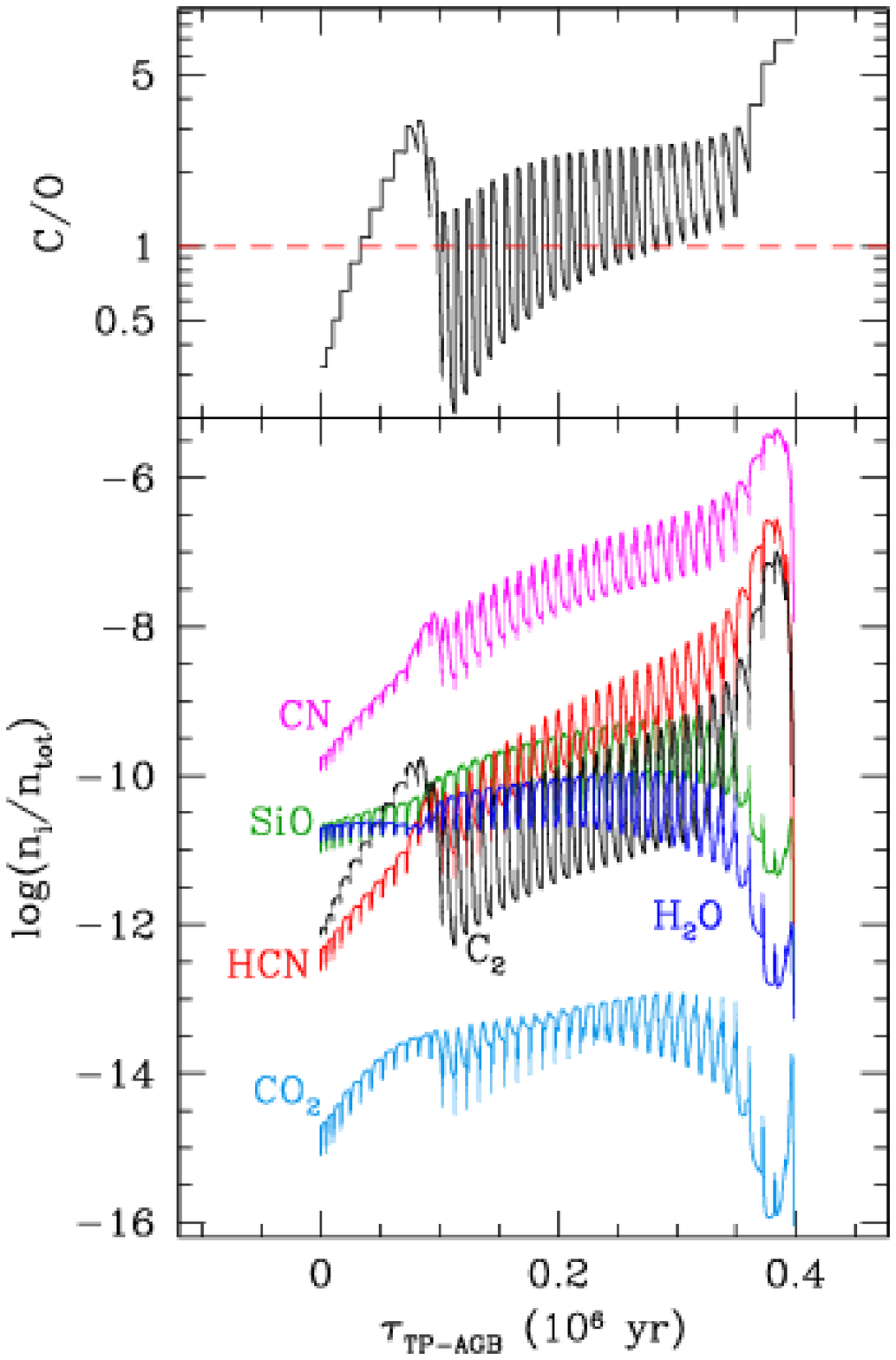}}
\end{minipage}
\caption{{\em Left-hand side panel:} Evolution of temperature, density, C/O ratio, 
and concentrations of the  most abundant molecular species in the gas phase at the
photosphere during the TP-AGB phase of a model with initial parameters 
$M_{\rm i}=2\,M_{\odot}$, $Z_{\rm i}=0.008$.
The star experiences several third dredge-up events so that
it is expected to become a carbon star. Note the huge dynamical range of
the molecular concentrations, up to $\simeq 25$ orders of magnitude!
{\em Right-hand side panel:}  Evolution of the photospheric C/O ratio 
and concentrations of six selected molecular species, among the most
abundant ones, during the whole TP-AGB phase of a model with initial parameters 
$M_{\rm i}=4\,M_{\odot}$, $Z_{\rm i}=0.0005$, experiencing both  very deep third 
dredge-up and efficient HBB. Note the key role of the C/O ratio in governing
the trends of the different molecules, as well as the several crossings at C/O$=1$.}
\label{fig_molchem}
\end{figure*}

For each \texttt{PARSEC} set of stellar tracks of given ($Z_{\rm i},Y_{\rm i}$) combination, 
we extract the initial conditions at the $1^{\rm st}$ TP 
for all the values of the initial stellar mass in the grid, 
ranging from  $\simeq 0.5\, M_{\odot} $ to $M_{\rm up}$, 
the latter being the maximum mass for a star 
to develop an electron-degenerate C-O core.
We deal typically with  $60-70$ low- and intermediate-mass  tracks for each 
initial chemical composition.

The core mass at $1^{\rm st}$ thermal pulse, $M_{\rm c,1}$, 
fixes a lower limit to the mass of the remnant white dwarf, and it is
closely connected to the initial-final mass relation.

Figure~\ref{fig_mc1} shows the \texttt{PARSEC} predictions for $M_{\rm c,1}$, 
as a function of the stellar mass for several choices of the initial 
metallicity.
The stellar mass, $M_1$,  is the value at the onset of the TP-AGB phase, 
so that, in principle,  one should correct for the amount of mass lost by low-mass stars 
$(M_{\rm i} \la 2\, M_{\odot})$ during 
the red giant branch (RGB) phase  in order to translate 
the $M_{\rm c,1}$ relation as a function of the initial stellar mass.

Two are the main features common to all the curves, namely: i)
 the almost constancy of $M_{\rm c,1}$ for stellar masses lower than 
$1.6-2.0\, M_{\odot}$ (depending on $Z_{\rm i}$), which simply reflects 
the fact that these stars develop He-cores of very similar mass due 
to the electron degeneracy after the main sequence; ii) the change of slope at stellar masses 
in the range $2.5-3.5\, M_{\odot}$ (depending on $Z_{\rm i}$) and the subsequent 
flattening of the $M_{\rm c,1}$ relations. This is 
the fingerprint of the occurrence of the second dredge-up 
during the Early-AGB of intermediate-mass stars, that causes a significant 
reduction of their core masses.

In Table~\ref{tab_mc1} of Appendix~\ref{app_fit} we present 
the fitting coefficients that we derive following the parametrization proposed 
by \citet{WagenGroen_98}, for several metallicities.
In each panel of Fig.~\ref{fig_mc1} the fitting curves 
are over-imposed to the \texttt{PARSEC} data for  $M_{\rm c,1}$.
We note, however, that our TP-AGB calculations use the true $M_{\rm c,1}$ values, and
not those derived from the formulas.  

\subsection{TP-AGB evolution}
\label{ssec_tpagbev}

For each stellar model with initial parameters $(M_{\rm i},Z_{\rm i})$
the characteristic quantities at the $1^{\rm st}$ thermal pulse 
(core mass, luminosity, effective temperature, envelope chemical composition), 
obtained from the {\em  PARSEC} database, are fed as initial conditions
to the \texttt{COLIBRI} code, which computes 
the  TP-AGB evolution until when almost the entire envelope is lost 
by stellar winds. Operatively the \texttt{COLIBRI} calculations are stopped
when the mass of the residual envelope falls below a limit of 
$0.002 M_{\odot} -0.005 M_{\odot}$. At this stage all evolutionary tracks are already
evolving off the AGB  towards higher effective temperatures, with  
a luminosity that depends mainly on the mass of the C-O core, and
the phase of the pulse cycle at which the last event of 
mass ejection took place (see Fig~\ref{fig_hr}).

For the present work we adopt a specific set of prescriptions for the mass loss
and the third dredge-up,
which we briefly outline below.
These models  will serve as a reference case for our ongoing TP-AGB calibration, 
and therefore the current parameters may be somewhat changed in future calculations.   
Anyhow, from various preliminary tests made with the present models, 
we expect that they already yield a fairly good description of the TP-AGB phase.
\paragraph*{Mass loss.}
It has been included under the hypothesis that it is driven by two main mechanisms, 
dominating at different stages. Initially, before radiation
pressure on dust grains becomes the main agent of stellar winds,
mass loss is described with the semi-empirical relation 
by \citet{SchroderCuntz_05}, which essentially assumes that the stellar wind originates from
magneto-acoustic waves operating below the stellar chromosphere. 
The corresponding mass-loss rates are indicated with 
$\dot M_{\rm pre-dust}$.

Later on the AGB the star enters the dust-driven wind regime, which is
treated with an approach similar to that developed by
\citet{Bedijn_88}, and recently adopted by \citet{Girardi_etal10}, to
which the reader is referred for all details.  Briefly, assuming that
the wind mechanism is the combined effect of two processes, i.e.,
radial pulsation and radiation pressure on the dust grains in the
outermost atmospheric layers, we adopt a formalism for the mass-loss
rate as a function of basic stellar parameters, mass $M$ and radius
$R$, expressed in the form $\dot M \propto e^{M^{a} R^{b}}$.  The free
parameters $a$ and $b$ have been calibrated on a sample of Galactic
long-period variables with measured mass-loss rates, pulsation
periods, stellar masses, radii, and effective temperatures.  More
details about the fit procedure will be given elsewhere.  We denote
the corresponding mass-loss rates with $\dot M_{\rm dust}$.

The key feature of this  formalism is that it predicts an exponential increase of the
mass-loss rates as the evolution proceeds along the TP-AGB, until typical
super-wind values, around $10^{-5}-10^{-4}\, M_{\odot} {\rm yr}^{-1}$, are 
eventually reached. 
The super-wind  mass loss is described in the same fashion as in 
\citet{VassiliadisWood_93},  and corresponds 
to a radiation-driven wind, $\dot M_{\rm sw}=L/c\, v_{\rm exp}$, where $c$ is the
speed of light and $v_{\rm exp}$ is the terminal velocity of the wind.

At any time during the TP-AGB calculations 
the actual mass-loss rate is taken as 
\begin{equation}
 \dot M = 
{\rm max}[\dot M_{\rm pre-dust}, {\rm min}(\dot M_{\rm dust}, 
\dot M_{\rm sw})].
\end{equation}

\paragraph*{The third  dredge-up.}
The onset of the third dredge-up is predicted according to the 
scheme described in Sect.~\ref{ssec_tbdred}. The minimum temperature parameter is set to
$\log(T_{\rm dup})=6.4$. This rather low value favours an  early
occurrence of the  third dredge-up episodes.
The efficiency $\lambda$ of the third dredge-up  is computed with 
the analytic fits provided by \citet{Karakas_etal02}, as a function of 
current stellar mass and metallicity. 

\begin{figure}
\resizebox{\hsize}{!}{\includegraphics{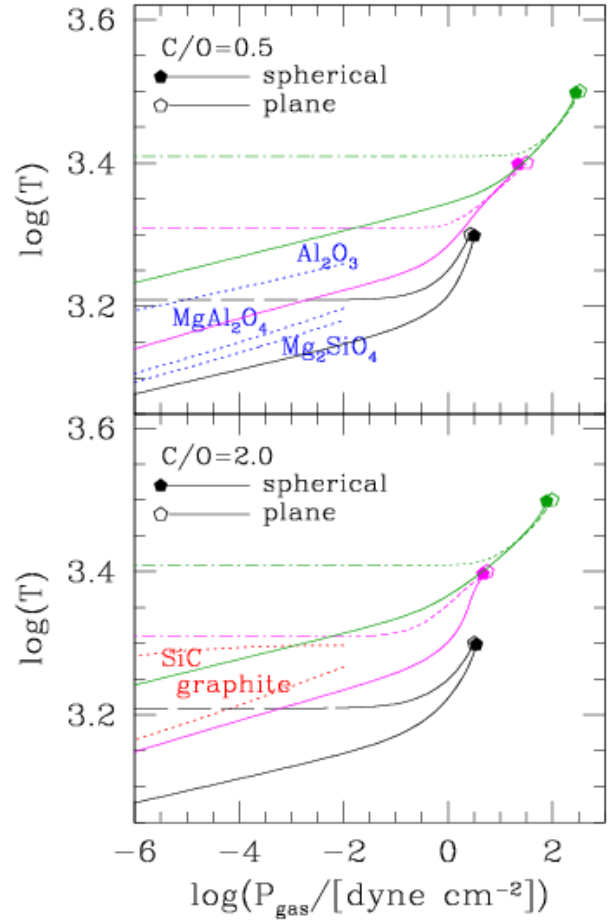}}
\caption{$P_{\rm gas}-T$ structure of static atmospheres corresponding to 
stellar model with $M=M_{\rm i}=1\,M_{\odot}$, $Z=Z_{\rm i}=0.008$, 
$X_{\rm i}=0.70$, $L=10^4\,L_{\odot}$,
and three choices of the effective temperature, i.e.  $\log(T_{\rm eff})=3.5,\,3.4,\,3.3$.
The thermodynamic stratification is shown
 for both plane-parallel (dot-dashed line) and spherically symmetric (solid line) geometries,
and assuming either C/O$=0.5$ (top panel), or C/O$=2.0$ (bottom panel). 
The photospheres  are indicated by pentagons.
The dotted lines correspond to the condensation temperatures at varying 
gas pressure for a few relevant species, namely: corundum (Al$_2$O$_3$), spinel
(MgAl$_2$O$_4$), and forsterite (Mg$_2$SiO$_4$) as predicted by  
\citet{LoddersFegley_99} for C/O$=0.5$  (top panel); graphite (C) and silicon carbide 
(SiC) from \citet{LoddersFegley_95} for C/O$=2.0$. }
\label{fig_atmo}
\end{figure}

Figure~\ref{fig_hr} illustrates a few selected evolutionary tracks of 
low- and intermediate-mass stars, zooming in their brightest portions
in the H-R diagram, that include the
whole TP-AGB computed with \texttt{COLIBRI} and some earlier evolution  
calculated with  \texttt{PARSEC}.
The transition from \texttt{PARSEC} to \texttt{COLIBRI} is not even 
distinguishable in most cases, except for the higher mass models with HBB 
($M_{\rm i}=5.0 M_{\odot},\,Z_{\rm i}=0.001$ and  
$M_{\rm i}=5.8 M_{\odot},\,Z_{\rm i}=0.01$) for which
 \texttt{COLIBRI} predicts somewhat cooler effective temperature at the
$1^{\rm st}$ TP compared to \texttt{PARSEC}.
This difference has been discussed in Sect.~\ref{ssec_teftest}, and can be partly
explained in terms of the small differences in molecular opacities  adopted by the two codes
(see Fig.~\ref{fig_dtefzvar}).

\begin{figure*}
\begin{minipage}{0.48\textwidth}
	\resizebox{\hsize}{!}{\includegraphics{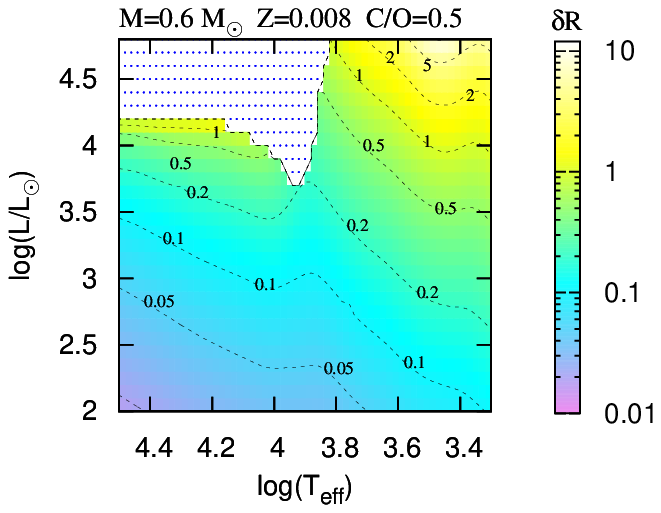}}
\end{minipage}
\begin{minipage}{0.48\textwidth}
	\resizebox{\hsize}{!}{\includegraphics{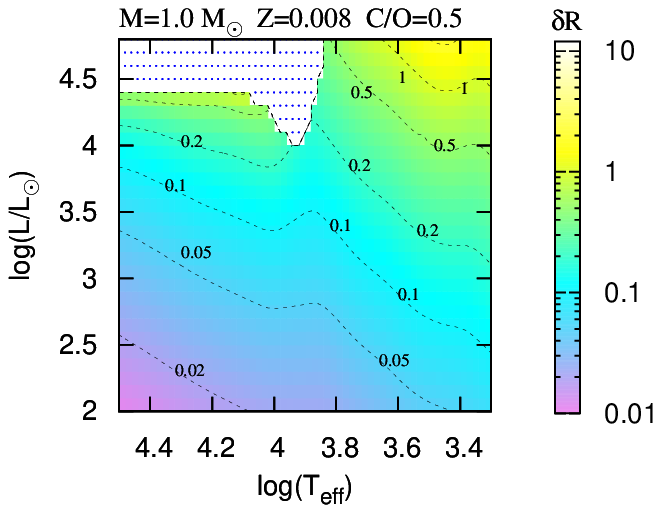}}
\end{minipage}
\vfill
\begin{minipage}{0.48\textwidth}
	\resizebox{\hsize}{!}{\includegraphics{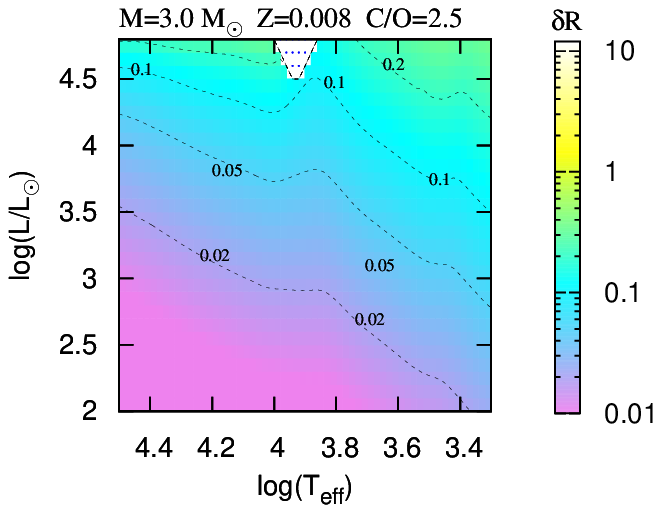}}
\end{minipage}
\begin{minipage}{0.48\textwidth}
	\resizebox{\hsize}{!}{\includegraphics{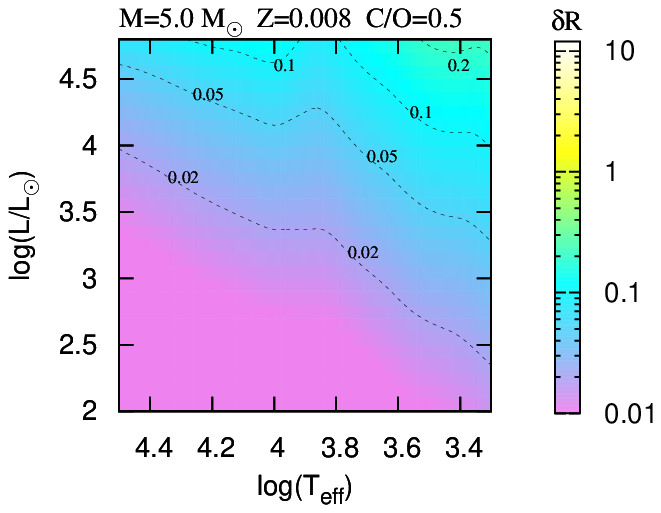}}
\end{minipage}
\caption{Maps of geometrical thickness $\Delta R$ of static stellar atmospheres 
in the H-R diagram, for four
choices of the stellar mass and metallicity $Z_{\rm i}=0.008$, as indicated. 
The radial extension, defined by 
Eq.~(\ref{eq_dr}), is referred to the outermost radius at which $P_{\rm gas}$ has decreased
to $10^{-4}$ dyne cm$^{-2}$. 
Contour lines of constant $\delta R$ (as indicated) are superimposed.
The dotted regions correspond to 
unbound atmospheres, i.e. in which the radiative acceleration exceeds 
the gravitational acceleration somewhere between $R$ and $R_0$, so that 
the Eddington factor $\Gamma$ (Eq.~\ref{eq_gamma}) becomes larger than unity.
 }
\label{fig_sphat}
\end{figure*}

We also note in  Fig.~\ref{fig_hr}  that low-mass models  
($M_{\rm i}=0.6 M_{\odot},\,1.0 M_{\odot}$) are characterised by quite narrow TP-AGB
tracks since at given metallicity, as long as the surface C/O$<1$, 
the effective temperature is mostly determined
by stellar mass and luminosity. 
Differently, models with larger masses ($M_{\rm i}=2.0 M_{\odot},\,3.0 M_{\odot}$), which are
expected to undergo the  transition to carbon stars, exhibit a pronounced
displacement towards lower effective temperatures, mainly driven by the increase
in molecular opacities. Finally, models with the highest masses  
($M_{\rm i}=5.0 M_{\odot},\,5.8 M_{\odot}$) present TP-AGB tracks  with the 
typical bell-shape modulated by the occurrence of HBB,  and with the peak
in luminosity reached when the envelope mass starts being drastically 
reduced by stellar winds.
These considerations apply in general to both metallicity cases here considered 
($Z_{\rm i}=0.001$ and $Z_{\rm i}=0.01$), with some systematic differences, i.e.
lower effective temperatures are expected at higher metallicities, again due to
surface opacity effects. A more detailed analysis of this aspect is given in 
Sect.~\ref{ssec_hayashi}. 

Finally, we note that in our TP-AGB calculations 
no particular convergence problem was met all the way to  
the complete ejection of the envelope,  whereas other studies, 
based on full TP-AGB 
calculations, report the divergence of the models in the late
stages of evolution \citep[e.g.,][]{WoodFaulkner_86, WagenhuberWeiss_94, Lau_etal12}. 
In the latter paper the authors suggest that 
the cause of the instability in the most massive TP-AGB models 
may be related to a local opacity maximum of Fe at the base of the 
convective envelope. At present we cannot identify the reason for
the different behaviour of \texttt{COLIBRI}, 
this delicate point will deserve a closer look in follow-up studies.
\section{Overview and analysis of the COLIBRI predictions}
\label{sec_results}

In the following we will discuss some relevant predictions of the \texttt{COLIBRI} code,
with the aim of understanding a few key dependencies 
of the various physical processes at work and their complex interplay, as well as 
giving a general overview of the \texttt{COLIBRI} predictive capability. 
\subsection{Molecular concentrations at the photosphere}
The {\em on-the-fly} use of the \texttt{\AE SOPUS} code  during the TP-AGB 
calculations enables us to predict, for the first time, the evolution
of the abundances of $\simeq 500$ molecular species in the outermost
layers of the envelope. In Fig.~\ref{fig_molchem} (left-hand side panel) 
we show the results at the photosphere 
of a  $M_{\rm i}=2\,M_{\odot}$, $Z_{\rm i}=0.008$ model. We see clearly how the
occurrence of thermal pulses produces large variations of the photospheric 
temperature and density (top panel), which in turn cause similar ``pulses'' 
in the concentrations of the molecules. 

\begin{figure*} 
\begin{minipage}{0.49\textwidth}
\resizebox{0.9\hsize}{!}{\includegraphics{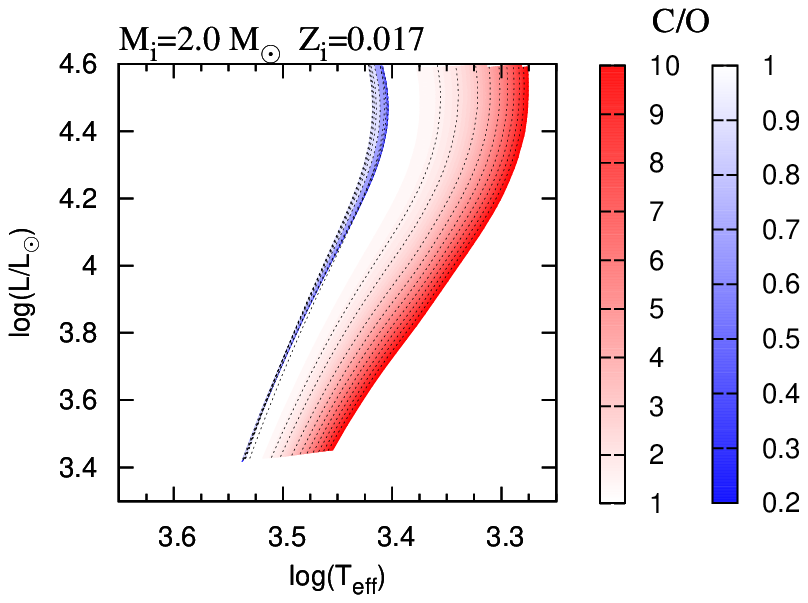}}
\end{minipage}
\begin{minipage}{0.49\textwidth}
\resizebox{0.9\hsize}{!}{\includegraphics{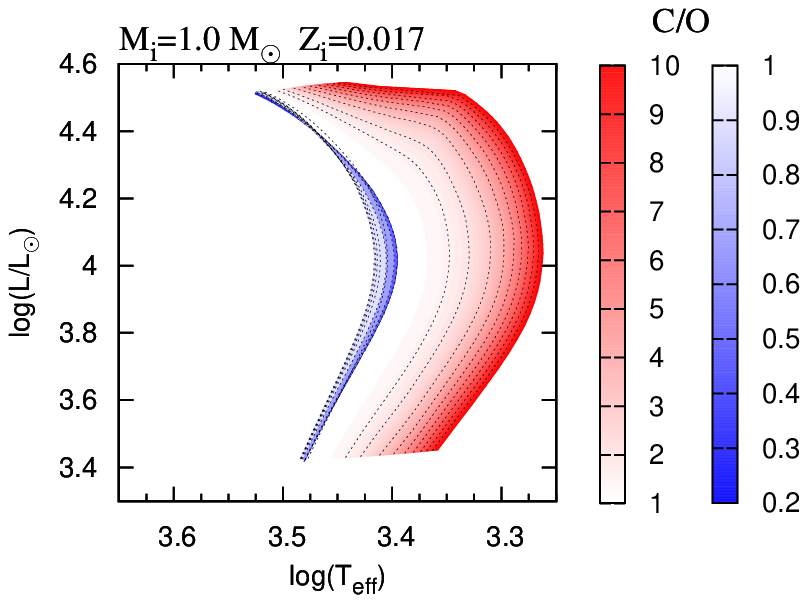}}
\end{minipage}
\begin{minipage}{0.49\textwidth}
\resizebox{0.9\hsize}{!}{\includegraphics{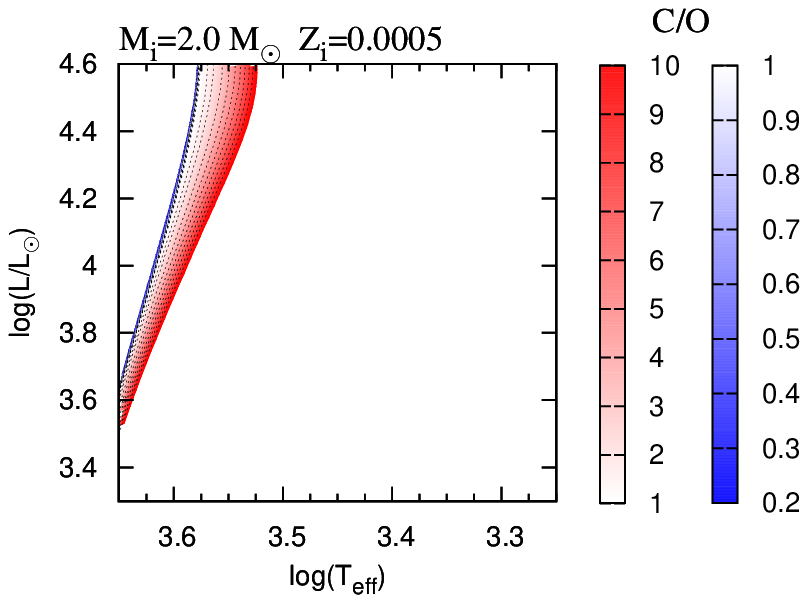}}
\end{minipage}
\begin{minipage}{0.49\textwidth}
\resizebox{0.9\hsize}{!}{\includegraphics{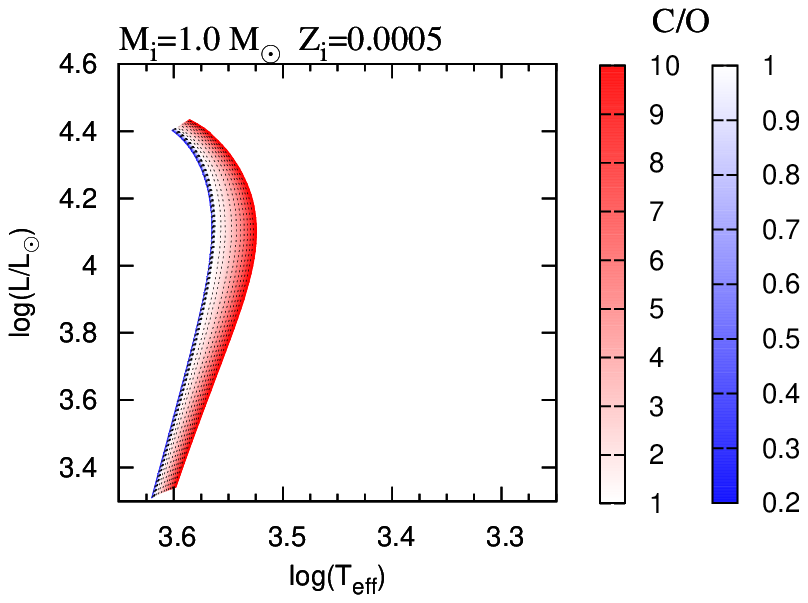}}
\end{minipage}
\caption{Hayashi lines on the AGB at increasing surface C/O ratio, color-coded
according to the scales at the right-hand side of each plot, i.e. shades of blue
for C/O$\le 1$, shades of red for C/O$> 1$.
Results are shown for constant stellar mass, 
$M=1.0\,M_{\odot}$ and  $M=2.0\,M_{\odot}$, 
and two choices of the initial metallicity $Z_{\rm i}=0.017$, and 
$Z_{\rm i}=0.0005$. Contour lines from C/O$=0.2$ to C/O$=10$, with
an incremental step $\Delta$(C/O)$=0.2$, are superimposed to guide the eye.
See the text for more details. }
\label{fig_hrco}
\end{figure*}

As amply discussed in 
Sects.~\ref{ssec_eos} and \ref{ssec_opac} 
the other critical factor determining the molecular chemistry is the surface C/O ratio.
The model under consideration experiences several third dredge-up 
episodes, that make the C/O ratio increase above unity (top panel). 
At the stage C/O$\approx 1$  we note an abrupt change in the molecular
equilibria: while the abundances of the O-bearing molecules drop (middle panel), 
the C-bearing molecules suddenly start dominating the atmospheric chemistry
(bottom panel). 
The abundance variations due to the increase of the C/O ratio 
are indeed remarkable, and they may span many orders of
magnitudes! In this respect we also acknowledge the numerical 
stability of \texttt{\AE SOPUS} code, which is able to handle molecular
species down to trace concentrations (e.g. SO$_2$ drops down to 
$\simeq 10^{-30}$ in the last TPs). 
At variance with the other molecules,  the concentration of the carbon monoxide (CO)
remains almost unperturbed by the evolution of the C/O ratio (except for
a modest increment following the increase of C), due to its extremely large bond 
energy.  

To better appreciate the role of the C/O ratio as the main driving factor of molecular
chemistry, Fig.~\ref{fig_molchem} (right-hand side panel)  zooms in the evolution
of just six molecules, among the most abundant ones,  during the TP-AGB phase
of a  $M_{\rm i}=5\,M_{\odot}$, $Z_{\rm i}=0.001$ star.  This model is predicted 
to suffer significant changes in its envelope chemical composition due to 
both the third dredge-up and HBB, which produce a complex 
evolution of the C/O ratio. We expect that the surface C/O follows a 
sawtooth trend crossing  the critical region around unity several times, even during the single TPs. 
This may happen under particular conditions such that one dredge-up episodes brings the 
C/O$>1$ and later, during the interpulse period, HBB is able to burn C into
N, hence lowering C/O below unity again.

In particular we note that the during the last TPs HBB is extinguished while
the  third dredge-up keeps on taking place, so that a significant 
increase of the C/O ratio is predicted in the last stages,
as already noted by \cite{Frost_etal98}.
Correspondingly, the molecular species exhibit quite drastic variations: 
the C-bearing molecules (CN, HCN, C$_2$) follow the steep increase of the 
C/O ratio, whereas those of the O-bearing molecules (SiO, H$_2$O, CO$_2$)
show a specular behaviour. Eventually the abundances of all molecules  
drop when the atmosphere starts warming up  as the star evolves off 
the AGB.

\subsection{Extended atmospheres in the H-R diagram}

Figure~\ref{fig_atmo}  displays the gas pressure -- temperature stratifications
of a few static atmosphere models (with the same stellar mass and luminosity),  
under the assumption of either
plane parallel or spherically-symmetric geometry (see Sect.~\ref{ssec_atmo}). 
Computations were carried
out for  three values of the effective temperature and two 
choices of the C/O ratio.

It is interesting to note that, at least for the models  under consideration, 
at given $T_{\rm eff}$ and C/O, the photospheric pressure is almost
insensitive to the geometry, while the separation between the thin
and the extended atmospheres grows  wider and wider at lower pressures.  
On the contrary a major effect is produced by the C/O ratio: at fixed $T_{\rm eff}$,
the photospheric pressure is lower  for C-rich than for O-rich models.
This will have a sizable impact on the inner envelope structure of  AGB stars with
different C/O ratios,  since the photosphere sets two of the four boundary conditions
for the envelope integrations described in Sect.~\ref{ssec_envmod}.  

By comparing the atmospheric structures for the two geometry options in 
Fig.~\ref{fig_atmo},  it is clear the  relevance of the dilution of the radiation field 
in the extended atmospheres
of AGB stars. For instance, for C/O$=2.0$ 
the plane-parallel model with $T_{\rm eff} \ge 3.4$ remains too cool and does 
not enter the condensation region of SiC and graphite, 
while the corresponding spherical model does it successfully.
On the other hand, almost all models with C/O$=0.5$ stay outside the condensation 
area even at the lowest $T_{\rm eff}=3.3$.
Indeed, a detailed analysis on the nucleation and growth of dust grains 
in the outer envelopes of AGB stars requires abandoning 
the static approximation in favour of an expanding 
envelope model. This important issue is beyond the scope of the present work, 
and is addressed in a forthcoming  paper \citep{Nanni_etal13}.

Figure~\ref{fig_sphat} illustrates the areas in the H-R diagram where  
AGB and post-AGB stars (cooler than $\sim 3 \times 10^4$ K) are expected 
to have extended atmospheres, i.e. the radial extension
of the atmosphere being a non-negligible fraction of the photospheric radius.
The  geometrical thickness $\delta R$ is defined according to Eq.~(\ref{eq_dr}).

First of all we note that, at given stellar mass, $\delta R$ increases at higher $L$ and lower $T_{\rm eff}$.
Giants with lower masses have thicker atmospheres (higher $\Delta R$), since smaller
$M/L$ values tend to reduce the effective gravitational acceleration, 
$g_{\rm eff}= (1-\Gamma) g$, 
by increasing the Eddington factor 
\begin{equation}
\Gamma = \frac{\kappa }{4 \pi G c}\frac{L}{M}
\label{eq_gamma}
\end{equation}
where $g=GM/R^2$ is the gravitational acceleration, $\kappa$
is the flux-averaged opacity, while the other constants have their usual meanings.
As shown in the top panels of Fig.~\ref{fig_sphat}, these conditions 
are preferably met by evolved M-type stars of low mass,  a circumstance already discussed
by e.g. \citet[][]{Schmid_etal81}, and \citet{Laskarides_etal90}. 

At higher $L$ and increasing $T_{\rm eff}$  atmospheres may even become gravitationally
unbound, as the Eddington factor rises above unity due  to 
the increasing opacity $\kappa$ in the outermost layers. 
In fact, for  temperatures $\log(T)\ga 3.8$ K,
the Rosseland mean opacity is expected to grow steeply due to the
increasing contributions of the hydrogen bound-free and free-free   
absorptions \citep[see e.g.][]{MarigoAringer_09}.
It follows that this condition may apply, for instance, to
post-AGB stars with high mass ($\ga\! 1 M_{\odot}$) (evolved from
more massive AGB stars with HBB) on their way towards the hotter 
regions of the H-R diagram (see the dotted area top-right panel 
of Fig.~\ref{fig_sphat}).

\subsection{Hayashi lines on the TP-AGB}
\label{ssec_hayashi}
Figure~\ref{fig_hrco} displays several sequences of AGB Hayashi lines,
with the aim of illustrating their basic dependencies on stellar mass,  envelope mass, 
metallicity and C/O ratio.
To this aim we consider two choices of the stellar mass, 
$1.0\,M_{\odot}$ and $2.0\,M_{\odot}$,  and
two values of the initial metallicity $Z_{\rm i}=0.0005,\,{\rm and}\,0.017$.

The surface C/O ratio is made vary from $0.1$ to $10$ 
in steps of $\Delta($C/O$)=0.2$, by increasing the C abundance, 
while keeping O constant (to mimic the effect of the third dredge-up). 
Therefore the actual metallicity $Z$ increases as C/O increases.

\begin{figure}
\resizebox{1\hsize}{!}{\includegraphics{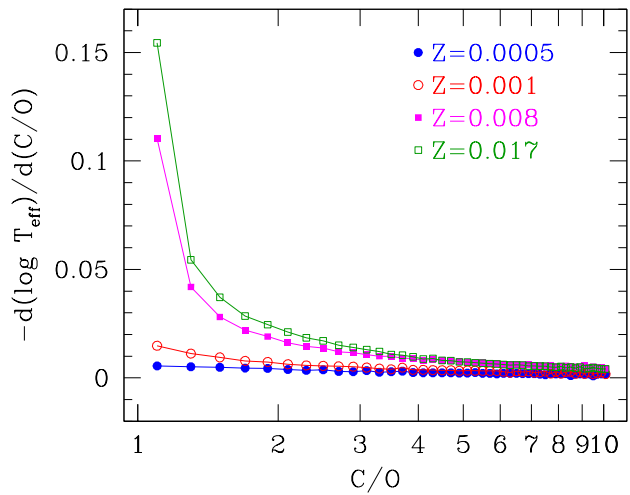}}
\caption{Cooling rate, measured by the derivative $|d(\log T_{\rm eff})/d({\rm C/O})|$,
as a function of increasing C/O ratio (in the regime $>1$) as predicted by envelope integrations
referring to a model with a mass of $2\, M_{\odot}$ and luminosity 
$\log(L/L_{\odot})=4$. Note the pronounced sensitivity to the 
initial metallicity.}
\label{fig_slope}
\end{figure}

For each value of C/O, the core mass $M_{\rm c}$ is made increase  from 
$0.5\,M_{\odot}$ in steps of $\Delta M_{\rm c}=0.1\,M_{\odot}$, until
either the luminosity reaches $\log(L/L_{\odot})=4.6$, or the envelope mass falls below
$10\%$ of the total stellar mass, i.e. $(M-M_{\rm c})/M<0.1$.
While the former condition is first met by the $2.0\,M_{\odot}$ sequences, the latter
applies to the $1.0\,M_{\odot}$ tracks, that are terminated  when   
$M_{\rm c} =0.94\, M_{\odot}$.

The effective temperature and the luminosity  are determined 
by complete integrations of envelope models  (see Sect.~\ref{ssec_envmod}), 
with gas opacities calculated on-the-fly consistently with the current chemical composition 
(and C/O ratio). 

We remark that these calculations are simply grids of envelope 
integrations and are meant to yield an overall picture of 
the Hayashi lines of C stars and their critical dependencies, 
but they cannot, by construction, be strictly representative 
of the TP-AGB evolution.
For instance, the over-luminosity effect due to a deep third 
dredge-up is not taken into account and the Hayashi lines  
in Fig.~\ref{fig_hrco} are those  corresponding  to a standard CMLR
(for $\lambda=0$).
As a consequence, at a given stellar mass, luminosity, and C/O ratio 
the ``actual'' effective temperature of an evolving  C star model should be
somewhat lower than that predicted in Fig.~\ref{fig_hrco}. 
This said, the following discussion is nevertheless instructive since 
the general trends remain valid.

Examining Fig.~\ref{fig_hrco} several features can be noticed.
As long as  C/O$<1$ the Hayashi lines have a steep slope 
and span a limited $T_{\rm eff}$ range, which becomes
narrower at decreasing metallicity. This interval defines the
expected location of M and S stars.
The value C/O$=1$ corresponds to the warmest Hayashi line,
due to a deep minimum in the molecular opacities 
\citep[nearly all C and O atoms are locked in the CO molecule; see][]
{MarigoAringer_09}.

As soon as C/O overcomes unity we expect a sudden jump 
of the Hayashi lines to lower effective temperatures, 
the amplitude of the temperature jump  being more pronounced at increasing
metallicity. 
The cooling rate, expressed by the derivative 
$|d(\log T_{\rm eff})/d({\rm C/O})|$, progressively decreases 
at increasing C/O ratio, so that larger and larger C/O ratios 
are required  to reach lower effective temperatures. 
This is evident by looking at the thickening of the iso-C/O curves 
in Fig.~\ref{fig_hrco} (dashed lines), which become gradually closer one to the next.  

It means that, above some critical C/O ratio, the atmospheric structure becomes
less and less sensitive  to a further increase of the carbon abundance.
This kind of ``saturation'' effect shows up at lower C/O  ratio for decreasing metallicity,
as can be better appreciated in Fig.~\ref{fig_slope}. 
We notice that at higher $Z_{\rm i}$ the cooling rate is large for C/O values slightly above 
$1$, then it decreases  until it flattens out to a nearly constant, small value.
This trend is found also at lower $Z_{\rm i}$, but with smoother features: the initial drop of 
$T_{\rm eff}$ becomes less pronounced and $|d(log T_{\rm eff})/d({\rm C/O})|$
levels off at  lower C/O ratios. Note, for instance, the extremely low cooling rate
at $Z_{\rm i}=0.0005$ all over the C/O ratio under consideration ($1 \le {\rm C/O} \le 10$).

\subsection{The core mass at the onset of the third dredge-up}
\label{ssec_mcmin}

As already mentioned in Sect.~\ref{ssec_tbdred}, we can determine the minimum
core mass for the occurrence of the third dredge-up $M_{\rm c}^{\rm min}$, 
checking if and when the $T_{\rm bce}$ exceeds a critical value 
$T_{\rm dup}$ at the stage of post-flash luminosity peak.
The quantity $T_{\rm dup}$ is assumed as a free parameter.

\begin{figure*}
\begin{minipage}{0.48\textwidth}
\resizebox{\hsize}{!}{\includegraphics{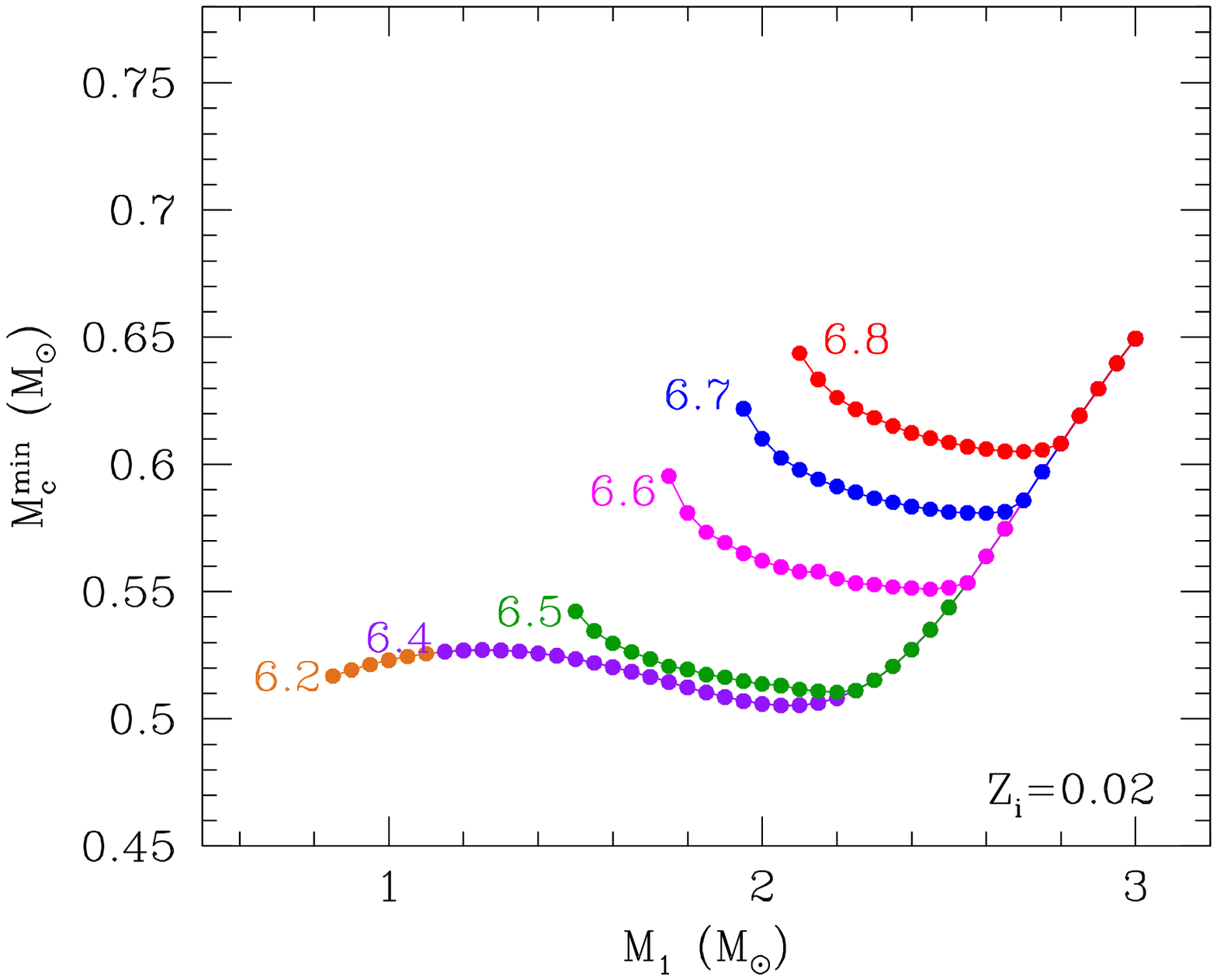}}
\end{minipage}
\hfill
\begin{minipage}{0.48\textwidth}
\resizebox{\hsize}{!}{\includegraphics{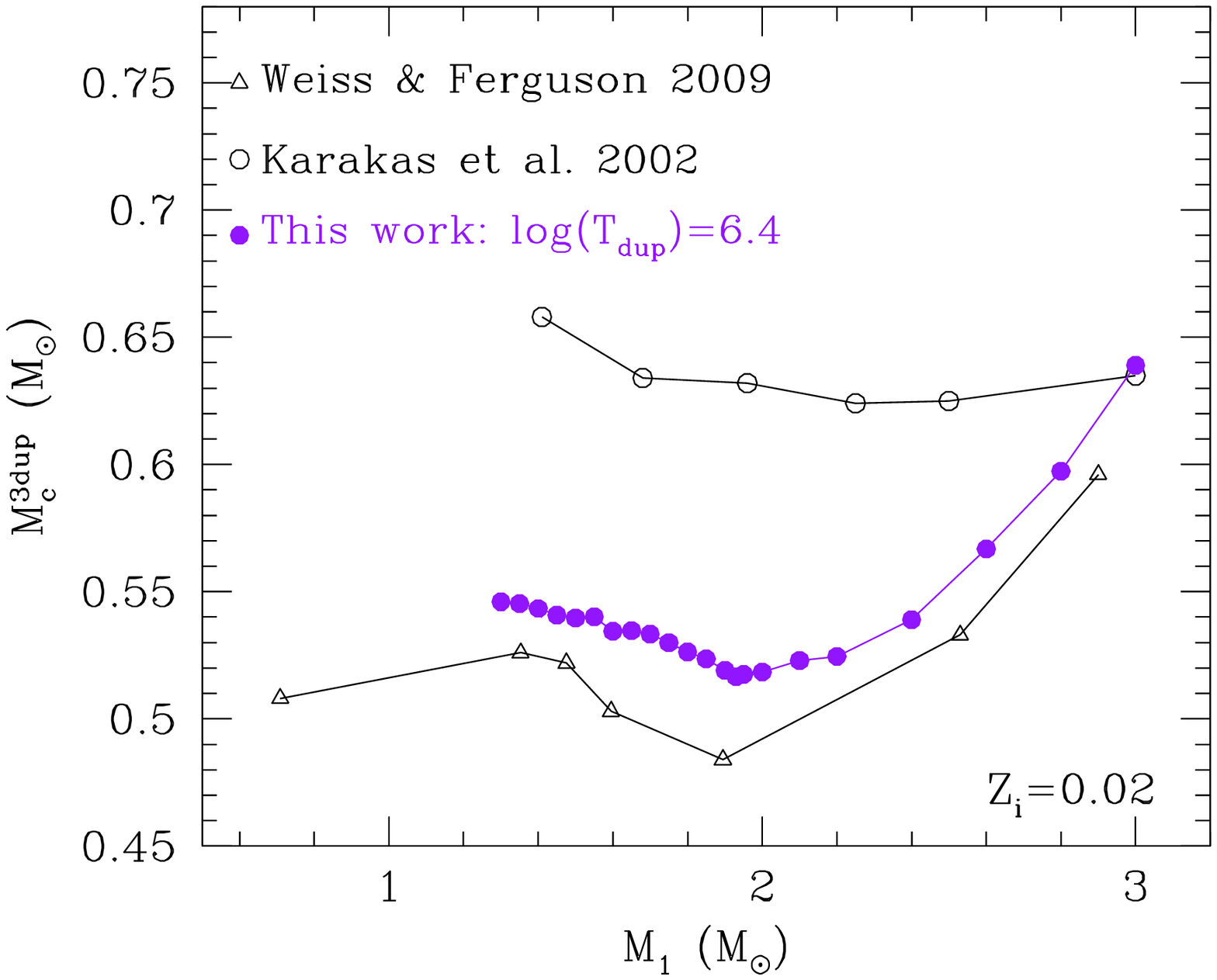}}
\end{minipage}
\hfill
\begin{minipage}{0.48\textwidth}
\resizebox{\hsize}{!}{\includegraphics{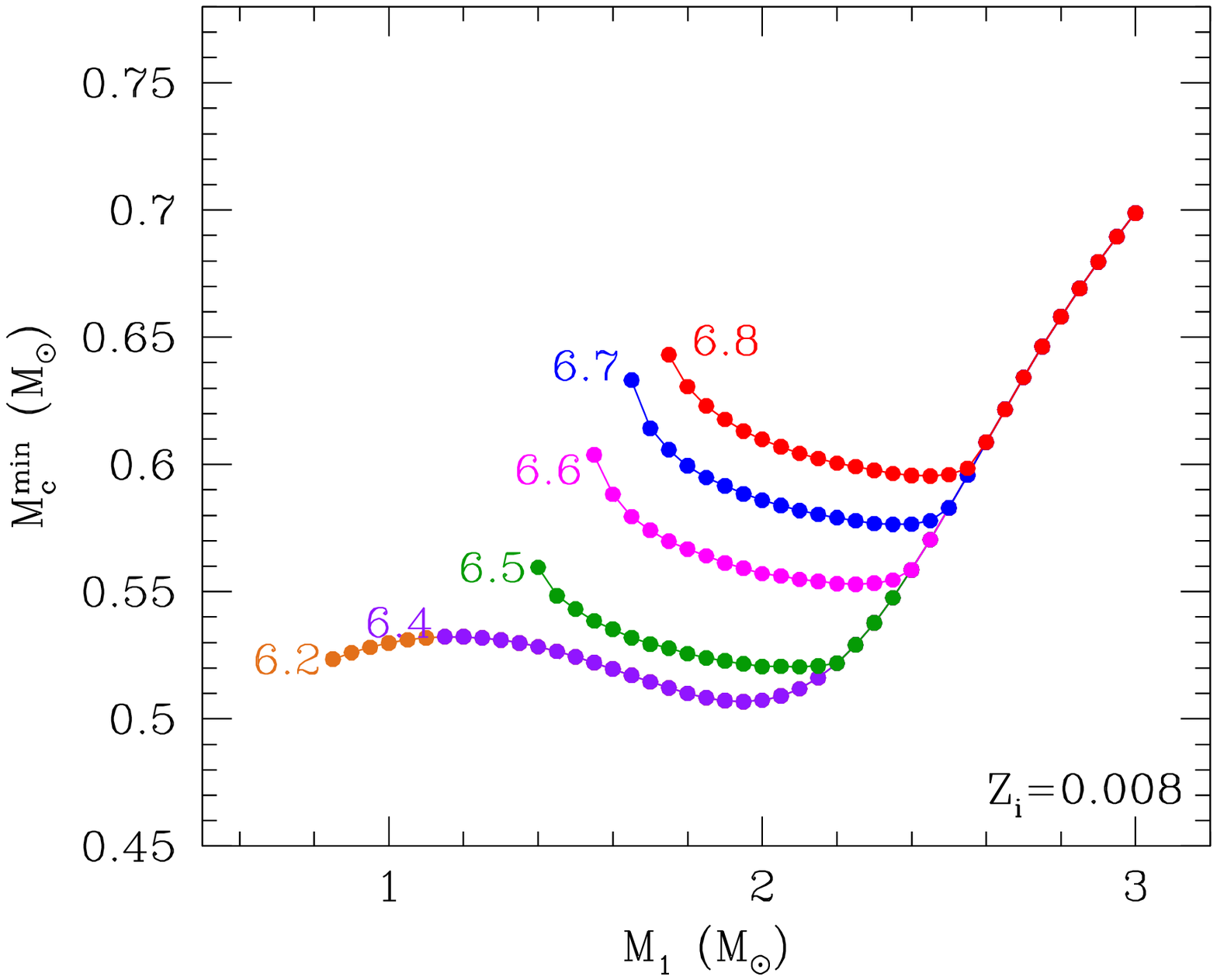}}
\end{minipage}
\hfill
\begin{minipage}{0.48\textwidth}
\resizebox{\hsize}{!}{\includegraphics{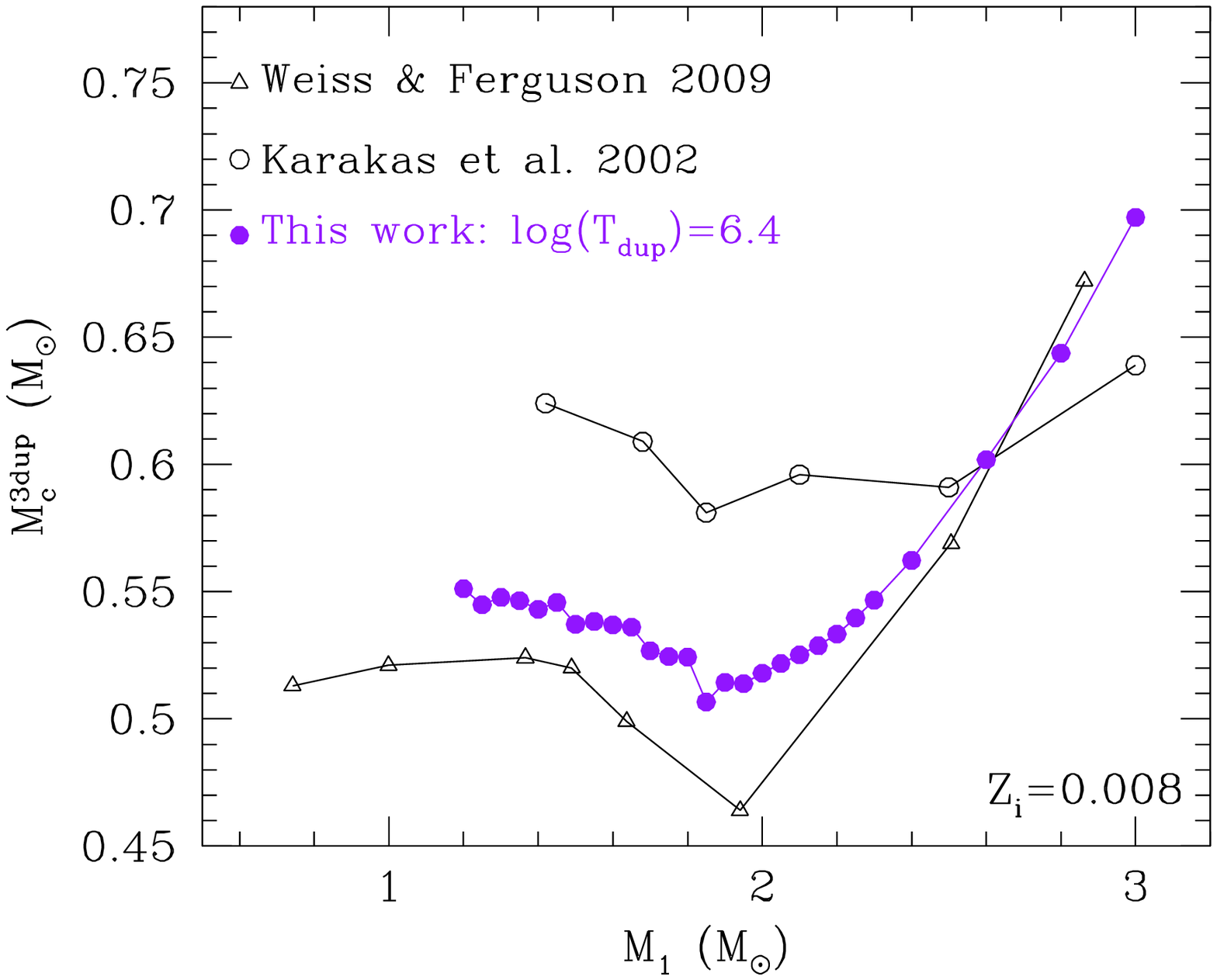}}
\end{minipage}
\hfill
\begin{minipage}{0.48\textwidth}
\resizebox{\hsize}{!}{\includegraphics{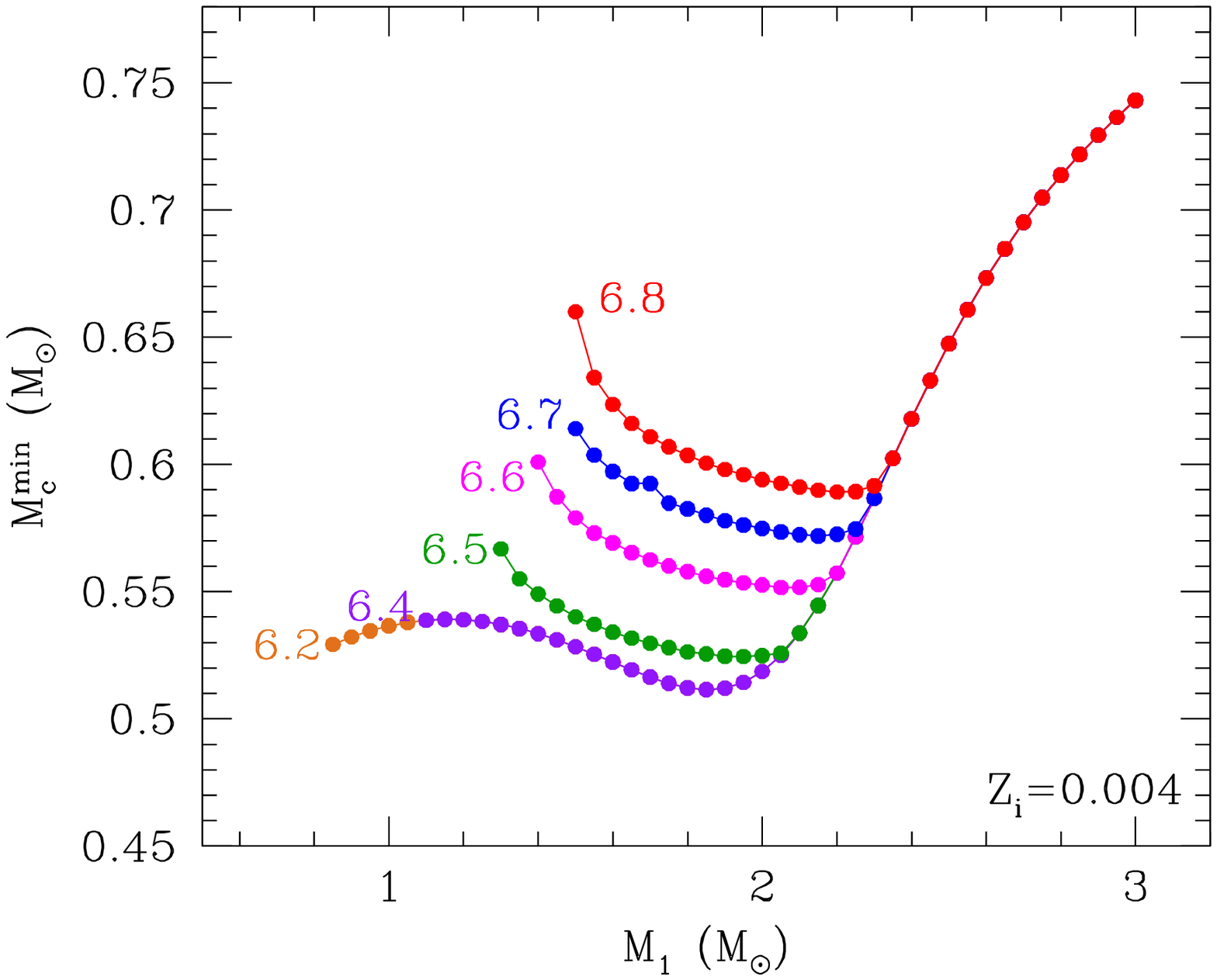}}
\end{minipage}
\hfill
\begin{minipage}{0.48\textwidth}
\resizebox{\hsize}{!}{\includegraphics{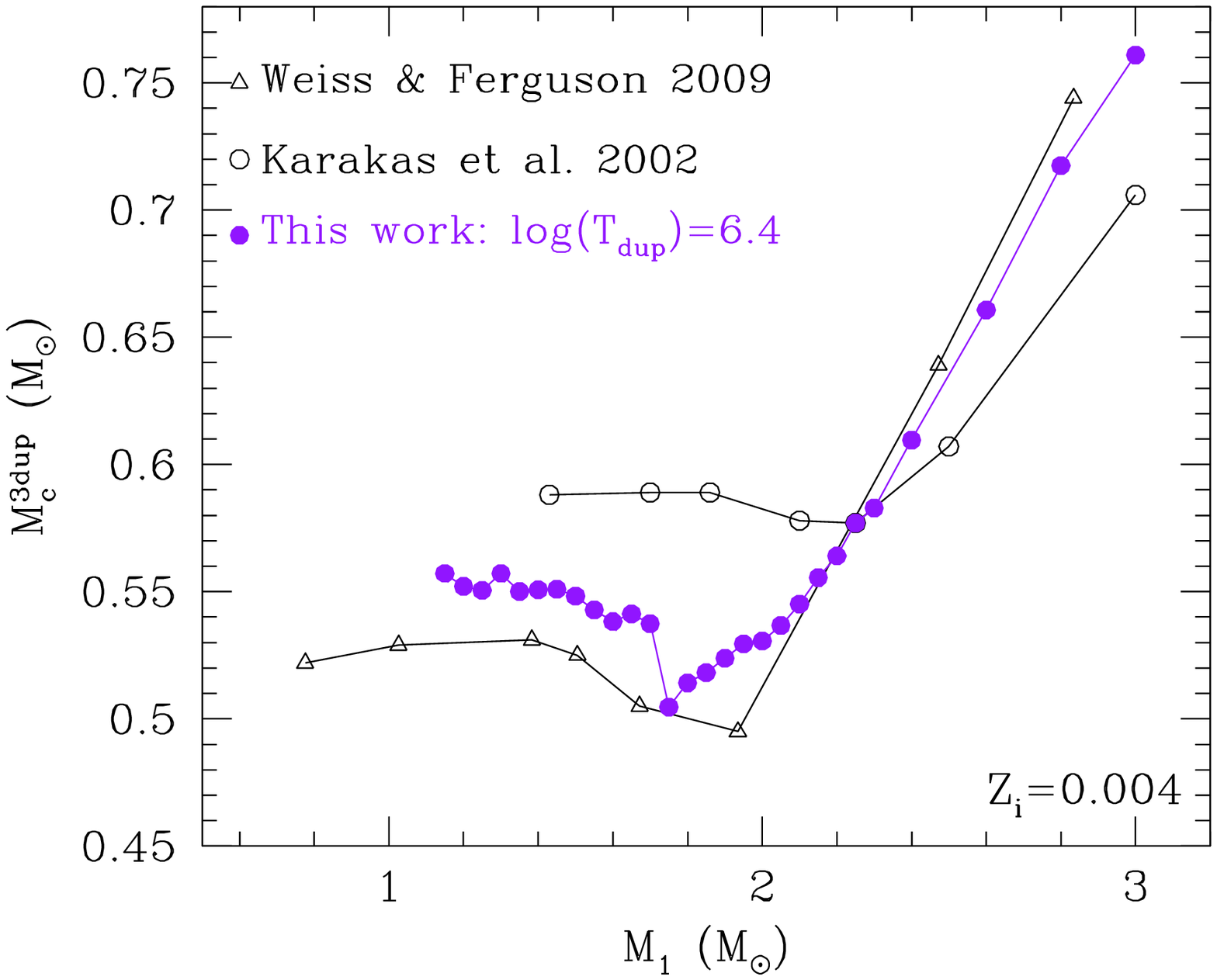}}
\end{minipage}
\caption{The minimum core mass for the third dredge-up
as a function of the stellar mass at the onset of the TP-AGB phase, and three
values of the initial metallicity as indicated.
{\em Left-hand side panels:} Predictions for $M_{\rm c}^{\rm min}$ are obtained from envelope
integrations, as detailed in Sect.~\ref{ssec_tbdred}, adopting  six values of the 
minimum temperature for the base of the convective envelope, namely: 
$\log(T_{\rm dup})= 6.2, 6.3, 6.4, 6.5, 6.6, 6.7, 6.8$.
{\em Right-hand side panels: } The core mass at the first occurrence of the  third 
dredge-up, $M_{\rm c}^{\rm 3dup}$, during the TP-AGB evolution of models of different masses and metallicities, as indicated.
We compare the results from two sets of full TP-AGB calculations 
\citep{WeissFerguson_09, Karakas_etal02}, with 
the \texttt{COLIBRI} predictions assuming $\log(T_{\rm dup})=6.4$.}
\label{fig_mc3}
\end{figure*}

In Figure~\ref{fig_mc3} the left-hand side panels display the $M_{\rm c}^{\rm min}$ 
predictions for $\log(T_{\rm dup})= 6.2, 6.4, 6.5, 6.6, 6.7, 6.8$ and three values 
of the initial metallicity, $Z_{\rm i}=0.02, \,Z_{\rm i}=0.008, \,{\rm and}\, Z_{\rm i}=0.004$.
The numerical method described in Sect.~\ref{ssec_tbdred} has been applied
 for stellar masses ranging from $1\,M_{\odot}$ to $3\,M_{\odot}$ in steps 
of $0.05\,M_{\odot}$.  In practice, once set the 
minimum temperature  $T_{\rm dup}$, 
for each initial stellar mass and chemical composition,
$M_{\rm c}^{\rm min}$ is the value of the core mass for which
$T_{\rm bce}=T_{\rm dup}$ is satisfied. The solution is found iteratively 
with envelope integrations adopting  
the Brent root-finding algorithm 
\citep[chapter IX of ``Numerical Recipes'';][]{Press_etal88}.
In each case $M_{\rm c}^{\rm min}$ is taken as the maximum  between the
value obtained by the envelope-integration method and the core mass at the first
thermal pulse, $M_{\rm c,1}$.
We do not show the results for $M> 3M_{\odot}$, since for the higher 
 masses the temperature criterion is always satisfied since the onset of the
TP-AGB, regardless of the value $T_{\rm dup}$.
We see that all the curves share the same trend. 
Starting from lower masses towards the higher ones,
$M_{\rm c}^{\rm min}$ slightly decreases, reaches a minimum  and
then steeply increases. It is interesting to note that the minimum
in $M_{\rm c}^{\rm min}$ corresponds exactly to the critical maximum mass,
$M_{\rm HeF}$, for a star to develop a degenerate He-core and 
experience the He-flash at the tip of the RGB. 
This reflects the same correspondence between $M_{\rm HeF}$ 
and the minimum of $M_{\rm c,1}$ (see Fig.~\ref{fig_mc1}),
already pointed out long ago by \citet[e.g.][]{Lattanzio86}.

For a given initial metallicity,  at decreasing  $T_{\rm dup}$, 
the sequences  move downward and reach lower stellar
masses, that is $M_{\rm c}^{\rm min}$ decreases  and the 
third dredge-up is expected to take place in stars of lower and lower masses.
We note that for $T_{\rm dup}\le 6.4$ the minimum core mass 
$M_{\rm c}^{\rm min}$ coincides with $M_{\rm c,1}$.

The values  of the core mass, $M_{\rm c}^{\rm 3dup}$,  
when the third dredge-up effectively occurs
for the first time during the TP-AGB evolution, are shown in the right-hand 
side panels of Figure~\ref{fig_mc3}. 
We note that, in general, 
$M_{\rm c}^{\rm 3dup} \ge M_{\rm c}^{\rm min}$, as expected.
The \texttt{COLIBRI} results for $M_{\rm c}^{\rm 3dup}$,  corresponding to 
$\log(T_{\rm dup})= 6.4$,   show a similar trend with 
the stellar mass  compared to  full TP-AGB models calculations.
At the same initial metallicity and stellar mass 
our predictions for  $\log(T_{\rm dup})= 6.4$ are
lower than \citet{Karakas_etal02},  but
somewhat larger than \citet{WeissFerguson_09}.

Clearly significant differences exist between the two sets of full calculations,
which supports the need to accurately calibrate $M_{\rm c}^{\rm min}$ with the
aid of observations of M and C giants of different ages and metallicities. 
This calibration is presently underway and will be presented in subsequent 
papers.

\begin{figure*} 
\begin{minipage}{0.48\textwidth}
 \resizebox{\hsize}{!}{\includegraphics{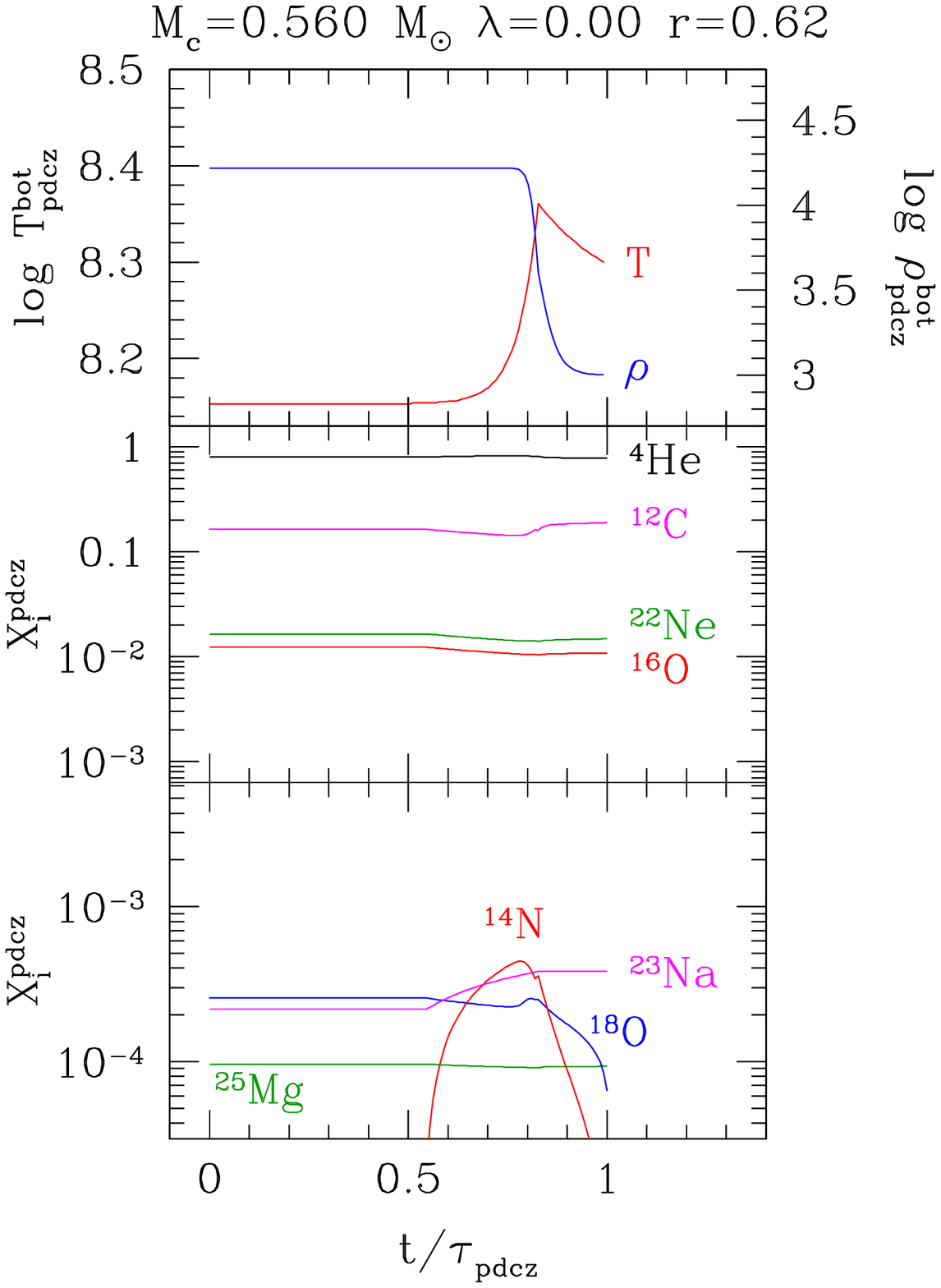}}
\end{minipage}
\hfill
\begin{minipage}{0.48\textwidth}
 \resizebox{\hsize}{!}{\includegraphics{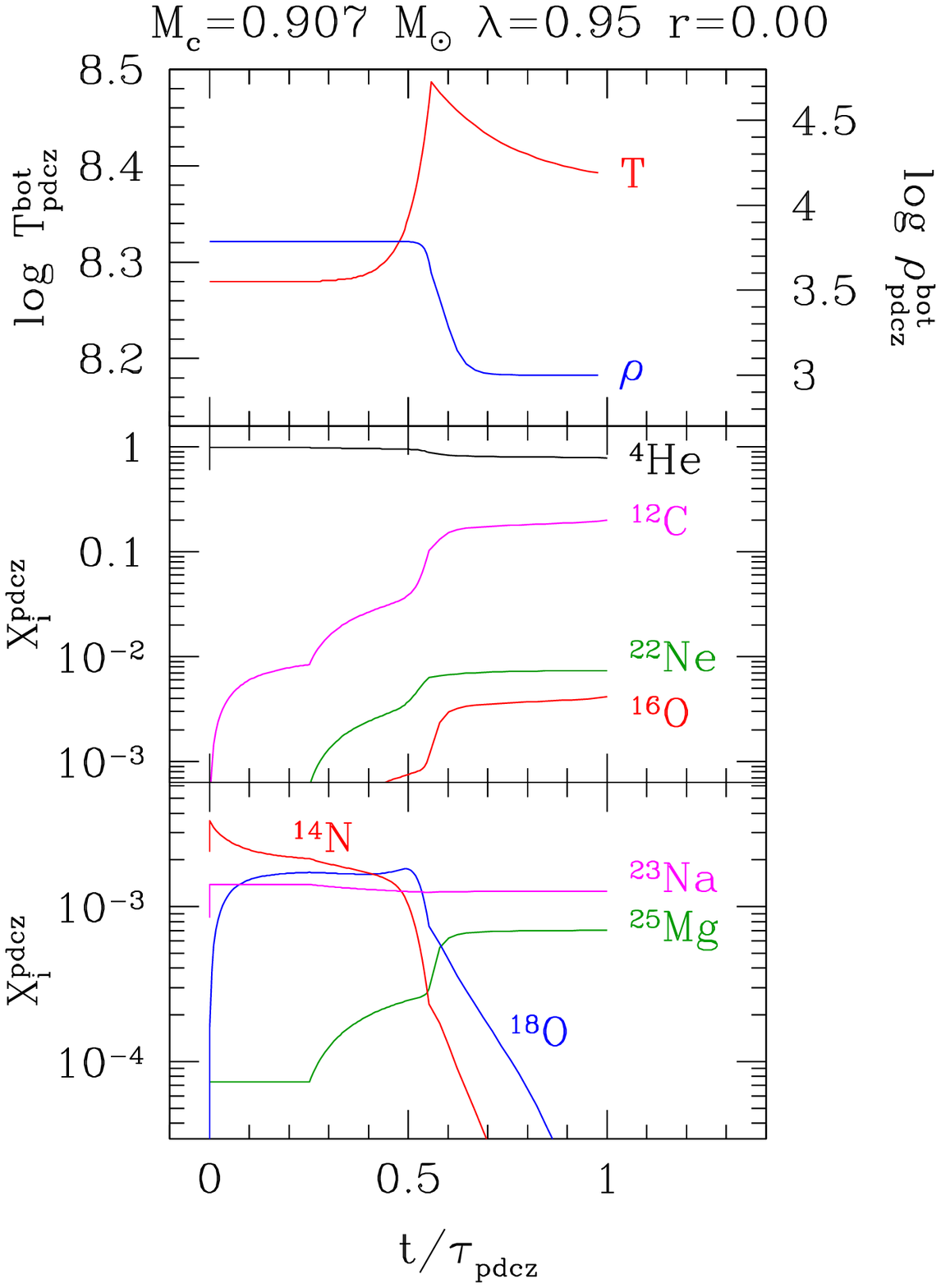}}
\end{minipage}
\caption{Structural characteristics of the pulse-driven convective zone, 
 corresponding to the $10^{\rm th}$ and $18^{\rm th}$ thermal pulse of the
$M_1 = 1.2\,M_{\odot},\,Z_{\rm i}=0.017$ (left panel) and 
$M_1=5.4\,M_{\odot},\,Z_{\rm i}=0.017$ (right panel) models, respectively. 
The corresponding values of the core mass ($M_{\rm c}$), 
efficiency of the third dredge-up ($\lambda$),
and degree of overlap ($r$) are indicated on  top of each plot.
 Quantities are presented as a function
of the phase $\phi=t/\tau_{\rm pdcz}$, from the onset of pulse
convection ($\phi=0$) to its disappearance  ($\phi=1$).
For each stellar model, the panels show
the evolution of temperature and density at the current
PDCZ base (top panel); the evolution of the PDCZ abundances (in mass fractions), 
homogenized
over the current PDCZ mass (middle and bottom panels). }
\label{fig_pdcztp}
\end{figure*}
\begin{figure*}  
\begin{minipage}{0.48\textwidth}
\resizebox{\hsize}{!}{\includegraphics{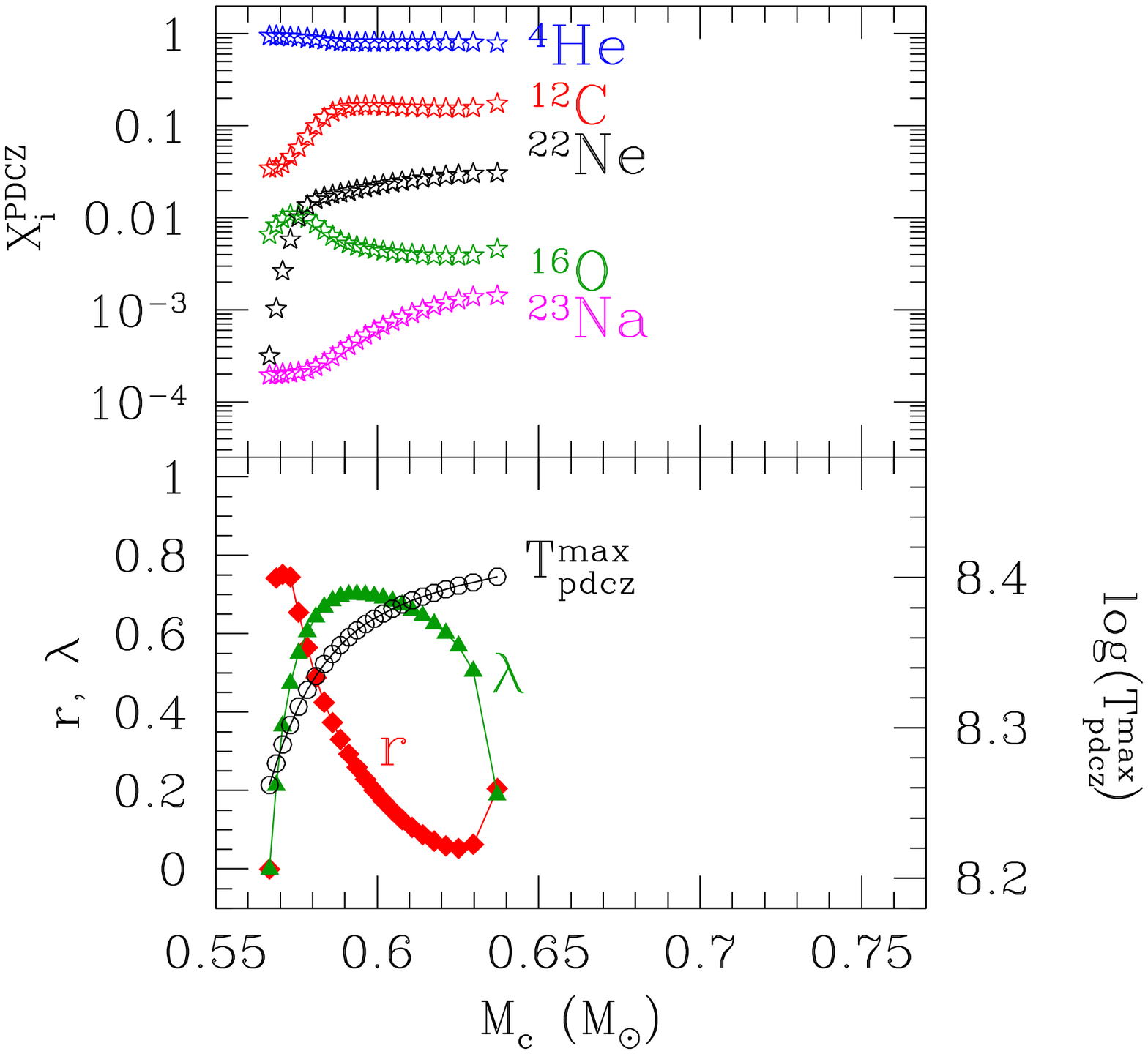}}
\end{minipage}
\hfill
\begin{minipage}{0.48\textwidth}
\resizebox{\hsize}{!}{\includegraphics{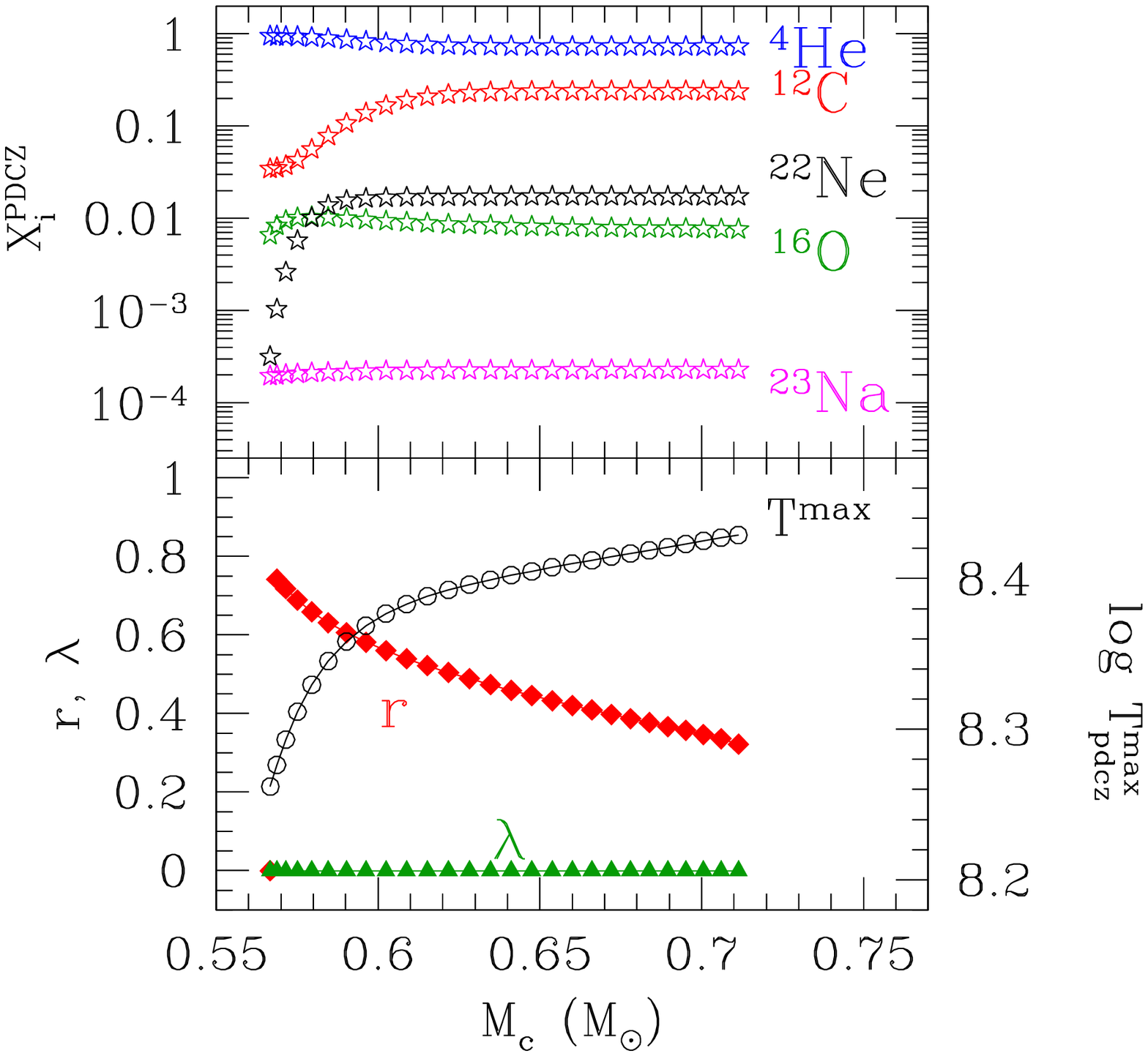}}
\end{minipage}
\caption
{{\em Top panels}: predicted intershell abundances as a function of the core mass 
during the entire TP-AGB phase of a ($M_{\rm i}=2.6,\,Z_{\rm i}=0.017$) model.
{\em Bottom panels}: Evolution of the maximum temperature $T^{\rm max}$ 
reached at the bottom of the PDCZ at each TP, efficiency $\lambda$ 
of the third dredge-up, and degree of overlap $r$ between two consecutive
PDCZs. The results in the right-hand side panels are obtained under the assumption 
that the third dredge-up does not occur ($\lambda=0$). 
Note that in this case the $^{23}$Na abundance in the PDCZ is lower than in
models with dredge-up (left-hand side panel).}
\label{fig_xpdcz2}
\end{figure*}

\subsection{Intershell abundances}
\label{ssec_xpdcz}

The standard  chemical composition of the intershell region,
 left after the development of a thermal pulse, amounts to
roughly $20\%-25\%$ of $^{12}$C, $1\%-2\%$ of $^{16}$O, $1\%-2\%$ of $^{22}$Ne,
with $^{4}$He essentially comprising all the rest 
\citep{Schoenberner_79, BoothroydSackmann_88b, Mowlavi_99a},
almost regardless of metallicity and core mass.

These standard values are presently debated.
\citet{Izzard_etal04} find  a lower value  for $^{16}$O, typically
amounting to $\approx\,0.5\%$, while the inclusion of convective
diffusive overshooting applied to all convective boundaries of the PDCZ
determines a substantial increase of the  $^{12}$C and $^{16}$O abundances
at the expense of $^{4}$He \citep{Herwig_etal97}.
\citet{Herwig_00} shows that with his calibrated parametric scheme 
for overshoot, the $^{12}$C and $^{16}$O intershell
abundances reach typical values of $0.45$, and $0.25$, respectively.

We will now discuss our predictions obtained from the semi-analytic
scheme detailed in Sect.~\ref{ssec_pdcz}.
Figure~\ref{fig_pdcztp} exemplifies the evolution of the main 
characteristics of the PDCZ during a thermal pulse, in two
models with different core masses.
Let us first analyse the results for the model with 
$M_{\rm c}=0.576\,M_{\odot}$
(left-hand side panels).

We see that while the density at the bottom of the PDCZ is
continuously dropping, the corresponding temperature first 
rises up to the maximum value, $T_{\rm pdcz}^{\rm max}$, and then
decreases (top panels). 

Before reaching the maximum temperature, 
the chemical composition of the PDCZ may vary mainly due to 
its growth in mass, as 
the ashes of the H-burning shell are reached by the expanding convection.
As a consequence the abundances of $^{4}$He,  $^{14}$N, and  
$^{23}$Na are expected to increase. The sharp rise of $^{14}$N is evident 
in the bottom left-hand side panel of Fig.~\ref{fig_pdcztp}. 
The increase of $^{14}$N is only temporary: 
as soon as the PDCZ heats up nitrogen is completely destroyed by 
the chain $^{14}{\rm N}(^4{\rm He},\gamma)^{18}{\rm F}(\beta^+ \nu)^{18}{\rm O}$.

In the short phase around the temperature 
maximum  the PDCZ reaches its widest mass extension. 
At this point the main $\alpha$-capture reactions are turned on, 
leading to the production of primary carbon 
via the  \reac{He}{4}{2\,^{4}He}{\gamma}{C}{12} reaction,
together with some synthesis of $^{16}$O from 
\reac{C}{12}{^{4}He}{\gamma}{O}{16},  and $^{22}$Ne
from \reac{O}{18}{^{4}He}{\gamma}{Ne}{22}.
Correspondingly the $^{4}$He abundance decreases.

Finally, when the PDCZ cools and the convection recedes the chemical
composition barely changes, so that  the entire intershell 
with mass $\Delta M_{\rm pdcz}$ is assigned the final mixture at $\phi=1$.

Basically the same analysis holds for the model with the higher core mass
(right-hand side panels), but for a few differences that are explained 
mainly by the higher $T_{\rm pdcz}^{\rm max}$, the shorter duration
$\tau_{\rm PDCZ}$ of the PDCZ, and by the previous dredge-up history.
As we discuss later, the intershell abundances do depend on the 
indirect interaction of one pulse with the preceding one, which can be 
quantified by the so-called ``degree of overlap'', denoted with $r$. 
We also note that in the model with higher $M_{\rm c}$ 
a higher $T_{\rm pdcz}^{\rm max}$ is attained so that
\reac{Ne}{22}{^{4}He}{n}{Mg}{25} is also activated. This reaction 
is recognized as a source of neutrons for the s-process nucleosynthesis
in more massive AGB stars \citep[e.g.][]{Busso_etal99, Pumo_etal09}.

Figure~\ref{fig_xpdcz2} shows the evolution of the final PDCZ abundances
left after each thermal pulse (bottom panels), 
during the entire TP-AGB evolution of the ($M_{\rm i}=2.6,\,Z_{\rm i}=0.017$) model. 
The left- and right-hand side panels compare the results in the cases
the third dredge-up takes place (left-hand side panel; $\lambda > 0$) 
or does not  (left-hand side panel; $\lambda = 0$). 
In the $\lambda > 0$ case, the efficiency of the third dredge-up
is described following the analytic relations presented 
by \citet{Karakas_etal02} which fit the results of their full AGB models 
(see also Sect.~\ref{ssec_tpagbev}),  the $\lambda = 0$ case is simply treated 
setting the efficiency to zero at each thermal pulse. This is equivalent to 
assume a high value for $T_{\rm dup}$.
Several remarks can be made.
\begin{figure}
\resizebox{0.9\hsize}{!}{\includegraphics{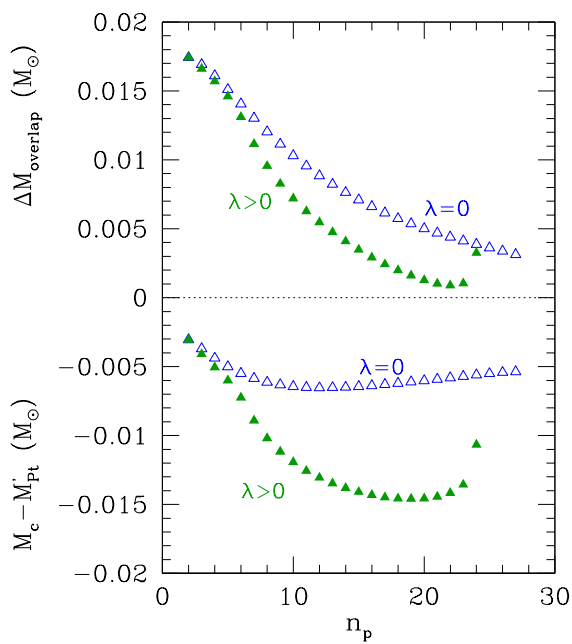}}
\caption{Mass of overlap, $\Delta M_{\rm overlap}$, between 
two consecutive PDCZs, and mass difference 
$(M_{\rm Pt}^{\prime}-M_{\rm c})$
between the top of the previous PDCZ, $M_{\rm Pt}^{\prime}$, 
and the current core mass, $M_{\rm c}$. 
Predictions are shown as 
a function of the pulse number, for the same evolutionary
sequences of Fig.~\ref{fig_xpdcz2}, computed with two different 
assumptions for the efficiency third dredge-up: 
$\lambda=0$ (empty blue triangles), 
or $\lambda>0$  (filled green triangles; according to the relations  
of \citet{Karakas_etal02}). Note the effect of the third dredge-up that
pushes $M_{\rm Pt}^{\prime}$ inward.}
\label{fig_moverlap}
\end{figure}
\begin{figure}  
\resizebox{\hsize}{!}{\includegraphics{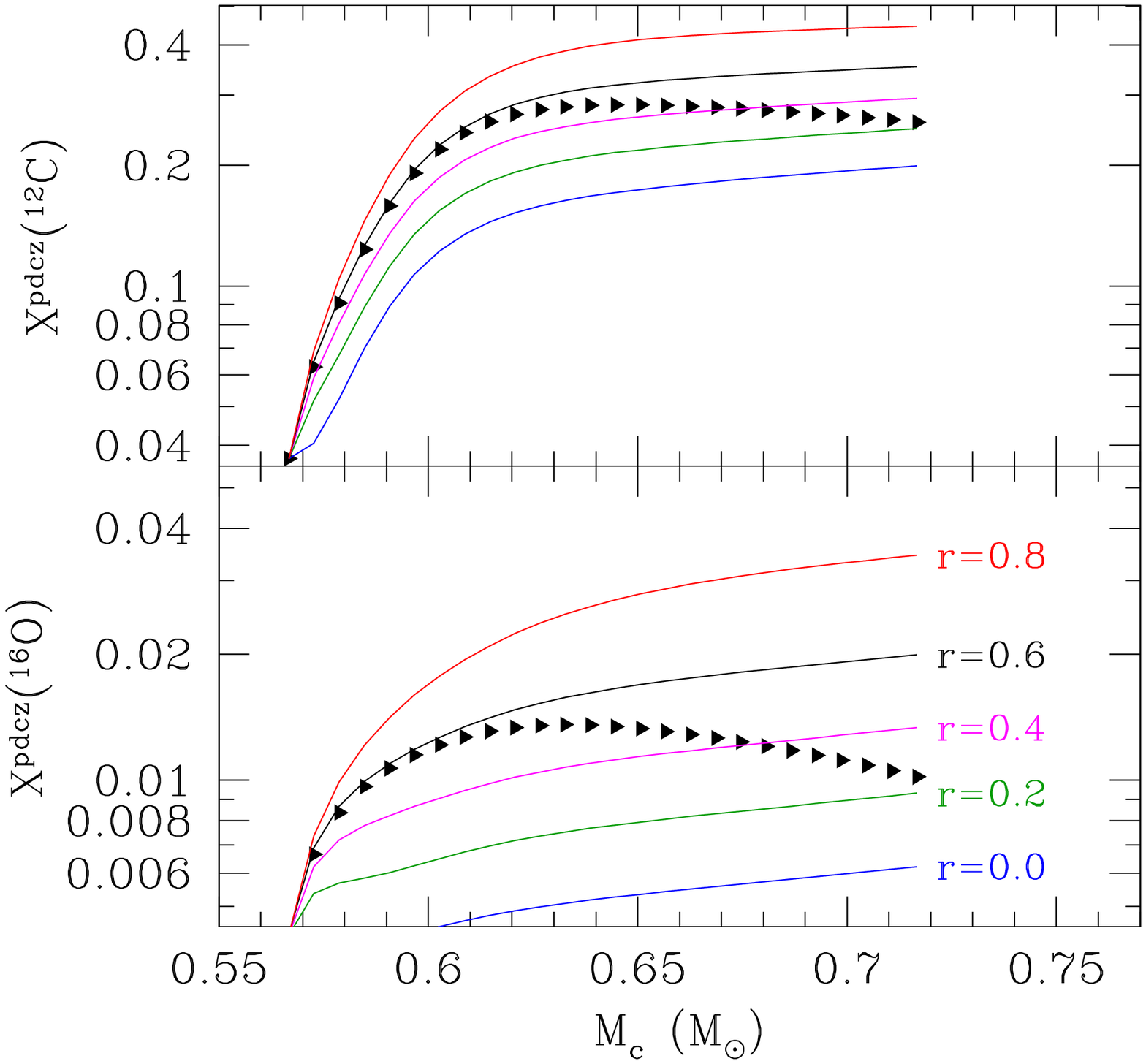}}
\caption
{Dependence of the intershell abundances of  $^{12}$C (top panel) and 
$^{16}$O (bottom panel) on the degree of overlap $r$ between
consecutive PDCZs.  The bunch of lines show the predictions for 
a constant value of $r$ (indicated in the bottom panel) 
between consecutive PDCZs. 
The sequences of triangles show the PDCZ abundances as a function
of the core mass during the evolution of a TP-AGB star with 
($M_{\rm i}=2.6 M_{\odot},\,Z_{\rm i}=0.017$) in which we assume the third does not
occur. 
In this test case the TP-AGB phase is computed without mass loss
to extend the calculations up to $M_{\rm c}=0.7\,M_{\odot}$.}
\label{fig_over}
\end{figure}
\paragraph*{The ``standard'' intershell abundances.}
Our intershell abundances of  
$^{4}$He, $^{12}$C, and $^{16}$O recover nicely  
the ``standard''  values obtained by the class of full AGB models 
\citep[e.g.][]{Schoenberner_79, BoothroydSackmann_88b, Izzard_etal04, 
KarakasLattanzio_07} in which
the borders of the PDCZ are determined by the classical 
Schwarzschild criterion applied to the temperature gradients.
We find typical values of  $\approx 20\,\%$ for $^{12}$C and 
$\approx 0.5\%-1\,\%$ for $^{16}$O (Figs.~\ref{fig_xpdcz2},
\ref{fig_over}, \ref{fig_rate}, \ref{fig_xpdcz}).
More specifically, our predictions
for $^{16}$O are in closer agreement with the lower abundances reported
by \citet{Izzard_etal04}, than the higher values of $1\% - 2 \%$ defining 
the ``standard'' intershell composition \citep{BoothroydSackmann_88b}.
This difference will be discussed below, being likely related to
the efficiency of the third dredge-up.

\paragraph*{Dependence on the degree of overlap.}
The degree of overlap $r$ is defined as the fraction of the matter
contained in a given PDCZ  that is incorporated
into the PDCZ produced at the next thermal pulse.
The reader may refer to Table~\ref{tab_mod}  
for the operative definition of $r$ 
in terms of  mass coordinates, and to Fig.~\ref{fig_pdcz} for a graphical representation.
The parameter  $r$ was originally introduced and discussed 
in early  studies \citep[e.g.][]{Ulrich_73, Iben_75, TruranIben_77, Iben_77}
to highlight the importance of the overlap between 
successive pulses to the slow-neutron capture nucleosynthesis of heavy
elements, especially in relation to the synthesis of $^{22}$Ne and its 
role in the release of neutrons via the 
\reac{Ne}{22}{^{4}He}{n}{Mg}{25} reaction. 

We also find a significant dependence of the intershell
abundances on the degree of overlap $r$  between consecutive  PDCZs.
This can be better appreciated by looking at the bottom panels of 
Fig.~\ref{fig_xpdcz2}. We see that the degree
of overlap tends in general to decrease from pulse to pulse, but the
occurrence of the third dredge-up ($\lambda > 0$) makes $r$ to drop
more steeply, eventually reaching zero in the last TPs.
The smooth decline of $r$, expected for $\lambda=0$, is mostly 
due to the inverse correlation between $\Delta M_{\rm pdcz}$ and $M_{\rm c}$
(see top panel of Fig.~\ref{fig_pdczfit}),
so that less massive PDCZs are produced in later TPs.
On the other hand, every time a dredge-up event takes place the upper border,
$M_{\rm Pt}$, of the PDCZ is shifted inward in mass coordinate,
by an amount which is larger at increasing $\lambda$.
This circumstance causes a further reduction of $r$.
We find that for $\lambda \ga 0.7$ the overlap $r \simeq 0$, implying
that the PDCZs are almost decoupled one from the next. 

This effect is clearly shown in Fig.~\ref{fig_moverlap}, 
where the mass difference, $M_{\rm Pt}-M_{\rm c}$, becomes more and more
negative when the third dredge-up is active ($\lambda >0$), 
at variance with the nearly constancy, or even small increase,  
expected if the process does not take 
place ($\lambda =0$). 
Consequently, the decrease 
of the overlap mass
between two consecutive PDCZs, $\Delta M_{\rm overlap}$, 
 is steeper at increasing $\lambda$.  
We note that in the TP-AGB model with $\lambda>0$ 
the overlap mass gradually reduces to almost zero, and then 
it grows again in the very last thermal pulses when the third dredge-up 
does not take place anymore. 

The consequences on the PDCZ nucleosynthesis are exemplified in 
Fig.~\ref{fig_xpdcz2}. While in absence of dredge-up events
$(\lambda=0)$ all intershell abundances tend to flatten out at nearly
constant values, when the third dredge-up  takes place this  pattern
is modified. In particular, as the third dredge-up
starts to occur we expect that the intershell
abundance of $^{16}$O  somewhat declines  levelling off 
in the last TPs, while those of $^{22}$Ne and $^{23}$Na increase, 
steadily. 
These findings for $^{22}$Ne and $^{23}$Na are 
in full agreement with  \citet{Mowlavi_99a, Mowlavi_99b}, to 
which the reader should also refer for a very detailed analysis.

The increase of $^{22}$Ne, 
that reaches up to $\simeq 2\% - 3 \%$ in the cases under consideration, 
is directly related to the increase of primary $^{12}$C in the envelope
composition caused by the third dredge-up.  
The more abundant the surface $^{12}$C is,
the larger amount of  $^{14}$N is synthesized during the 
interpulse period by the CNO-cycle 
operating in the radiative H-burning shell. 
In turn, the more abundant $^{14}$N is, the 
more $^{22}$Ne will be produced by the chain of reactions 
$^{14}{\rm N}(^4{\rm He},\gamma)^{18}{\rm F}(\beta^+ \nu)^{18}{\rm O}(^4{\rm He}, \gamma)^{22}{\rm Ne}$ occurring inside the PDCZ.

The increase of  $^{23}$Na is related to the larger envelope 
abundance of $^{22}$Ne, that we expect as a consequence of the third 
dredge-up. We recall that the $^{23}$Na in the PDCZ is not 
synthesized in situ during the pulse, but is it inherited 
as part of the material 
processed by the radiative H-burning shell, where the conversion 
$^{22}{\rm Ne}(p, \gamma)^{23}{\rm Na}$ took place.

The trends of the $^{12}$C and $^{16}$O intershell abundances, 
mainly synthesized as primary products during the TPs, 
are also affected by the third
dredge-up, hence by the degree of overlap between  consecutive PDCZs.

To better investigate this aspect, we have performed a 
few test calculations, assuming each time a different value of the overlap 
parameter $r$,  which is kept constant along a pre-determined sequence 
of thermal pulses. 
Given a selected  value of $\widehat{r}$, at each thermal pulse 
the mass coordinate of the top of the previous PDCZ, 
$M_{\rm pdcz}^{\prime}$, is 
artificially varied such that the condition 
$\Delta M_{\rm overlap}=M_{\rm Pt}^{\prime}-M_{\rm He} 
= \widehat{r}\, \Delta M_{\rm pdcz}$ is fulfilled (see Fig.~\ref{fig_pdcz}). 
This  is equivalent to suitably adjusting the maximum depth of the 
third dredge-up, hence its efficiency $\lambda$.

The results are presented in Fig.~\ref{fig_over}, together
with a TP-AGB sequence computed without the third dredge-up 
(black triangles). 
From the intersections with the bunch of lines we can read out the
corresponding values of the degree of overlap $r$, which is found to
decrease slowly from roughly $0.8$ to $0.4$.

As for the PDCZ abundances, 
we notice that, after the first pulses, 
the curves tracing the evolution of the intershell 
abundances at constant $r$,  run almost
parallel at increasing core mass. For instance, at 
$M_{\rm c}=0.65\, M_{\odot}$, passing from $r=0.8$ to $r=0.0$ the 
$^{12}$C ($^{16}$O) concentration decreases from $\simeq 41\,\%$ 
($\simeq 2.7\, \%$) to $\simeq 17\,\%$ ($\simeq 0.5\, \%$).
The relative change with $r$ appears larger for  $^{16}$O 
($approx$ a factor of six) than for $^{12}$C ($approx$ a factor of two).

From these results we suggest that the lower $^{16}$O intershell abundances
($< 1\%$) 
reported by \citet{Izzard_etal06}, compared to the standard values ($1\%-2\%$) 
found by \citet{BoothroydSackmann_88b} reflect the larger efficiency of 
the third dredge-up (i.e. higher $\lambda$ hence lower $r$) found in the
more recent works compared to the past.
\begin{figure}  
\resizebox{\hsize}{!}{\includegraphics{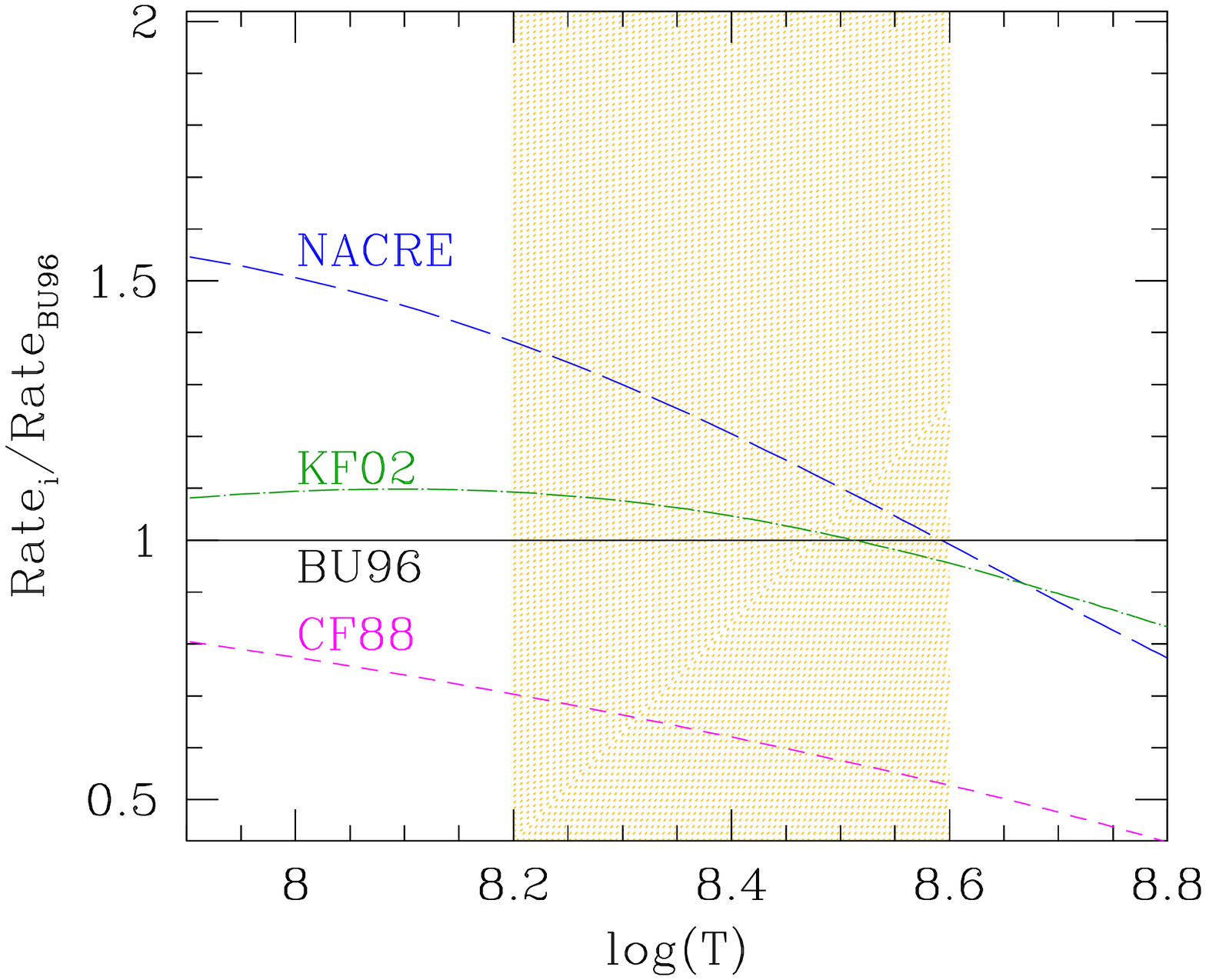}}
\caption
{Comparison of a few versions for the 
$^{12}{\rm C}(^4{\rm He},\gamma)^{16}{\rm O}$ rate, taken from 
the JINA reaclib database \citep{Cyburt_etal10}.
They correspond to \citet[][CF88]{CaughlanFowler_88}, 
\citet[][NACRE]{Angulo_99}, \citet[][BU96]{Buchmann_96}, 
\citet[][KF02]{Kunz_etal02}. We plot the ratio of each rate relative to
BU96, which is our default choice. The hatched area corresponds 
to the relevant range of the temperature attained  
at the bottom of the PDCZ (see Fig.~\ref{fig_pdczfit}). }
\label{fig_rate_oxy}
\end{figure}

\begin{figure}  
\resizebox{\hsize}{!}{\includegraphics{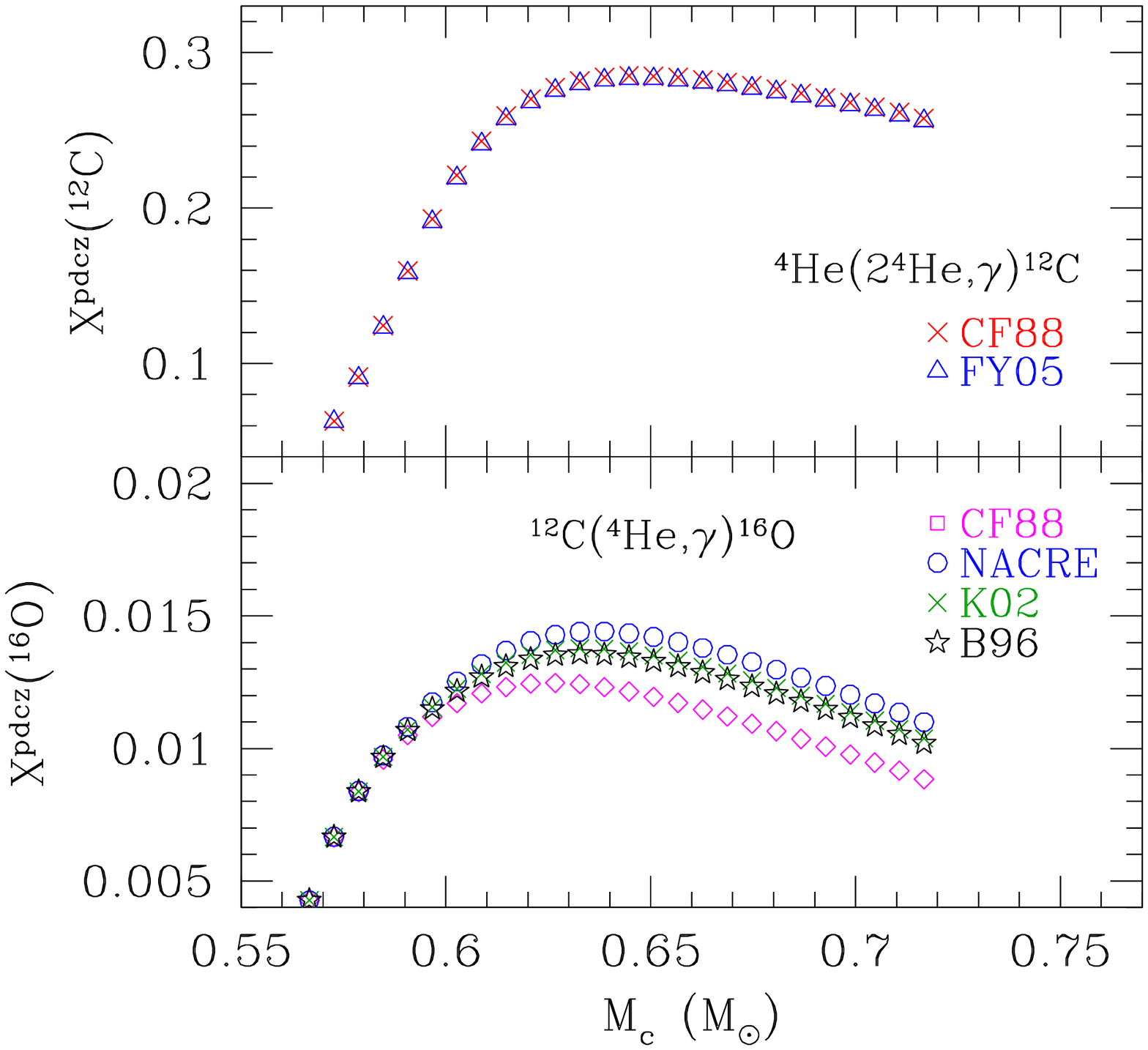}}
\caption
{Dependence of the intershell abundances of $^{12}{\rm C}$ 
(top panel) and 
$^{16}{\rm O}$ (bottom panel) on the nuclear reaction rates
 \reac{He}{4}{2\,^{4}He}{\gamma}{C}{12}, and 
\reac{C}{12}{^{4}He}{\gamma}{O}{16}, respectively.
Labels in the top panel stand for the    
\citet[][CF88]{CaughlanFowler_88} and the \citet[][FY05]{Fynbo_etal05} rates,
while labels in the bottom panel are the same as in Fig.~\ref{fig_rate_oxy}. 
The calculations refer to the same TP-AGB model with 
($M_{\rm i}=2.6,\,Z_{\rm i}=0.017$), as in Fig.~\ref{fig_over}.}
\label{fig_rate}
\end{figure}

\begin{figure}  
\resizebox{\hsize}{!}{\includegraphics{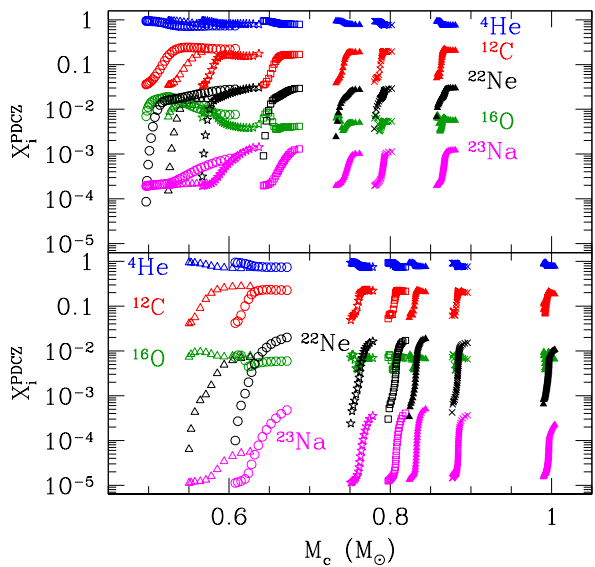}}
\caption
{Predicted intershell abundances as a function of the core mass 
overt the entire TP-AGB evolution for a few models with 
various choices of the initial mass ($1.2\,M_{\odot}$: empty triangles; 
$2.0\,M_{\odot}$: empty circles; $2.6\,M_{\odot}$: stars; 
$3.0\,M_{\odot}$: empty squares; $4.0\,M_{\odot}$: crosses; 
$2.0\,M_{\odot}$: filled  triangles.
The initial metallicity is $Z_{\rm i}=0.017$ (top panel) and
$Z_{\rm i}=0.001$ (bottom panel).}
\label{fig_xpdcz}
\end{figure}

\paragraph*{Dependence on the nuclear reaction rates.}
We have investigated the robustness of the ``standard'' intershell 
composition against reasonable changes in two key nuclear reaction rates,
namely $^4{\rm He}(2\,^4{\rm He}, \gamma)^{12}{\rm C}$ and 
$^{12}{\rm C}(^4{\rm He}, \gamma)^{16}{\rm O}$.
A few versions for the latter rate are compared in Fig.~\ref{fig_rate_oxy}.
The results for the PDCZ abundances of  
$^{12}{\rm C}$ and $^{16}{\rm O}$ are shown in Fig.~\ref{fig_rate}.
There is an almost perfect overlap of the $^{12}$C predictions obtained
with the \citet{CaughlanFowler_88} and the \citet{Fynbo_etal05} rates.
This is not surprising since the two versions are quite  similar 
(with a relative difference always below $1\%$)  in the temperature range
of interest for the pulse nucleosynthesis, i.e. $2\times 10^8 {\rm K} \la T  \la 4 \times 10^8 {\rm K}$.

The results for $^{16}$O exhibit a somewhat larger dependence, but still modest,
on the assumed $^{12}{\rm C}(^4{\rm He}, \gamma)^{16}{\rm O}$.
A comparison of four rates for this reaction is displayed in Fig.~\ref{fig_rate_oxy}.
In the relevant temperature range the largest difference reaches roughly a factor of $2$
between the \citet{CaughlanFowler_88} and the \citet{Fynbo_etal05} rates, while
the variation in  the intershell abundance of 
$^{16}$O remain quite small, $\approx 10\%$.  
The rather low sensitivity of the $^{16}$O abundance PDCZ to significant 
 changes of the $^{12}{\rm C}(^4{\rm He}, \gamma)^{16}{\rm O}$ rate was already 
noticed by \citet{BoothroydSackmann_88b} and is essentially explained by the 
fact that the proper temperature conditions are kept for too short a time to allow
a sizable conversion of $^{12}{\rm C}$ into $^{16}$O.

\paragraph*{Dependence on stellar mass and metallicity.} 
Figure~\ref{fig_xpdcz} shows the evolution of the final PDCZ abundances
left after each thermal pulse, during the entire TP-AGB evolution of 
models with  a few values of initial stellar masses and two choices
of the initial metallicity $Z_{\rm i}=0.017$ and $Z_{\rm i}=0.001$.
As already mentioned, our predictions are essentially consistent with the recent results from
full stellar models without overshooting applied to the PDCZ
\citep{Mowlavi_99a, Karakas_etal02, Izzard_etal04}. In particular the 
$^{12}$C abundance evolves towards an asymptotic value of $\simeq 20\%$, 
independent of mass and metallicity, while in most cases
the $^{16}$O abundance settles down around a value 
of $\simeq 0.005-0.008$, in any case lower than $2\%$ reported by \citet{BoothroydSackmann_88b}.

The abundance of $^{22}$Ne is nearly always  larger than that of $^{16}$O, reaching up
to $\approx 2\% -3 \% $ in models with $Z_{\rm i}=0.017$, while lower values up to 
$\approx 1\% -2 \% $ are attained for  $Z_{\rm i}=0.001$. 
However we note that, relative to its value at the first TP,  the PDCZ 
concentration of $^{22}$Ne shows a larger increase in lower metallicity models,
while at larger metallicity the increment is by one order of magnitude at most.
This result is likely related to the fact that at lower metallicity we
expect a more efficient enrichment of primary $^{12}$C, hence of the
total CNO abundance, in the envelope caused by the third dredge-up.
In this way the synthesis of $^{22}$Ne is favoured, as it is the end product
of a chain of $\alpha$-capture reactions that just start with $^{14}$N, 
the most abundant product of the CNO cycle
(after $^{4}$He) operating in the H-burning shell.

A similar trend characterizes the evolution of the $^{23}$Na intershell  
abundance, which depends on the proton capture reactions occurring in the
H-burning shell during the quiescent interpulse periods.
High-metallicity models show, in general, higher values of  $^{23}$Na
ingested in the PDCZ, up to $\simeq 10^{-3}$, but the relative increase
over the TP-AGB evolution is larger in low-metallicity models. 
 \begin{figure}  
\resizebox{\hsize}{!}{\includegraphics{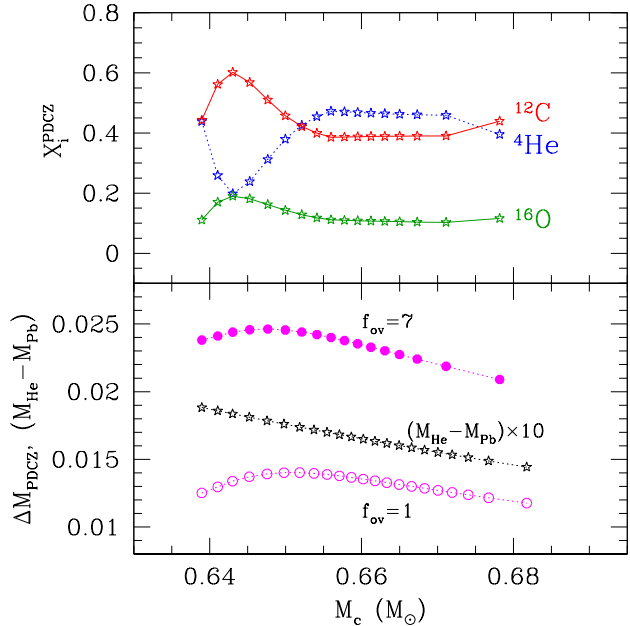}}
\caption
{{\em Top panel}: PDCZ abundances as a function of the core mass at each TP
of a model with initial mass $M_{\rm i}=3\,M_{\odot}$ and metallicity 
$Z_{\rm i}=0.02$.
Predictions are obtained with a parameter $f_{\rm ov}=7$ that mimics the 
inclusion of convective overshoot at the base of the PDCZ. This value of
$f_{\rm ov}$ allows to nicely reproduce the results of detailed calculations
by \citet{Herwig_00} for the same $(M_{\rm i}, Z_{\rm i})$ combination.
{\em Bottom panel}:  Mass of the PDCZ
at its maximum extension predicted with $f_{\rm ov}=1$
(empty circles), and with $f_{\rm ov}=7$ (filled circles).
The amount of helium burning products ingested by the PDCZ, 
corresponding to the mass difference
$(M_{\rm He}-M_{\rm Pb})$, multiplied by a factor of ten, is also shown. 
All masses are in solar units.
See the text for more explanation.}
\label{fig_herwig}
\end{figure}
\paragraph*{Dependence on overshoot.}
The scheme depicted in Fig.~\ref{fig_pdcz} for the PDCZ can be easily modified 
to account for overshoot applied to the  base of the convective pulse.
As a test case, we have considered the results obtained by \citet{Herwig_00},
who applied an exponential diffusive overshoot at the convective boundaries of the 
PDCZ. One major consequence is a depletion of helium and enhancement of carbon and
oxygen in the intershell abundance distribution.
Typical abundances (by mass) are $0.4 -0.5$ for $^{12}$C, $0.15-0.20$ for  $^{16}$O, 
and $0.30-0.40$ for $^{4}$He, obtained by \citet{Herwig_00} 
with his calibrated overshoot parameter.

We underline that in our model the physical structure of the PDCZ is described 
via analytic fits to the results of full TP-AGB model (see Sect.~\ref{ssec_pdcz}), 
so that we cannot perform physical tests of stability against convection at the
borders of the convective intershell. Nonetheless, we can simulate the effect of 
different prescriptions with the aid of a simple parametric approach. 
To mimic the effect of overshoot applied to 
the PDCZ boundaries, we shift inward the mass coordinate 
of its bottom, adopting the parametrization:
\begin{equation}
  M_{\rm Pb}^{\rm oversh} = M_{\rm He} - f_{\rm ov} (M_{\rm He}-M_{\rm Pb})\,,
\end{equation}
where $f_{\rm ov}\ge 1$ is an adjustable factor.
For $f_{\rm ov}=1$ we recover 
the typical intershell chemical composition that 
is predicted by full TP-AGB models when using the 
Schwarzschild criterion, while the effect of convective overshoot 
is simulated adopting $f_{\rm ov} >1$.  As mentioned by  \citet{Herwig_00}, 
there is no noticeable effect of overshoot at the top of the PDCZ, so that we 
keep the mass coordinate $M_{\rm Pt}$ unchanged.
 
The mass difference ($M_{\rm He}-M_{\rm Pb}$), 
derived from full TP-AGB calculations 
as a function of the core mass,  is plotted in Fig.~\ref{fig_herwig}. 
We find that \citet{Herwig_00} results are reasonably well reproduced with 
$f_{\rm ov} \simeq 7$, in terms of both PDCZ mass and abundances 
(see his figures 7d and 11 for the 
$(M_{\rm i}=3 M_{\odot}, Z_{\rm i}=0.02)$ model).
\begin{figure*} 
\begin{minipage}{0.49\textwidth}
\resizebox{0.85\hsize}{!}{\includegraphics{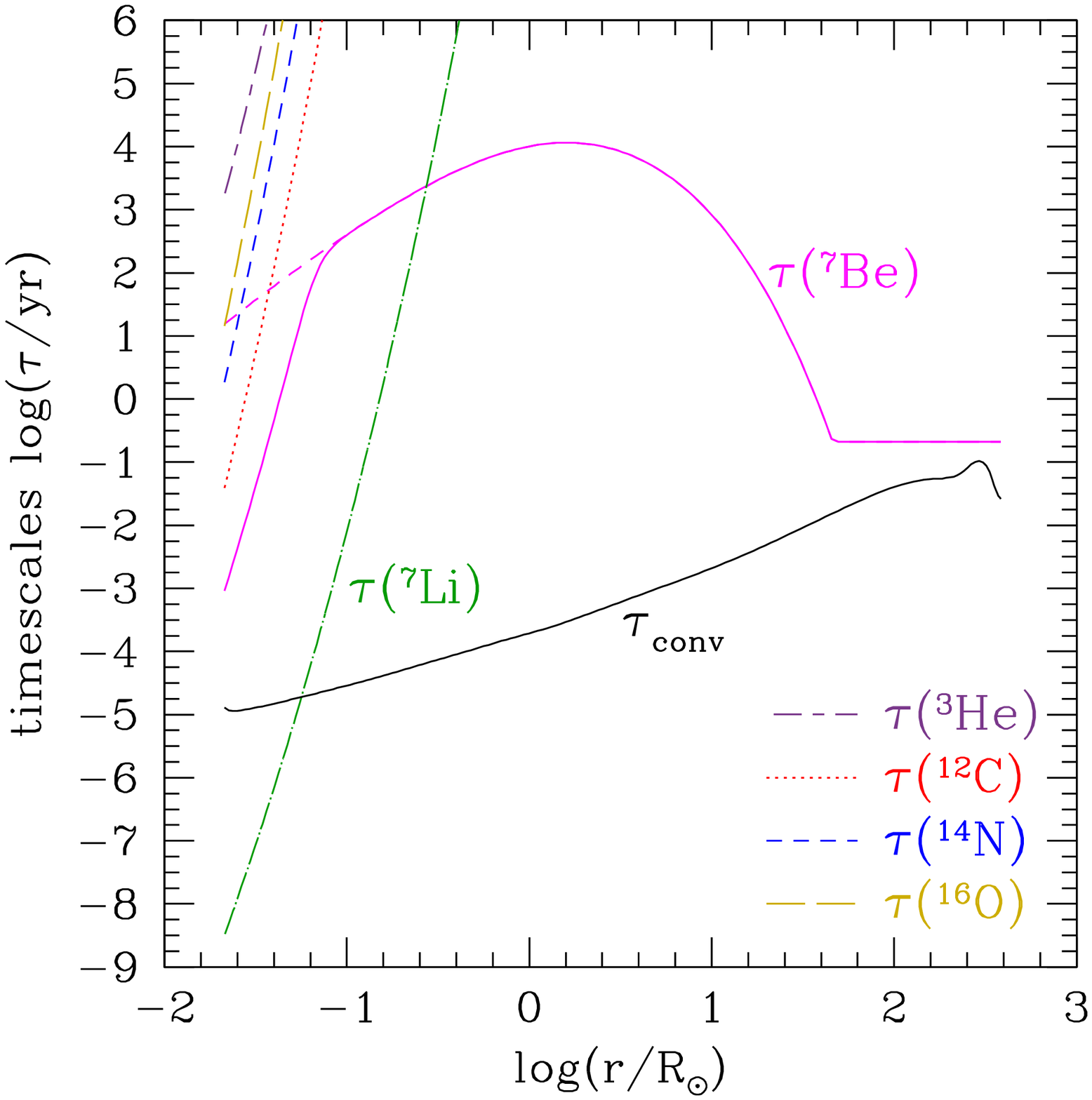}}
\end{minipage}
\hfill
\begin{minipage}{0.49\textwidth}
\resizebox{0.85\hsize}{!}{\includegraphics{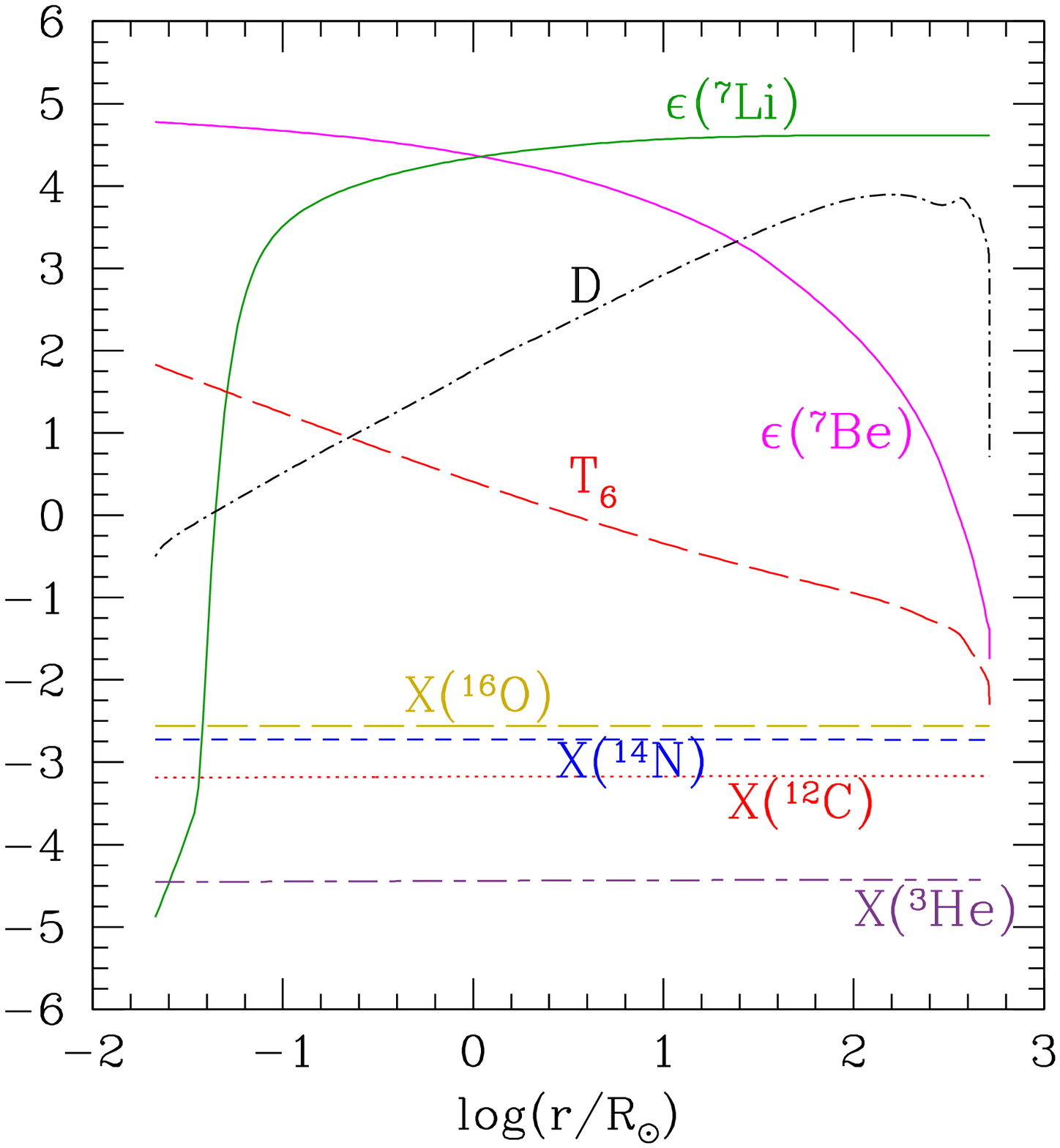}}
\end{minipage}
\caption{Thermodynamic and abundance profiles across the 
deep convective envelope
of an intermediate-mass TP-AGB model experiencing HBB, 
with $M_{\rm i}=5.4\, M_{\odot}$ and $Z_{\rm i}=0.008$,
taken at its maximum $^{7}$Li surface enrichment, when 
$M_{\rm bol}=  -6.48$ and $M_{\rm c} = 0.96\, M_{\odot}$.
{\em Left panel}: Nuclear timescales of a few relevant species against
proton captures and electron captures (only for $^{7}$Be), and 
convective timescale $\tau_{\rm conv}$.
{\em Right panel}: Logarithmic profiles of abundances 
for a few selected species, expressed 
either in mass fraction (for  $^3$He, $^{12}$C, $^{14}$N, $^{16}$O), 
or with the spectroscopic notation $\epsilon_i = \log[n_i/n({\rm H})] + 12$,
where $n$ corresponds to the number density of atoms
(for $^7$Li and $^{7}$Be).
The temperature $T_6=\log(T/10^6 {\rm K})$ and the diffusion coefficient 
$\log(D/R_{\odot}^2\, {\rm yr}^{-1})$, defined by Eq.~(\ref{eq_difco}), are also 
shown.
}
\label{fig_tauli}
\end{figure*}
\begin{figure*}  
\resizebox{0.65\hsize}{!}{\includegraphics{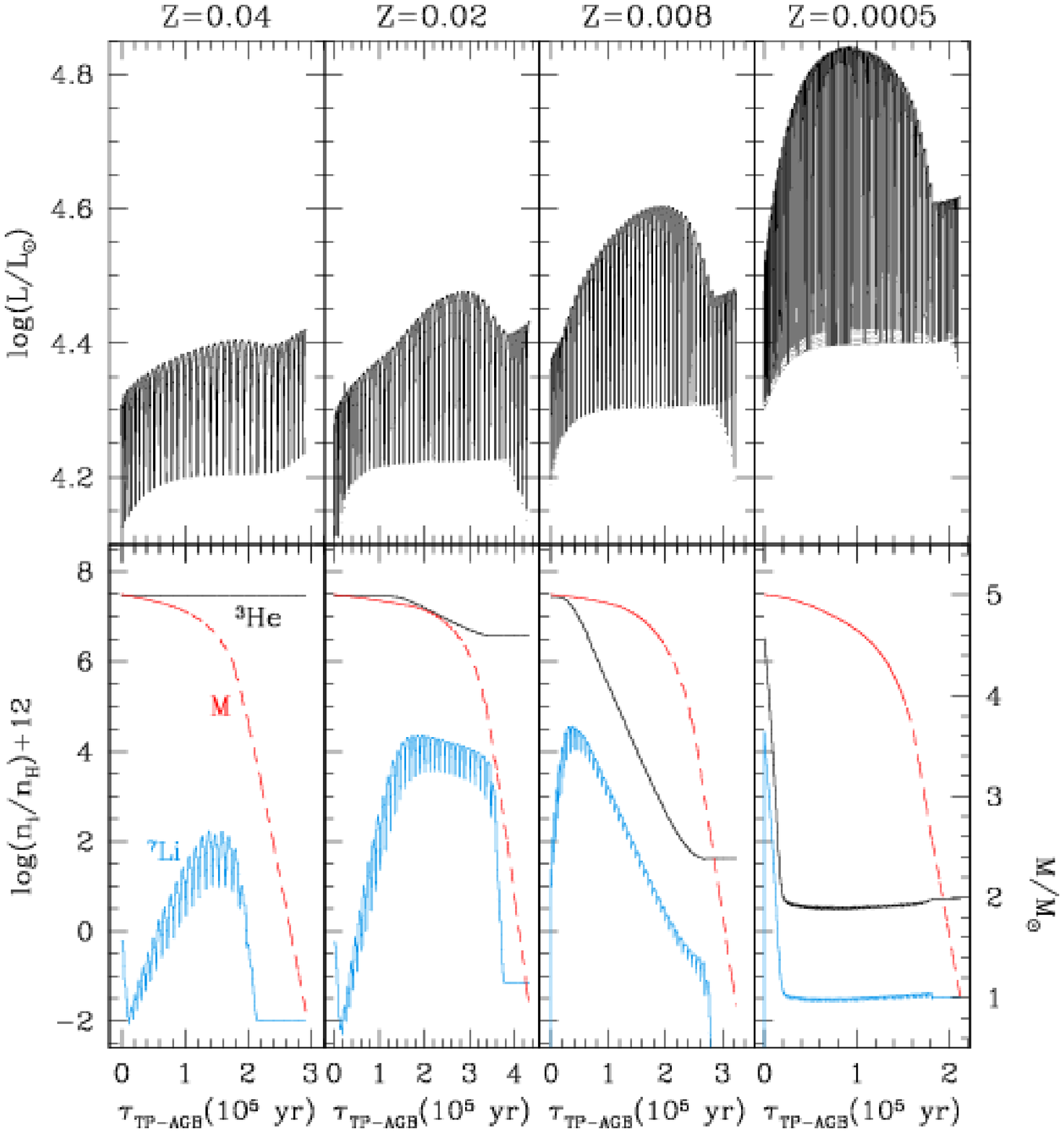}}
\caption{Hot-bottom burning and synthesis of lithium via the 
{\em Cameron-Fowler beryllium transport mechanism} in TP-AGB models 
with initial mass $M_{\rm i}=5\,M_{\odot}$ and varying metallicity.
{\em Top panel}: Evolution of the luminosity during the TP-AGB
phase. Note the larger HBB over-luminosity at decreasing metallicity. 
{\em Bottom panel}: Evolution 
of the surface abundances of $^{7}$Li, $^{3}$He, and of the current stellar 
mass being reduced by stellar winds.}
\label{fig_hbbz}
\end{figure*}

Without pretending to investigate in more detail 
complex aspects of the PDCZ nucleosynthesis (e.g. the formation
of the $^{13}$C pocket is not considered here), we underline 
that this simple parametric approach may be useful
to explore the impact of the overshoot option on the formation and
evolution of carbon stars, by comparing population synthesis simulations 
including overshoot with observations, an important test 
which is still to be done to our knowledge.
An example of  test calculation is discussed in the next 
Sect.~\ref{ssec_hbbres}, and illustrated in Fig.~\ref{fig_hbbnuc}. 
\subsection{Hot-bottom burning nucleosynthesis}
\label{ssec_hbbres}
Figure~\ref{fig_tauli} demonstrates the importance of including 
{\em a time-dependent convective diffusion algorithm} to treat the  synthesis of lithium 
in intermediate-mass AGB stars with HBB. As thoroughly discussed by 
\citet{SackmannBoothroyd_92}, such an approach is necessary when the usual 
{\em instantaneous mixing}
\footnote{The instantaneous mixing approximation
is based on the assumption $\tau_{\rm conv} \ll \tau_{\rm nuc}$, 
that is the convective timescale,  $\tau_{\rm conv}$, 
is much shorter than the nuclear lifetime $\tau_{\rm nuc}$, such that any
element produced by nucleosynthesis is immediately homogenized all over the convective region. This brings a big simplification 
in nucleosynthesis calculations: nuclear reactions rates 
are mass-averaged throughout the convective region, which can be then treated as a  single radiative zone.}
approximation is no longer valid. 
This is the case for nuclei, like $^7$Li and $^{7}$Be, whose lifetimes may become
shorter or comparable to the convective timescale in some parts of the convective envelope.
The circumstance $\tau_{\rm conv} \approx \tau_{\rm nuc}$  occurs 
 in the inner regions for $^{7}$Li, and in the external layers for $^{7}$Be 
(see Fig.~\ref{fig_tauli}, left panel).
As a consequence, the abundances of these species may vary considerably across
the convective envelope, at variance with the concentrations 
of other nuclei (e.g. $^{3}$He, C, N, O) made homogeneous by the
rapid convective mixing.

In particular, the convective envelopes of 
intermediate-mass AGB stars present the suitable thermodynamic conditions 
to put the  {\em Cameron-Fowler beryllium transport mechanism} \citep{CameronFowler_71} 
at work: $^7$Li is efficiently produced and sustained in the outermost 
layers by electron captures on  $^{7}$Be nuclei until either
 the reservoir of $^{3}$He (involved in the reaction 
\reac{He}{4}{^{3}He}{\gamma}{Be}{7}) is exhausted, or 
HBB is extinguished due to envelope ejection by stellar winds.

The model displayed in Fig.~\ref{fig_tauli} shows   
the envelope structure of a TP-AGB star
with  $M_{\rm i} = 5.4,\, Z_{\rm i}=0.008$, that may be considered  as 
representative of the most luminous M giants in the Large Magellanic Cloud.
The structure is  taken at the maximum surface Li enrichment corresponding to
$\epsilon(^7{\rm Li}) \simeq  4.6$,  and $M_{\rm bol} \simeq -6.5$, 
in nice agreement with  the luminosities and the 
highest measured values of Li in the LMC super-rich Lithium stars 
\citep{SmithLambert_89, SmithLambert_90, Smith_etal95}.
Note the mirror behaviours of $^{7}$Be  and $^{7}$Li abundances: 
towards the surface $^{7}$Li is efficiently produced by electron captures on $^{7}$Be
nuclei. 
\begin{figure*} 
\begin{minipage}{0.49\textwidth}
\resizebox{\hsize}{!}{\includegraphics{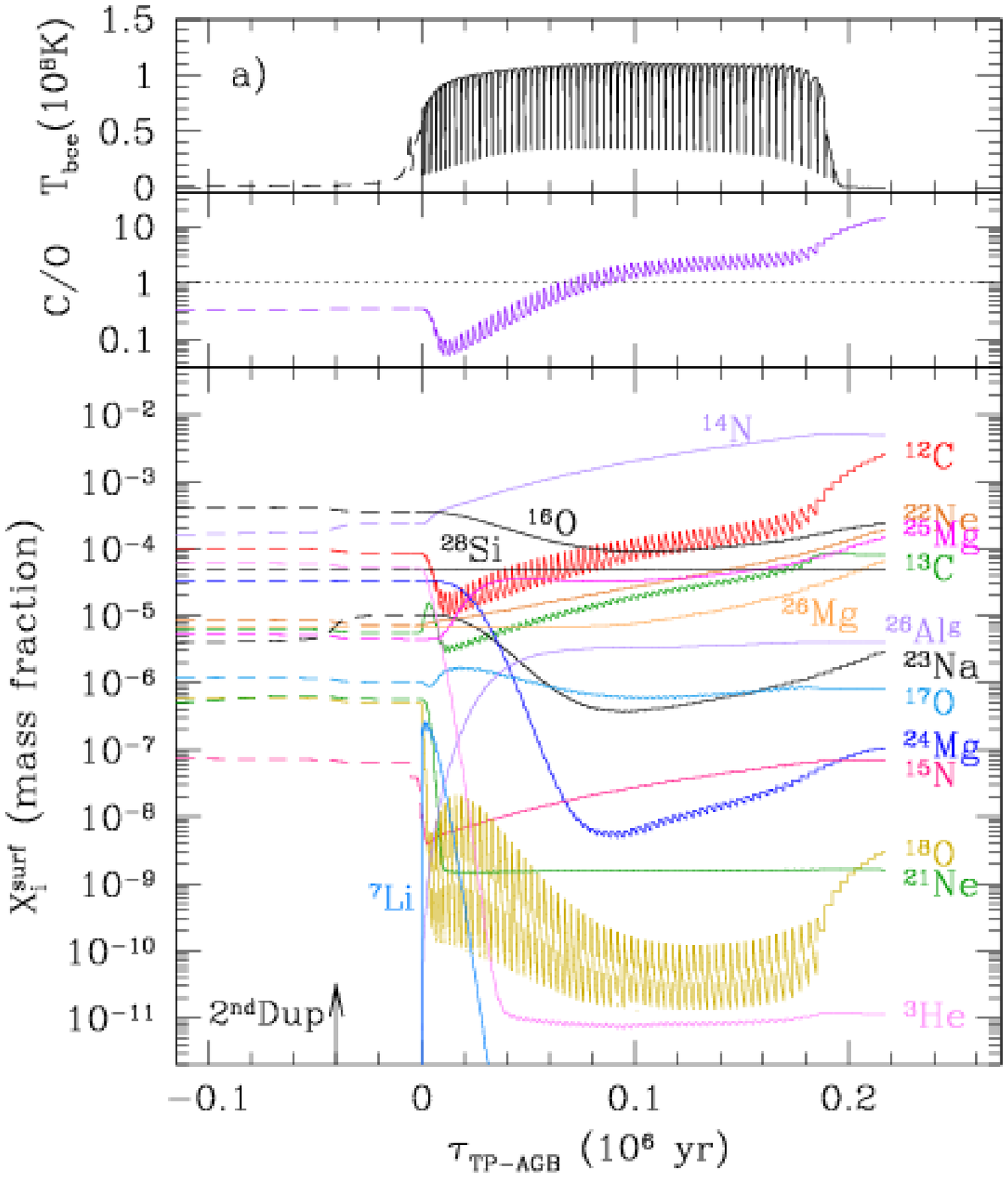}}
\end{minipage}
\begin{minipage}{0.49\textwidth}
\resizebox{\hsize}{!}{\includegraphics{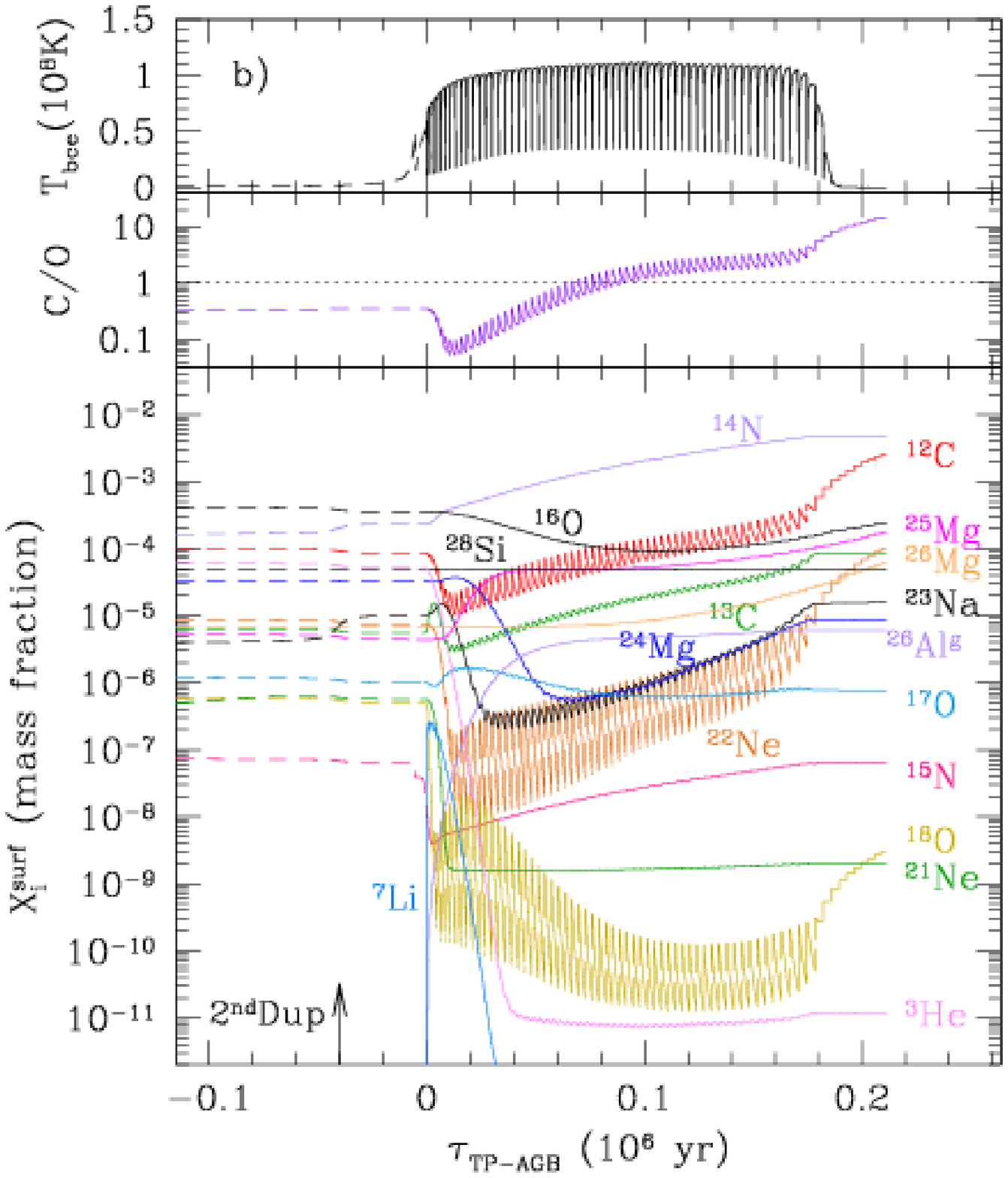}}
\end{minipage}
\vfill
\begin{minipage}{0.49\textwidth}
\resizebox{\hsize}{!}{\includegraphics{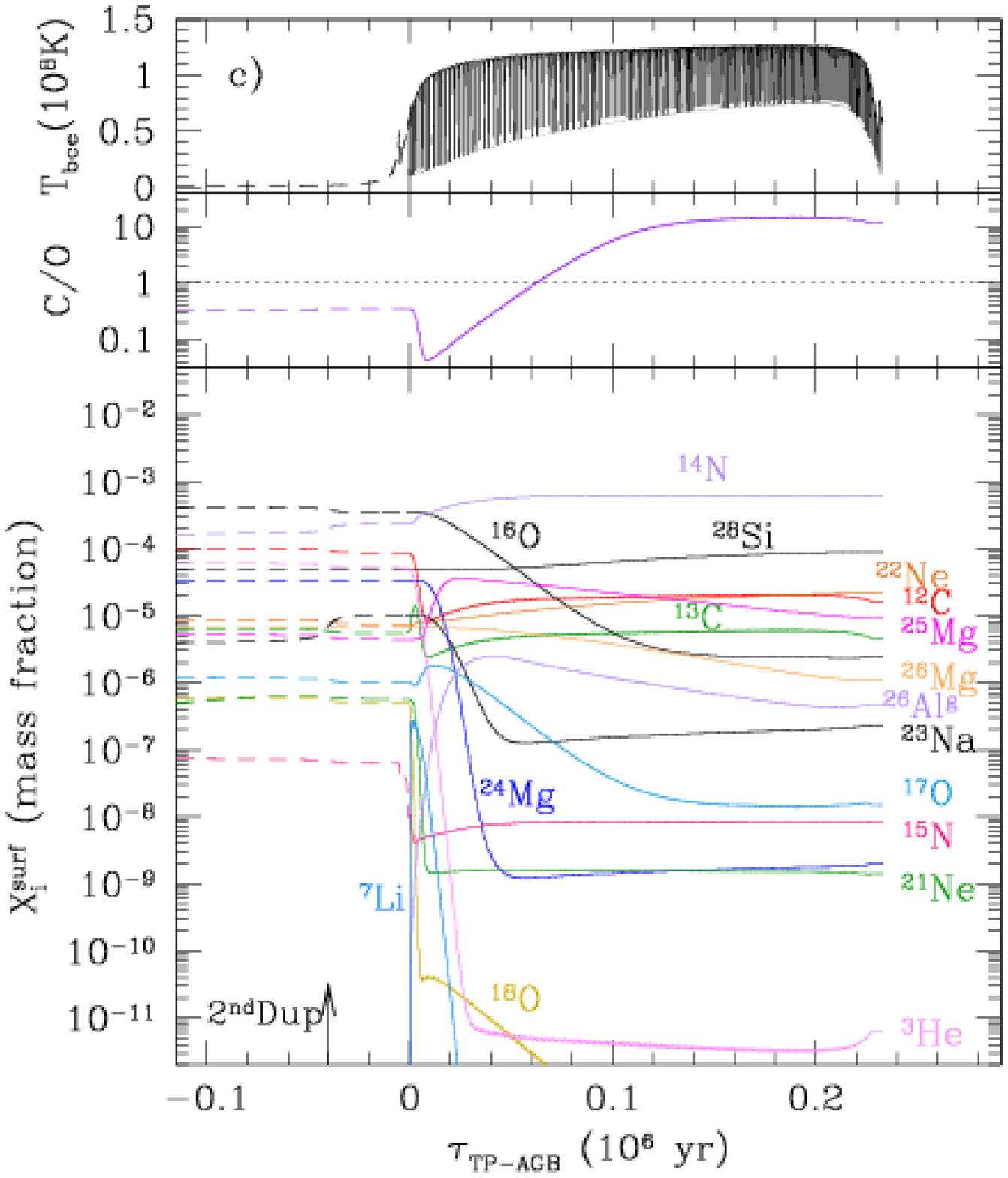}}
\end{minipage}
\begin{minipage}{0.49\textwidth}
\resizebox{\hsize}{!}{\includegraphics{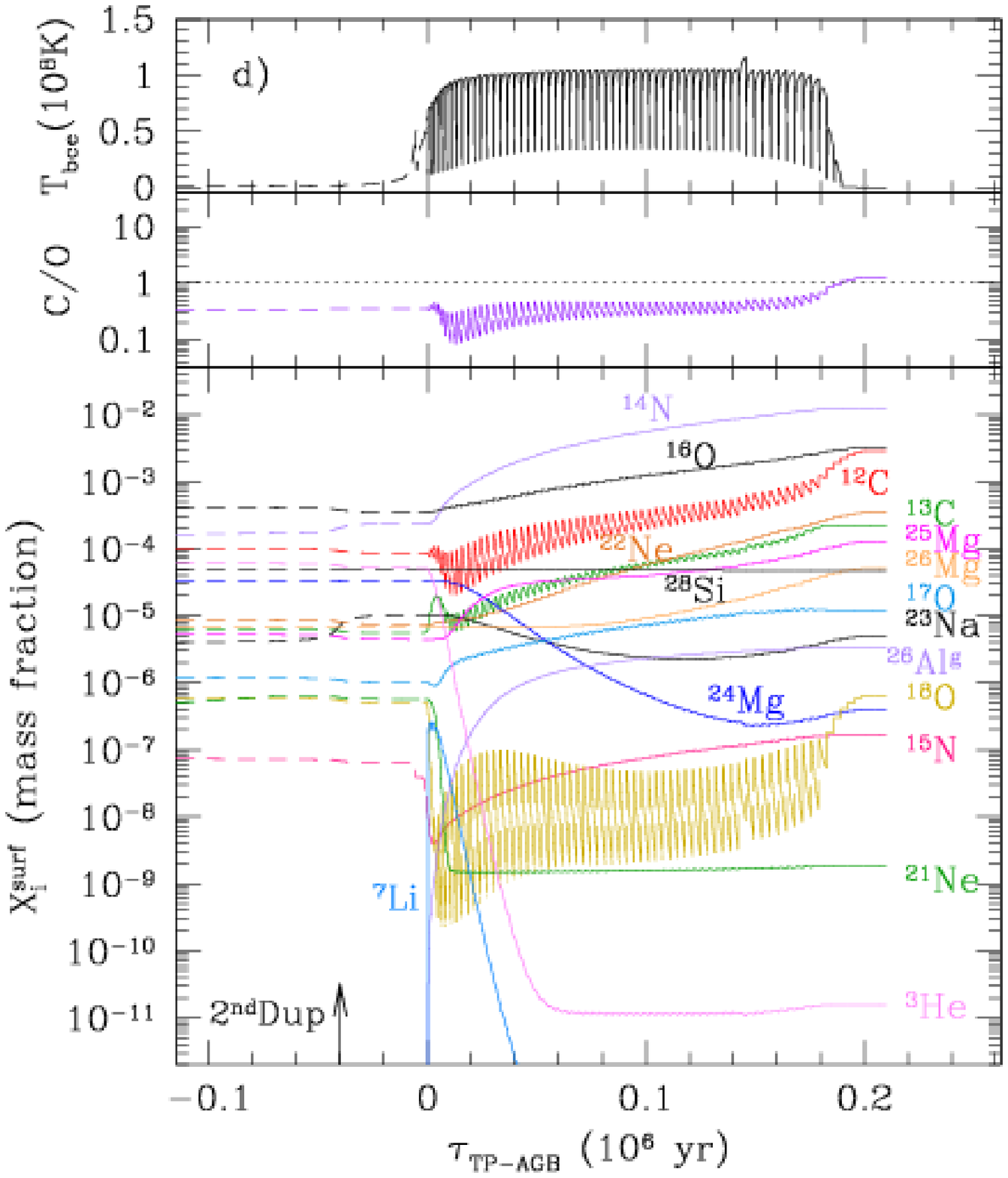}}
\end{minipage}
\caption{Evolution of the temperature at the base of the convective 
envelope, surface C/O ratio and elemental abundances  
during the whole TP-AGB phase of a 
$M_{\rm i}=5\,M_{\odot}, Z_{\rm i}=0.001$ model (solid lines) 
computed with the \texttt{COLIBRI} code.  
The nucleosynthesis of all species is  coupled in time and in 
space with a diffusive description 
of convection. Time is counted since the first TP. 
We show also a  portion of the previous evolution (negative times) 
during the early-AGB predicted by  the \texttt{PARSEC} stellar evolution 
code (dashed lines).  The arrow indicates the approximate
stage at which the second dredge-up takes place.
{\em Panel a)} corresponds to the reference model computed with the default
set of prescriptions, while panels b), c), and d) show the results 
obtained changing selected parameters.
{\em Panel b}: different rates for the nuclear reactions 
\reac{Ne}{22}{p}{\gamma}{Na}{23}, 
\reac{Na}{23}{p}{\gamma}{\,^4He + ^{20}\kern-2.0pt{Ne}}{}, and  
\reac{Na}{23}{p}{\gamma}{Mg}{24} (corresponding to the ``ths8'' version of
the JINA REACLIB database).
{\em Panel c}: suppression of the third dredge-up ($\lambda=0$).
{\em Panel d}: carbon- and oxygen-enhanced chemical composition of the
intershell.
Note that the nucleosynthesis is not computed via  
a post-process technique 
(i.e. assuming a fixed temperature and density stratification), but in all cases the  
chemical and thermodynamic structure of the envelope is solved at each time
step throughout the TP-AGB phase.
}
\label{fig_hbbnuc}
\end{figure*}

Figure~\ref{fig_hbbz} compares the evolution of luminosity 
and surface $^{7}$Li abundance   
in TP-AGB stars with the same initial mass of $5 M_{\odot}$ but different metallicities.
A few points are worth noting.
Since at decreasing $Z$ higher temperatures at the base of the envelope 
are reached,  the brightening of stars with HBB along the TP-AGB 
becomes steeper at lower metallicity, so that  
the classical Paczy{\'n}ski limit\footnote{In the old-fashion terminology the Paczy{\'n}ski limit, also known as ``AGB limit'', 
corresponds to the maximum luminosity that an AGB star,
 complying with the CMLR, may reach when its core mass
has grown up to the Chandrasekhar limit, $M_{\rm c} \simeq 1.4\, M_{\odot}$. Its 
physical meaning has been dismissed since the prediction of the break-down of the 
CMLR by hot-bottom burning in massive AGB stars.} \citep{Paczynski_70},
at $M_{\rm bol} ~\simeq -7.1$, may be even exceeded, like the 
$M_{\rm i}=5 M_{\odot}, \,Z_{\rm i}=0.0005$ 
model does.
In fact,  because of the break-down of the CMLR in stars with HBB,
the Paczy{\'n}ski limit is no longer a true upper bound to the AGB luminosity
\citep{BloeckerSchoenberner_91, BoothroydSackmann_92}, so that AGB stars brighter than $M_{\rm bol} ~\simeq -7.1$ 
could be effectively be observed with a core mass $M_{\rm c} < 1.4\, M_{\odot}$.

At the same time, the synthesis of lithium is more efficient at lower metallicity due
to the larger amounts of $^{7}$Be produced in the innermost layers
of the envelope by the \reac{He}{4}{^{3}He}{\gamma}{Be}{7} reaction.
But for very high metallicities, e.g. $Z_{\rm i}=0.04$, at which 
the Li production remains quite modest (left-hand side panels of Fig.~\ref{fig_hbbz}),
in the other cases under consideration a maximum value around 
$\log[n(^{7}{\rm Li})/n({\rm H})]+12 \simeq 4-4.5$ is reached, 
that is only moderately dependent on $Z_{\rm i}$.
This limiting value is in full agreement with earlier computations 
by \citet{SackmannBoothroyd_92}, and it is  
the result of the high temperature sensitivity of 
$\tau(^{7}{\rm Be})$ from one side, and of similar temperature conditions 
for the maximum Li synthesis in envelope models, on the other side.

Finally, we note that there should be a limited range
of metallicity for which we expect AGB stars to contribute to the 
lithium enrichment of the interstellar medium. Comparing the trends of the
$^{7}$Li abundance and the current stellar mass  
(bottom panels of  Fig.~\ref{fig_hbbz}), we see that for only models with $Z_{\rm i}=0.02$ 
significant mass loss takes place when the  surface $^{7}$Li is high, 
while at higher and lower metallicities, the ejecta
are practically $^{7}$Li free. In fact at  $Z_{\rm i}=0.04$ the $^{7}$Li synthesis 
is just a small and short-lived event, whereas  at  $Z_{\rm i}=0.008$ and $Z_{\rm i}=0.0005$ the $^{7}$Li 
production is quite efficient but confined to the earliest stages of the AGB evolution,
so that when the super-wind regime of mass loss is attained, practically whole
 $^{7}$Li has been destroyed, following the progressive exhaustion of the
$^{3}$He reservoir. These conclusions are drawn  for a particular set of
stellar models, while a more general analysis should be extended also to other
values of the stellar mass, which will be done a future investigation.

Figure~\ref{fig_hbbnuc} exemplifies the results of the nucleosynthesis calculations
made by \texttt{COLIBRI}  over the entire TP-AGB evolution of a $M_{\rm i}=5.0, \,Z_{\rm i}=0.001$ model,
corresponding to a low-metallicity star experiencing strong HBB.

\begin{figure*}  
\begin{minipage}{0.69\textwidth}
\resizebox{\hsize}{!}{\includegraphics{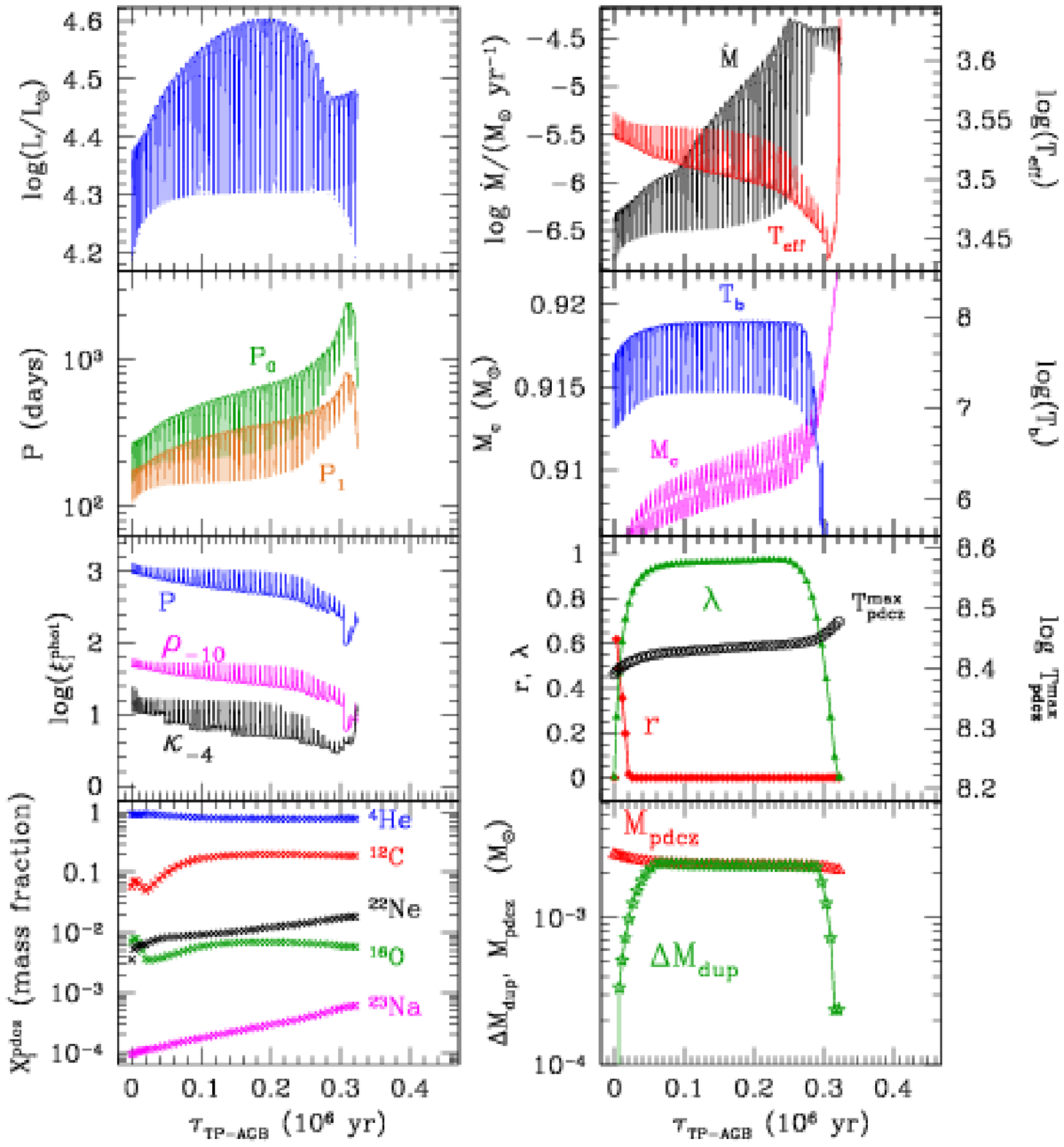}}
\end{minipage}
\hfill
\begin{minipage}{0.30\textwidth}
\caption{Sample output provided by the \texttt{COLIBRI} code. The
  evolution of several quantities, over the whole TP-AGB phase, is
  shown for a $(M_{\rm i}=5\, M_{\odot}, \,Z_{\rm i}=0.008)$ model, which may be
  taken as representative of the most luminous M-giants in the LMC.
  The entire TP-AGB evolution, consisting of 48 thermal pulse cycles,
  has been calculated in roughly $40$ minutes, using a standard
  2.2~GHz CPU. Each quantity is quoted with either [{\em n}] or [{\em a}],
  depending on whether it is predicted 
  by {\em numerical} integrations of envelope models and/or nuclear networks,
  or it is derived from  {\em analytic} fitting relations. 
  From top-left to bottom-right the eight panels show the evolution of
  (i) surface luminosity $L$ [{\em n}], (ii) mass-loss rate $\dot M$ [{\em a}] 
  and effective
  temperature $T_{\rm eff}$ [{\em n}], (iii) fundamental and first-overtone
  pulsation periods $P_0,\, P_1$ [{\em a}], 
  (iv) core mass $M_{\rm c}$ [{\em n}] and
  temperature at the base of the convective envelope $T_{\rm bce}$ [{\em n}], 
  (v) photospheric values of pressure $P$ [{\em n}], density 
$\rho_{-10}=\rho/(10^{-10}\,{\rm gr\, cm}^{-3})$ [{\em n}], 
  and Rosseland mean opacity $\kappa_{-4}=\kappa/10^{-4}$ [{\em n}]:
 (vi) efficiency $\lambda$ [{\em a}]  of the third dredge-up, maximum
  temperature $T_{\rm pdcz}^{\rm max}$ [{\em a}] at the bottom of the
  pulse-driven convective zone, and degree of overlap $r$ [{\em n}] between
  consecutive zones; (vii) inter-shell abundances [{\em n}]; (viii) 
  mass of the pulse-driven convective zone $\Delta M_{\rm pdcz}$ [{\em a}], 
  and dredged-up mass $M_{\rm Dup}$ [{\em a}] at each thermal pulse.}
\label{fig_stru}
\end{minipage}
\end{figure*}
\begin{figure*}  
\begin{minipage}{0.69\textwidth}
\resizebox{0.9\hsize}{!}{\includegraphics{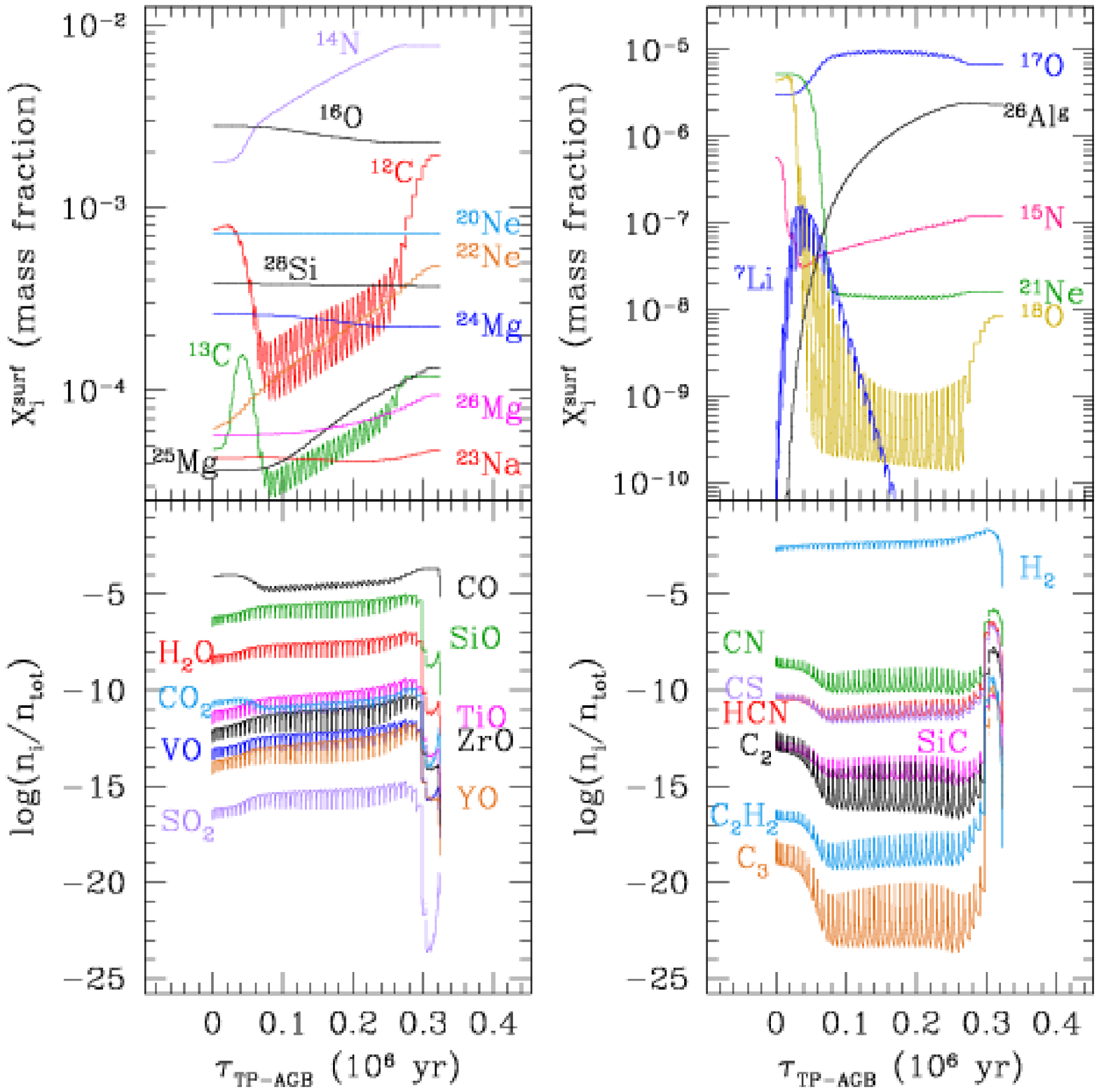}}
\end{minipage}
\hfill
\begin{minipage}{0.3\textwidth}
\caption{The same as in Fig.~\protect{\ref{fig_stru}}, 
but referred to  chemical 
properties at the photosphere. 
{\em Top panels}: evolution of the surface abundances [{\em n}]
of several light elements,  modulated by the occurrence of the 
third dredge-up and HBB. For all species HBB nucleosynthesis is 
followed by coupling the nuclear network to a 
diffusive description of convection.
{\em Bottom panels}: evolution of the 
photospheric concentrations of few molecular species [{\em n}]
(O-bearing species on the left, C-bearing species on the right),
which play a major role in determining the spectral features of AGB stars.
Predictions are obtained with on-the-fly  \texttt{\AE SOPUS} computations 
for the molecular chemistry, consistently coupled with the varying 
envelope abundances. Note the abrupt change in the molecular pattern
over the very last TPs, when the surface C/O increases from below to above
unity as a consequence of the third dredge-up.}
\label{fig_chem}
\end{minipage}
\end{figure*}
 
The nucleosynthesis of the CNO, NeNa and MgAl cycles at low metallicities
is of particular interest, in relation to the possible role of
primordial AGB (and Super-AGB) stars as polluters of the gas out of which 
the old stars,  presently observed in Galactic Globular Clusters (GGCs),   
may have formed \citep{VenturaDantona_08, Pumo_etal08}.
In this so-called self-enrichment scenario the HBB nucleosynthesis in metal-poor AGB
(and Super-AGB) stars could have left its signatures in the
prominent chemical anti-correlations (C-N, O-Na, Mg-Al) currently detected in
GGC stars \citep{Carretta_etal09}.

Indeed, our \texttt{COLIBRI} code may be fruitfully employed to investigate 
the several debated issues about the AGB chemical yields in the low $Z$ regime. 
An example is given in Fig.~\ref{fig_hbbnuc}, where we compare
the results of four sets of  computations obtained with exactly the same set of parameters, 
but varying a few key assumptions that should sample the spread in the predictions
of current TP-AGB models.  The effects on the predicted evolution of several
light elements is remarkable. 
The results of our reference model, computed with the default set of input prescriptions, 
are shown in panel a) of Fig.~\ref{fig_hbbnuc}.

In the first test case (panel b),  we have changed        
the rates of three nuclear reactions, namely  
\reac{Ne}{22}{p}{\gamma}{Na}{23}, 
\reac{Na}{23}{p}{\gamma}{\,^4He + ^{20}\kern-2.0pt{Ne}}{}, and 
\reac{Na}{23}{p}{\gamma}{Mg}{24}, replacing those quoted in Table~\ref{tab_rates} 
with the theory rates labeled ``ths8'' in the JINA REACLIB database, 
that were calculated with the NON-SMOKER
code$^{\rm WEB}$
version 5.0w developed by T. Rauscher\footnote{Online code 
NON-SMOKER$^{\rm WEB}$, version 5.0w and 
higher available at \url{http://nucastro.org/websmoker.html}} and presented 
in \citet{Cyburt_etal10}.
At the typical temperatures $T_{\rm bce}\ga 10^8$ K, 
the ``ths8'' rates are higher than the default ones.
In particular, the ``ths8'' destruction rate 
\reac{Ne}{22}{p}{\gamma}{Na}{23} 
can be larger by up to 3 orders of magnitude!
The large impact is evident by comparing the abundance trends of $^{22}$Ne, $^{23}$Na,
and $^{24}$Mg in panels a) and b).

In the second test case (panel c), we  assume that no  third dredge-up
takes place, i.e. $\lambda = 0$  at each TP, 
 a condition found, for instance, in the recent models of super-AGB stars by
\citet{Siess_10}, where the absence of extra-mixing at the edge 
of the convective boundaries prevents the development of dredge-up episodes.
The evolution of the elemental abundances in the envelope is simply regulated by the 
CNO, NeNa, and MgAl cycles. A very significant depletion of $^{16}$O 
is responsible for the transition to C/O$>1$. 
At the same time we see that, compared to the other models, 
the lack of carbon enrichment in the envelope 
favours the attainment of higher base temperatures $T_{\rm bce}$.

In the third test case (panel d), we mimic the effect of convective 
overshoot at the bottom of the PDCZ following the scheme described 
in Sect.~\ref{ssec_xpdcz}. As a consequence, the intershell abundance distribution 
becomes carbon- and oxygen-enhanced 
compared to the classical composition, resembling the findings by \citet{Herwig_00}
(see Fig.~\ref{fig_herwig}). The differences with respect to the standard 
model shown in panel a) are sizable. The enrichment of $^{16}$O due to the  
third dredge-up prevails over the its destruction by HBB, producing a continuously 
increasing surface abundance of $^{16}$O. The C/O ratio remains lower than one 
 for most of the TP-AGB evolution. Moreover, we note 
that the large increase of the metallicity due to the very efficient  
third dredge-up contributes to reach lower temperature
$T_{\rm bce}$.
\section{Closing remarks}
\label{sec_finalsum}
\subsection{Summary of \texttt{COLIBRI}'s features}

In this paper we have presented the main improvements and novelties characterizing  
the \texttt{COLIBRI} code for the computation of the TP-AGB phase. 
They are briefly recalled below.

Compared to purely synthetic TP-AGB codes,  \texttt{COLIBRI} 
relaxes a significant part of their analytic formalism in favour of a detailed
physics which, applied to a complete envelope model, allows to predict self-consistently: 

\begin{itemize}
\item the {\em effective temperature}, and more generally the convective 
envelope and atmosphere structures, suitably coupled to the changes in 
the surface chemical abundances and gas opacities;
\item the {\em CMLR and its possible break-down due to 
the occurrence of HBB} in the most massive AGB stars, by 
taking  properly into account the nuclear energy generation 
in the H-burning shell and in the deepest layers of the convective envelope;
\item the {\em HBB nucleosynthesis} via the solution of a complete nuclear network 
coupled to a diffusive description of mixing, in which the current 
stratifications of temperature and density are derived from integrations
of complete envelope models;
\item the {\em intershell abundances} left by each thermal pulse
 via the solution of a complete nuclear network applied to a simple model 
of the pulse-driven convective zone;
\item the {\em onset and quenching of the third dredge-up},
with a temperature criterion that is tested, at each thermal pulse, with the
aid of envelope integrations at the stage of the post-flash luminosity 
peak. 
\end{itemize}

At the same time \texttt{COLIBRI} pioneers new techniques in the
treatment of the physics of stellar interiors, not yet adopted
in full TP-AGB models.
Compared to present-day full stellar evolutionary codes,  the  
prerogatives of \texttt{COLIBRI} are related to 1) the computation of the
equation of state and opacities, and 2) computation requirements, as below summarized.
\begin{itemize}
\item \texttt{COLIBRI} is able to perform the first ever {\em on-the-fly} 
accurate computation of the {\em equation of state}
for roughly 800 atoms, ions, molecules,  and of the Rosseland mean {\em opacities} 
throughout the atmosphere and the deep envelope.
This has been accomplished by incorporating  the 
\texttt{\AE SOPUS} code \citep{MarigoAringer_09} and the {\em Opacity Project }  
software package \citep{Seaton05} as  internal routines
of the \texttt{COLIBRI} code.  Avoiding  the preliminary preparation 
of static tables and their subsequent interpolations, 
the new approach assures a complete consistency, step by step, 
of both EoS and opacity with the evolution of the chemical abundances
 caused by  the third dredge-up and HBB.
For the first time we show  the evolution of the photospheric  
molecular concentrations during the TP-AGB phase, and their modulation
driven not only by changes in the chemical compositions 
but also by the periodic occurrence of the TPs.
\item {\em Flexibility and optimized computation requirements}.
\texttt{COLIBRI} is competitive in terms of low computing-time requests.
Tests made with a standard 2.2~GHz CPU processor 
have shown that \texttt{COLIBRI}, on average, computes one complete pulse-cycle in
${0.5-1.0}$ min against the ${60-90}$ min taken  by full evolution
codes, e.g.\ \texttt{PARSEC} \citep{Bressan_etal12}, with 
a gain factor of ${\approx\,100}$.
This characteristic makes \texttt{COLIBRI} an agile tool suitable to carry out
extensive calculations of the TP-AGB evolutionary tracks covering
large and dense grids of stellar masses and metallicities.
\end{itemize}

Figures \ref{fig_stru} and \ref{fig_chem} collect a representative
sample of the most significant quantities that can be predicted 
by \texttt{COLIBRI} 
throughout the entire TP-AGB evolution of a star with given initial mass 
and chemical composition.
The quantity of available information is indeed large, including both
structural and chemical properties. We plan to keep the same 
level of richness also in the stellar isochrones we are going to
construct from the \texttt{COLIBRI} tracks.
\subsection{Ongoing and planned applications}
\label{ssec_future}

It should be mentioned that the present set of TP-AGB models is a
preliminary release, since we are currently working to a global TP-AGB
calibration as a function of stellar mass and metallicity, 
aimed at reproducing a large number of AGB observables at
the same time (star counts, luminosity functions, C/M ratios,
distributions of colors, pulsation periods, etc.) in different star
clusters and galaxies.  Since the calibration is still ongoing the
current parameters (e.g. efficiency of the third dredge-up 
and mass loss) of the TP-AGB model may be changed in future
calculations.

Anyhow, various tests indicate that the present version of the 
\texttt{COLIBRI} models 
already yields a fairly good description of the TP-AGB phase.
 Compared to our previously calibrated sets 
\citep{MarigoGirardi_07, Marigo_etal08, Girardi_etal10} the new TP-AGB models
yield somewhat shorter, but still comparable,  
TP-AGB lifetimes, and they successfully recover various observational constraints dealing with 
e.g. the Galactic initial--final mass relation (Kalirai et al., in prep.),
spectro-interferometric determinations of AGB stellar parameters
\citep{Klotz_etal13}, the correlation 
between mass-loss rates and pulsation periods, and the trends 
of the effective temperature with the C/O ratio observed in 
Galactic M, S and C stars.  

Further important support comes from the results of our new model for 
the condensation and growth of dust grains in the outflows
of AGB stars \citep{Nanni_etal13}, which has been applied 
to the  \texttt{COLIBRI} TP-AGB tracks. 
The results are extremely encouraging as they are found to nicely reproduce other independent 
sets of key observations, i.e. the correlation between 
expansion velocities and mass-loss rates/pulsation periods of Galactic AGB stars.
\section*{Acknowledgments}
It is a pleasure to thank Julianne Dalcanton  and Luciana Bianchi for 
their strong encouragement to this work,  Phil Rosenfield and Marco Gullieuszik  
for their contribution to test the preliminary versions 
of the new TP-AGB tracks. Warm thanks go to Anita and Alessio 
for having inspired the name of the code.
We acknowledge financial support from contract ASI-INAF n.~I/009/10/0, 
and from {\em Progetto di Ateneo 2012}, University of Padova, 
ID: CPDA125588/12.

\appendix
\section{Fitting relations}
\label{app_fit}
\subsection{Properties of the pulse-driven convection zone}
Here we present
relations for characteristic quantities of the PDCZ, based on full TP-AGB calculations 
by \citet{Wagenhuber_96, Karakas_etal02, KarakasLattanzio_07}. All masses are expressed in solar units, $\tau_{\rm pdcz}$
is given in years, and $T_{\rm pdcz}^{\rm max}$ in Kelvin degrees, $Z_{\rm i}$ denotes the initial metallicity.
\begin{eqnarray}
\label{eq_taupdcz}
 \log (\tau_{\rm pdcz}) & = & a_1 + a_2 Z_{\rm i}+ (a_3+a_4 Z_{\rm i}) M_{\rm c}\\
\nonumber
& & +10^{\displaystyle(a_5 + a_6 M_{\rm c} + a_7 \Delta M_{\rm c, nodup})}
\end{eqnarray}

\begin{eqnarray}
\label{eq_tpdcz}
\log (T_{\rm pdcz}^{\rm max}) & =  & (b_1 +b_2 \log(Z_{\rm i}))  + (b_3+b_4 \log(Z_{\rm i})) M_{\rm c}\\
\nonumber
  & & - 10^{\displaystyle( b_5 + b_6  \Delta M_{\rm c, nodup})} 
\end{eqnarray}

\begin{eqnarray}
\label{eq_rhopdcz}
\log(\rho_{\rm pdcz}^{\rm max}) = {\rm max}(3.7, c_1 + c_2 Z_{\rm i} +c_3 M_{\rm c})
\end{eqnarray}

\begin{eqnarray}
\label{eq_mpdcz}
\log (\Delta M_{\rm pdcz}) & =  & d_1 +d_2 M_{\rm c} + d_3 M_{\rm c}^2+d_4 \log(Z_{\rm i})\\
\nonumber
& & -10^{\displaystyle(d_5 + d_6 M_{\rm c} + d_7  \Delta M_{\rm c, nodup})} \\
\nonumber 
& & + d_8 M_{\rm c} \log(Z_{\rm i})
\end{eqnarray}

\begin{eqnarray}
\label{eq_tqti}
x_q =\tau_{\rm q}/\tau_{\rm pdcz}& = &(e_1+e_2 Z_{\rm i})M_{\rm c}+e_3 Z_{\rm i} +e_4 \\
\nonumber
& & -10^{\displaystyle(e_5 M_{\rm c,1} + e_6 \Delta M_{\rm c, nodup})}
\end{eqnarray}
\subsection{The core mass at the $1^{\rm st}$ thermal pulse}
We follow the parametrization proposed by \citet{WagenGroen_98}, 
where  $M$ denotes the stellar mass at the onset of the TP-AGB phase.
All masses are expressed in solar units.
Coefficients are obtained by fitting the predictions from the \texttt{PARSEC} sets 
of stellar models \citep{Bressan_etal12}.
\begin{eqnarray}
\label{eq_mc1}
M_{\rm c,1} &  =  & [-p_{1} (M - p_{2})^2 + p_{3}]f\\
\nonumber
            &     &  + (p_{4}M + p_{5})(1-f)\,, \\
\nonumber
   f &  =  & \left(1 + {\rm exp}^{\frac{M  - p_{6}}{p_{7}}} \right)^{-1} 
\end{eqnarray} 
\begin{table*}
\caption{Fitting coefficients of analytic relations
for a few key properties of the pulse-driven convection zone.}
\begin{tabular}{lllllllll}
\hline
\multicolumn{3}{l}{Eq.~(\ref{eq_taupdcz}): PDCZ duration} \\
\cline{1-3}
\multicolumn{1}{c}{$a_1$} & \multicolumn{1}{c}{$a_2$} & \multicolumn{1}{c}{$a_3$} & \multicolumn{1}{c}{$a_4$} & \multicolumn{1}{c}{$a_5$} & \multicolumn{1}{c}{$a_6$} & \multicolumn{1}{c}{$a_7$} & \multicolumn{1}{c}{}\\
\hline
4.675 & -18.56 & 3.793 & 22.65 & -2.451 & 2.216 & 116.7 & \\
\hline
\multicolumn{8}{l}{Eq.~(\ref{eq_tpdcz}): PDCZ maximum temperature} \\
\cline{1-4}
\multicolumn{1}{c}{$b_1$} & \multicolumn{1}{c}{$b_2$} & \multicolumn{1}{c}{$b_3$} & \multicolumn{1}{c}{$b_4$} & \multicolumn{1}{c}{$b_5$} & \multicolumn{1}{c}{$b_6$} & \multicolumn{1}{c}{} & \multicolumn{1}{c}{}\\
\hline
8.037 & -0.06876 & 0.5697 & 0.07701 & -0.8459 & -22.18 & & \\
\hline
\multicolumn{4}{l}{Eq.~(\ref{eq_rhopdcz}): PDCZ maximum density} \\
\cline{1-4}
\multicolumn{1}{c}{$c_1$} & \multicolumn{1}{c}{$c_2$} & \multicolumn{1}{c}{$c_3$} & \multicolumn{1}{c}{} & \multicolumn{1}{c}{} & \multicolumn{1}{c}{} & \multicolumn{1}{c}{} & \multicolumn{1}{c}{}\\
\hline
4.96 & - 2.4 & - 1.25 & & & & & \\
\hline
\multicolumn{8}{l}{Eq.~(\ref{eq_mpdcz}): PDCZ maximum mass} \\
\cline{1-4}
\multicolumn{1}{c}{$d_1$} & \multicolumn{1}{c}{$d_2$} & \multicolumn{1}{c}{$d_3$} & \multicolumn{1}{c}{$d_4$} & \multicolumn{1}{c}{$d_5$} & \multicolumn{1}{c}{$d_6$} & \multicolumn{1}{c}{$d_7$} & \multicolumn{1}{c}{$d_8$}\\
\hline
-1.134 &  0.2884 & -1.898 & -0.08295 & -2.171 & 1.429 & -21.55 & 0.09189 \\
\hline
\multicolumn{8}{l}{Eq.~(\ref{eq_tqti}): Ratio of the quenching time over PDCZ duration} \\
\cline{1-6}
\multicolumn{1}{c}{$e_1$} & \multicolumn{1}{c}{$e_2$} & \multicolumn{1}{c}{$e_3$} & \multicolumn{1}{c}{$e_4$} & \multicolumn{1}{c}{$e_5$} & \multicolumn{1}{c}{$e_6$} & \multicolumn{1}{c}{} & \multicolumn{1}{c}{}\\
\hline
0.8220 & 0.9602 & 5.481 & - 0.4321 & -0.8632 & -26.23 & &\\
\hline
\end{tabular}
\label{tab_fpdcz}
\end{table*}
\begin{table*}
\caption{Fitting coefficients of Eq.~(\ref{eq_mc1}) for the core mass at the $1^{\rm st}$ thermal pulse.}
\begin{tabular}{llllllll}
\hline
\multicolumn{1}{c}{$Z_{\rm i}$} & \multicolumn{1}{c}{$p_1$} & \multicolumn{1}{c}{$p_2$} & \multicolumn{1}{c}{$p_3$} & \multicolumn{1}{c}{$p_4$} & \multicolumn{1}{c}{$p_5$} & \multicolumn{1}{c}{$p_6$} & \multicolumn{1}{c}{$p_7$}\\
\hline
    0.0005&   9.616573E-02  &   1.300268E+00  &  5.567979E-01&    9.204736E-02 &   5.204188E-01 &    1.947073E+00 &   1.607459E-01 \\
    0.001 &    1.173875E-01 &   1.188889E+00  &  5.505528E-01 &   9.301397E-02 &   5.100448E-01 &   1.954574E+00 &   1.670251E-01  \\
    0.004 &    1.074609E-01 &   1.150773E+00  &  5.389349E-01 &   9.559346E-02 &   4.645270E-01 &   2.170495E+00 &   1.949511E-01  \\
    0.006 &    9.772655E-02 &   1.148381E+00 &   5.347831E-01 &   9.128342E-02 &   4.641443E-01 &   2.254396E+00 &   2.278098E-01  \\
    0.008 &     9.020493E-02 &   1.156664E+00 &   5.318839E-01 &    8.671702E-02 &   4.719326E-01 &   2.319841E+00 &   2.560683E-01  \\
    0.01  &    7.480933E-02 &   1.193024E+00 &   5.300704E-01  &  9.499056E-02 &   4.257837E-01 &   2.365426E+00 &   2.470678E-01  \\
    0.014 &   7.496712E-02 &   1.189756E+00 &   5.286927E-01 &   9.300582E-02 &   4.175395E-01 &   2.375119E+00 &   2.651535E-01 \\
    0.017 &   6.956924E-02 &   1.227015E+00 &   5.275279E-01 &   8.479260E-02 &   4.427424E-01 &   2.477161E+00 &   2.505828E-01 \\
    0.02  &   6.530806E-02 &   1.243030E+00 &   5.269612E-01 &   8.581963E-02 &   4.315992E-01 &   2.459101E+00  &  2.572425E-01 \\
    0.03 &    5.160226E-02 &   1.249103E+00 &   5.268402E-01 &   7.668322E-02 &   4.601484E-01 &   2.516399E+00 &   2.637952E-01  \\
    0.04 &    4.661234E-02 &   1.274814E+00&    5.324125E-01 &   7.903245E-02 &   4.494590E-01 &   2.481670E+00 &   2.438550E-01  \\
    0.05  &   5.827199E-02 &   1.337793E+00 &   5.441922E-01 &   8.204387E-02 &   4.402451E-01 &   2.389034E+00 &    2.424820E-01 \\
\hline
\end{tabular}
\label{tab_mc1}
\end{table*}
\section{Accuracy tests}
\label{sec_dtests}
\subsection{Effective temperature}
\label{ssec_teff}
A fundamental check is to compare our determination of the
effective temperatures, based on envelope integrations 
($T_{\rm eff}^{\rm env}$; the method
is detailed in Sect.~\ref{ssec_envmod}),  against
the results of full stellar models ($T_{\rm eff}^{\rm full}$).

\begin{figure} 
\resizebox{1.0\hsize}{!}{\includegraphics{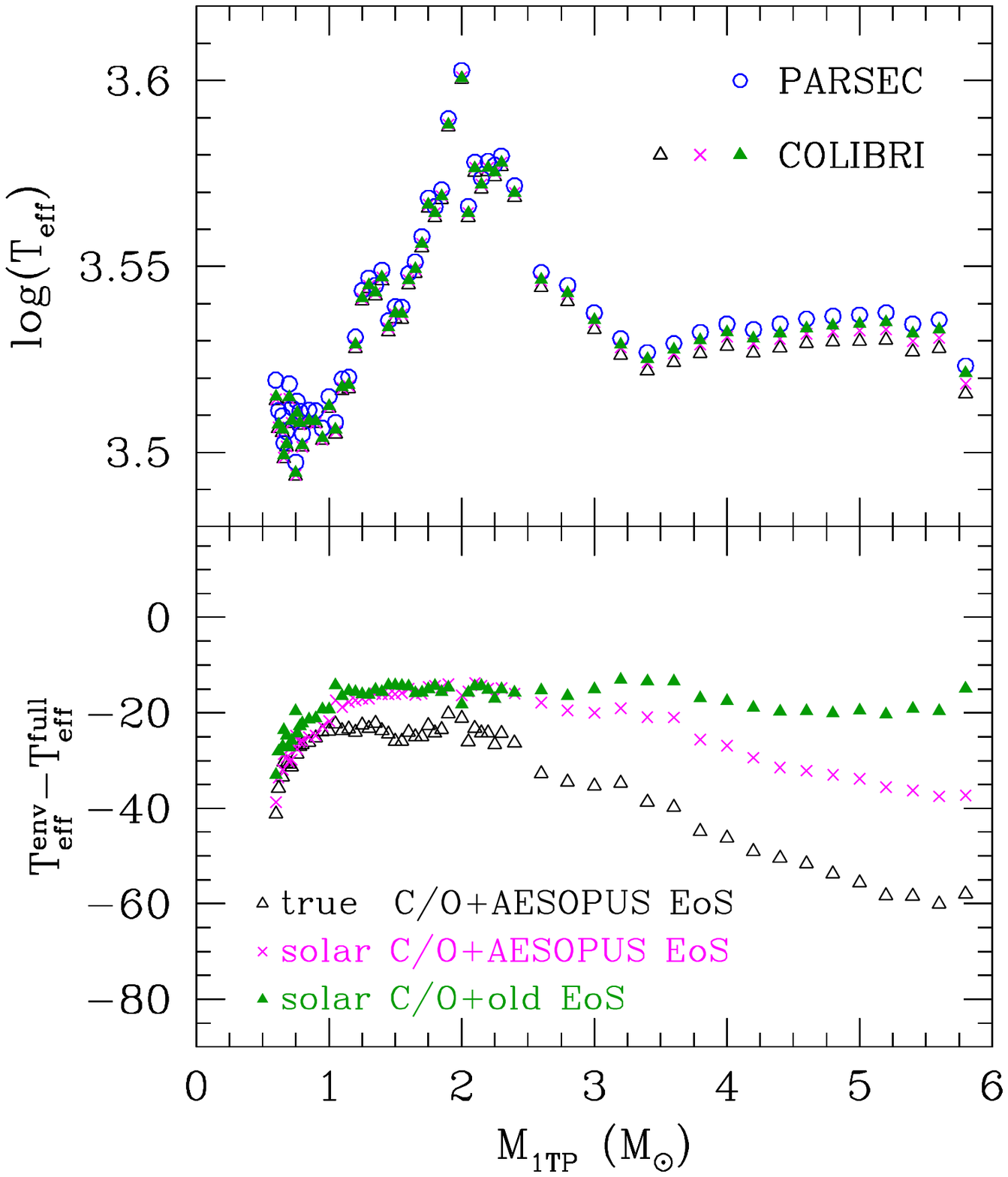}}
\caption{Accuracy tests on the effective temperature.
{\em Top panel}: $T_{\rm eff}$ values 
as a function of the stellar mass at the $1^{\rm st }$ thermal pulse  
for the set with ($Z_{\rm i}=0.01,\, Y_i=0.267$), as predicted 
by \texttt{PARSEC} full stellar models ($T_{\rm eff}^{\rm full}$) 
and by \texttt{COLIBRI} envelope-integration method ($T_{\rm eff}^{\rm env}$) 
for different assumptions of the abundance distribution and the EoS.
{\em Bottom panel}: Differences $T_{\rm eff}^{\rm full} - T_{\rm eff}^{\rm env}$ 
in Kelvin degrees.}
\label{fig_dtefz01}
\end{figure}

In Fig.~\ref{fig_dtefz01} we show the results
for the set of stellar evolutionary tracks 
with initial chemical composition ($Z_{\rm i} =0.01$, $Y_{\rm i}=0.267$),
computed with \texttt{PARSEC} \citep{Bressan_etal12}.
In the top panel we compare directly 
the effective temperatures, $T_{\rm eff}^{\rm full}$ and $T_{\rm eff}^{\rm env}$,
relative to the quiescent pre-flash luminosity maximum at the  $1^{\rm st }$ thermal pulse.
We can already see that the agreement is very good for all stellar masses 
here considered. We also note that $T_{\rm eff}^{\rm env}$ is systematically 
lower than $T_{\rm eff}^{\rm full}$ by a small amount, which appears to 
increase somewhat with the stellar mass.
Considering that part of the differences is likely due 
to unavoidable numerical effects impossible to be disentangled, 
we have also investigated other possible physical causes that may explain 
some systematic trends. 
In particular we have considered the effects due to different descriptions
of the EoS and the opacities in the \texttt{PARSEC} and \texttt{COLIBRI} codes.

In the bottom panel of Fig.~\ref{fig_dtefz01} we zoom in the difference
$T_{\rm eff}^{\rm env} - T_{\rm eff}^{\rm full}$ (in $K$ degrees), as a function of the 
stellar mass. The three sequences are obtained with three 
combinations of the EoS and low-$T$ opacities used in the \texttt{COLIBRI} code.
The lowest sequence (black empty triangles),  showing 
the largest deviations from  \texttt{PARSEC}, 
corresponds to the  $T_{\rm eff}^{\rm env}$ predictions with 
the optimal configuration of all input physics in \texttt{COLIBRI}. Specifically, envelope
integrations have been carried out with both the EoS and the Rosseland mean 
opacities computed with \texttt{\AE SOPUS} on-the-fly according 
to the actual chemical mixture of all elements.

This implies that the molecular chemistry is accurately
solved, exactly  complying with the true surface C/O ratio that characterizes each
stellar model at the onset of the TP-AGB phase. In fact, the surface C/O 
ratio may have decreased, compared to its initial value at the main sequence
(${\rm C/O} < {\rm C/O}_{\rm initial} = ({\rm C/O})_{\odot}\simeq 0.55$ 
for the scaled-solar case under consideration),
as a consequence of the first dredge-up and, in stars with $M > 4\,M_{\odot}$,
because of the second dredge-up.

In contrast, in \texttt{PARSEC} the opacities are derived through
interpolations on pre-computed opacity tables as a function of temperature, density, 
hydrogen abundance, and current metallicity $Z$, while keeping the distribution 
of metals fixed to the initial configuration, 
$X_i/Z=X_{i,\odot}/Z_{\odot}$. In particular this means that that no change
in the C/O ratio is considered, i.e. ${\rm C/O}=({\rm C/O})_{\odot}$ 
is assumed in all opacity tables.

To test the effect produced on the effective temperatures 
by low-$T$ opacities with a fixed chemical partition, we have performed a second run 
of envelope integrations 
setting the metals partition in the \texttt{\AE SOPUS}  chemistry routine frozen 
to the scaled-solar one ($X_i/Z=X_{i,\odot}/Z_{i,\odot}$), as in \texttt{PARSEC}. 
The differences $(T_{\rm eff}^{\rm env} - T_{\rm eff}^{\rm full})$ are now smaller, 
as one can see in Fig.~\ref{fig_dtefz01} comparing  the sequence of 
magenta crosses  with that of black triangles. In this case the temperature differences are  
mostly comprised within $25$ K, and in all cases lower than $40$ K.
The fact the assumed solar C/O ratio is higher than 
the actual values at the $1^{\rm st}$ TP, 
implies that a smaller excess of oxygen atoms, (O-C),  
is available to form the H$_2$O molecule, the most efficient
opacity source at the atmospheric temperatures under consideration.
The effect seems to be somewhat larger at increasing stellar mass.

Finally, we have explored  possible additional EoS effects.
At this stage we cannot obtain a quantitative comparison with respect to
\texttt{PARSEC}, in which the EoS is solved with the 
\texttt{FreeEOS} code\footnote{\texttt{FreeEOS} is a  software package developed by A.W. Irwin, and freely
available under the GPL licence at \url{http://freeeos.sourceforge.net/}}, 
since these  latter is not implemented in our \texttt{COLIBRI} code. 
Anyway, to obtain an order-of-magnitude estimate, we have carried  
out a third run of envelope integrations, 
switching the EoS option from the \texttt{\AE SOPUS} routine 
to an older and simpler EoS description based on \citet{Kippenhahn_etal65}.
We see that now the deviations $T_{\rm eff}^{\rm env} - T_{\rm eff}^{\rm full}$ reduce
further, keeping of the order of $\approx 20\,K$ or lower.
Therefore we may conclude that the EoS treatment may also explain part 
of the differences $T_{\rm eff}^{\rm env} - T_{\rm eff}^{\rm full}$, by
an amount that is comparable to that driven by the opacities.
\begin{figure} 
\resizebox{1.0\hsize}{!}{\includegraphics{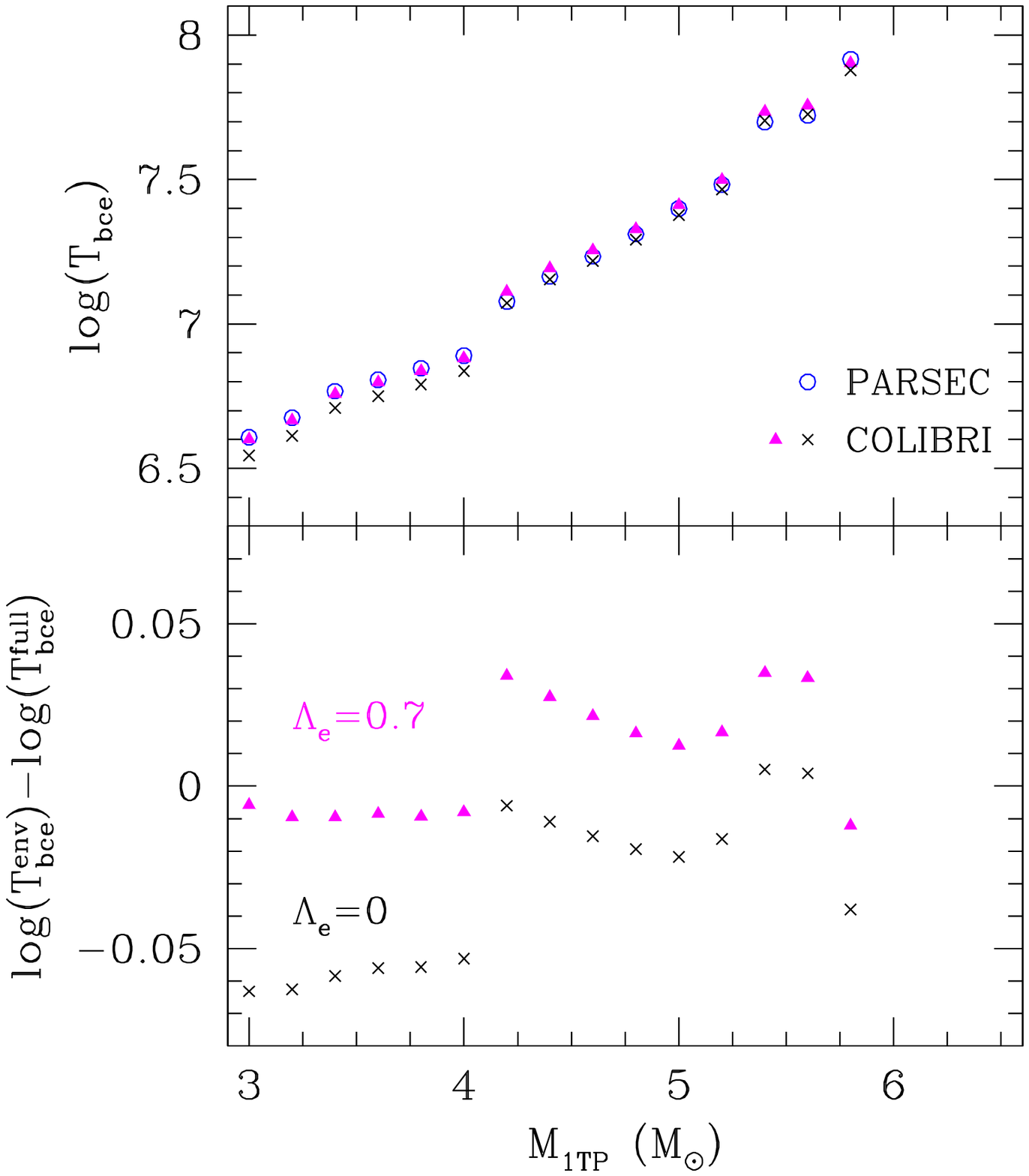}}
\caption{Accuracy tests on the temperature at the
bottom of the convective envelope. 
{\em Top panel}: $T_{\rm bce}$ values a
as a function of the stellar mass at the $1^{\rm st }$ thermal pulse  
for the set with ($Z_{\rm i}=0.01,\, Y_i=0.267$), as predicted 
by \texttt{PARSEC} full stellar models ($T_{\rm eff}^{\rm bce}$) 
and by \texttt{COLIBRI} envelope-integration method ($T_{\rm bce}^{\rm env}$) 
without  and with convective overshooting beyond the formal Schwarzschild
border.
{\em Bottom panel}: Logarithmic difference 
$\log(T_{\rm bce}^{\rm full}) - \log(T_{\rm eff}^{\rm env})$.}
\label{fig_dtbotz01}
\end{figure}
\subsection{Temperature at the base of the convective envelope}
\label{ssec_tbot}
The quantity $T_{\rm bce}$
provides an additional performance test of our envelope-integration method, and
it is particularly relevant for massive AGB models ($M > 4\,M_{\odot}$)
as it measures the efficiency of hot-bottom burning.
 
In full stellar models calculated with \texttt{PARSEC} convective overshoot is applied
to the formal Schwarzschild border of the envelope, with an efficiency 
parameter\footnote{The radial extension of the overshooting region 
is given by $\Lambda_{\rm e}\times H_P$, where $H_P$ is the local pressure
scale height at the Schwarzschild border.} 
$\Lambda_{\rm e}=0.05$ for $M<M01$ and  $\Lambda_{\rm e}=0.7$ for
$M>M02$. The transition masses, with approximate values 
$M01\approx 1.0-1.5\,M_{\odot}$ and $M01\approx 1.5-2.0\,M_{\odot}$, are
 operatively defined in Bressan et al. (2012) and 
depend on chemical composition.  

We apply the same scheme to our envelope integrations and then compare
the predictions for $T_{\rm bce}$ as a function of stellar mass and metallicity.
Results are shown in Fig.~\ref{fig_dtbotz01}.
We have verified that variations in the EoS and opacities, as those discussed
in Sect.~\ref{ssec_teff}, produce almost negligible changes in  $T_{\rm bce}$
for the models under considerations, 
so that we do not show the corresponding results.  

The effect of convective overshoot on $T_{\rm bce}$ is illustrated in 
Fig.~\ref{fig_dtbotz01} for the set with initial chemical composition 
$Z_{\rm i}=0.01, Y_i=0.267$.
As a general rule models with $\Lambda_{\rm e} > 0$ tend to have
higher $T_{\rm bce}$ since the base of the convective envelope penetrates
more deeply inward.
For masses $M<M01$ the differences
in  $T_{\rm bce}^{\rm env}$ remain small
among models with or without overshoot, 
with  $[\log T_{\rm bce}^{\rm env}
(\Lambda_{\rm e}=0.05)- \log T_{\rm bce}^{\rm env}(\Lambda_{\rm e}=0)] 
\la 0.006$, reflecting the little overshoot efficiency adopted in
since this mass range.
In all cases 
$\log(T_{\rm bce}^{\rm full}) - \log(T_{\rm bce}^{\rm env})$ keep positive,
i.e. the envelope-integration method yields somewhat higher temperatures
than full stellar models.

Larger differences in $T_{\rm bce}^{\rm env}$ arise instead for masses
$M>M02$, depending on whether we assume or not convective overshoot.
We see that passing from $\Lambda_{\rm e}=0.7$
to $\Lambda_{\rm e}=0$
in our envelope integrations the differences 
$\log(T_{\rm bce}^{\rm full}) - \log(T_{\rm bce}^{\rm env})$ tend to
become negative, i.e. the envelope-integration method yields lower  
temperatures than full stellar models.
A systematic decrease of $[\log T_{\rm bce}^{\rm env}
(\Lambda_{\rm e}=0.7)- \log T_{\rm bce}^{\rm env}(\Lambda_{\rm e}=0)] 
\simeq 0.03-0.05$ is predicted for these models. 

In general the deviations from the full stellar models are larger than
those for the effective temperatures, with 
$|\log(T_{\rm bce}^{\rm full}) - \log(T_{\rm eff}^{\rm env})|$ reaching up 
to a few hundredths of a dex.
Part of the  reason likely resides in the operative definition
of the convective border and the adopted mass meshing across the envelope.
In our \texttt{COLIBRI} code the classical Schwarzschild border
is determined by the equality between the radiative and adiabatic
temperature gradients, $\nabla_{\rm rad}=\nabla_{\rm ad}$,  and
all physical quantities are derived from interpolation between the last
convective mesh and the first radiative one during the inward envelope 
integration. In \texttt{PARSEC} the  Schwarzschild border is assumed to coincide 
with the last convective mesh, without interpolation in temperature
gradients. 

Limiting to the \texttt{COLIBRI} models with $\Lambda_{\rm e} > 0$, we note 
that larger deviations  from $T^{\rm full}_{\rm bce}$ are found at larger stellar masses 
($M > 4\, M_{\odot}$) where HBB starts to be operative.
Part of these differences are likely related to the arrangement 
of the mesh points across the envelope; in fact the base of the convective envelope
locates inside an extremely thin (in mass) 
region characterised by very steep gradients of all thermodynamic
quantities ($T,P,\rho$, etc.), As a consequence, even small
differences in mass resolution in this region may produce somewhat 
appreciable differences in the thermodynamic profile of the
innermost layers of envelope.

We  conclude that our envelope-integration method
yields a description of the deepest envelope layers which is in satisfactory
agreement with full stellar models, but unavoidable differences exist mainly
due numerical and technical details. The size of such deviations 
are in any case lower than the current differences  between various 
sets of AGB models, the latter reflecting the uncertainties of a still  
ill-defined theory of convection in stars.

\end{document}